\documentclass[useAMS,usenatbib]{mn2e}
\usepackage{graphicx,amsmath,amssymb,ifthen}
\usepackage{verbatim}
\newboolean{astroph}
\setboolean{astroph}{true}

\newboolean{mn2e}
\setboolean{mn2e}{false}

\newboolean{aastex}
\setboolean{aastex}{false}

\newcommand{\Msun}{\ensuremath{\mathrm{M}_\odot}}

\newcommand{\kmps}{\ensuremath{\mathrm{km\ s}^{-1}}}

\newcommand{\Mpc}{\ensuremath{\mathrm{Mpc}}}

\newcommand{\degree}{\ensuremath{^{\circ}}}

\newcommand{\Reff}{\ensuremath{R_{\mathrm{eff}}}}
\newcommand{\tc}{\ensuremath{t_{\mathrm{cross}}}}

\newcommand{\figComplete}{
  \begin{figure}
    \includegraphics[width=3in]{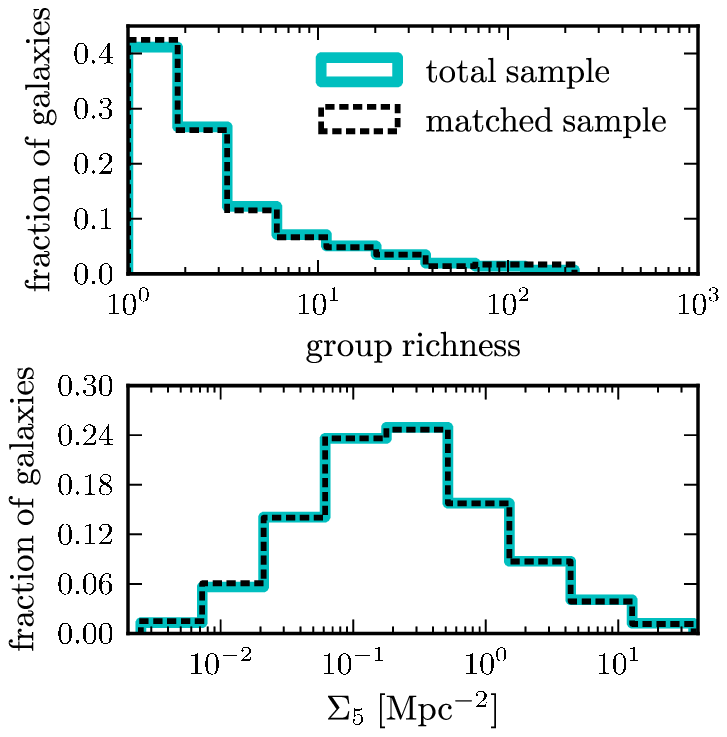}
    \caption{{\it Top panel:} The distribution of galaxy group
      sizes for the total group catalogue from \citet{Berlind2006} (blue
      solid line) and the subsample with  bulge-disc decompositions
      (black dashed line).  {\it Lower panel:} The
      distributions of $\Sigma_5$ for the same samples. These two
      distributions are indistinguishable at the $2\sigma$ level.
    }
    \label{fig:complete}
  \end{figure}
}
\newcommand{\figMagBerlind}{
  \begin{figure}
    \includegraphics[width=3in]{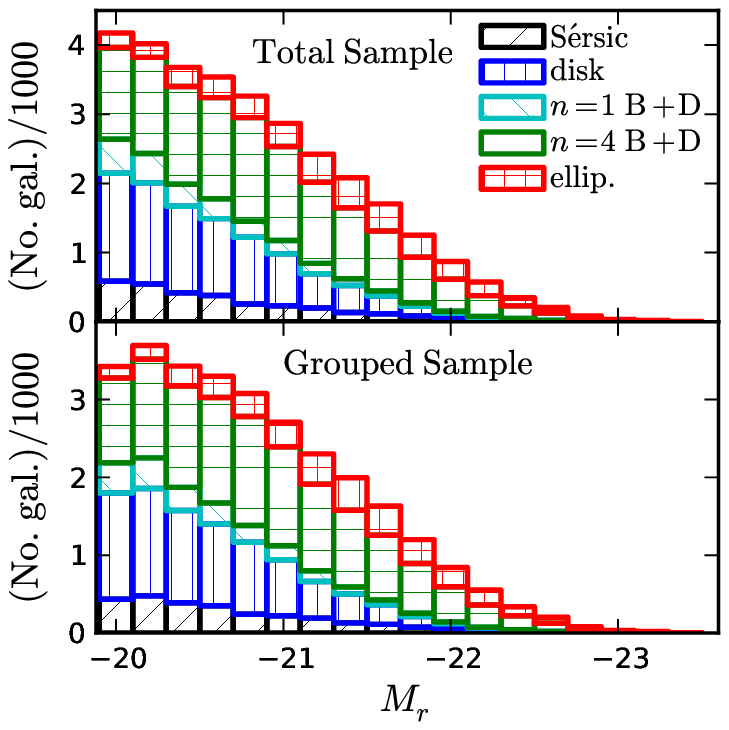}
    \caption{{\it Top panel:} The distribution of galaxy
      absolute magnitude for the galaxies from L12 with 
      $M_r<-19.77$. The different colours represent the different
      categories of galaxies. The number of galaxies in each category
      is the sum of the probabilities calculated in \S
      \ref{sssec:classify}. {\it Bottom panel:} The same
      distribution for galaxies from L12 matched to the group
      catalogue, which has a limiting magnitude of $M_{^{0.1}r} = -19.77$. }
    \label{fig:magBerlind}
  \end{figure}
}

\newcommand{\figGoto}{
\begin{figure}
\includegraphics[width=3in]{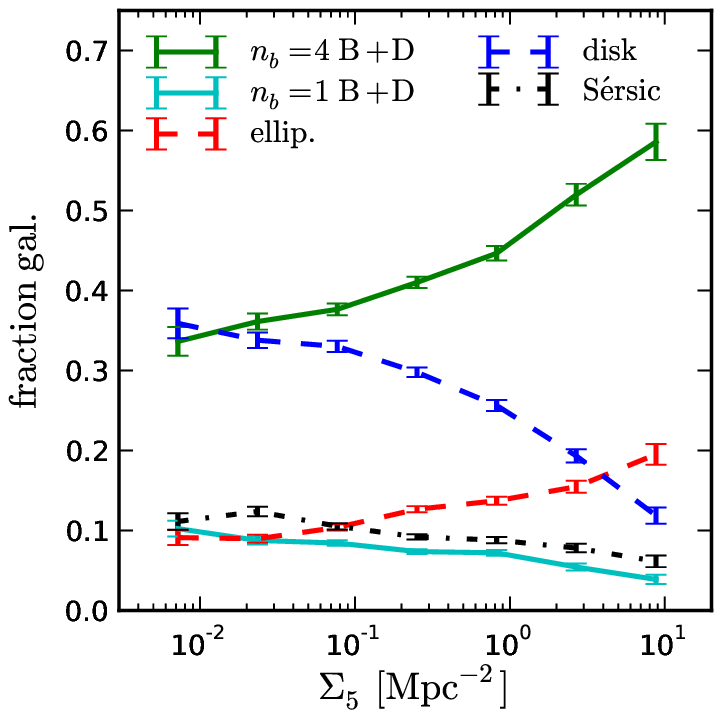}
\caption{The distribution of morphological types as a function of
  local density, $\Sigma_5$. Classical B$+$D galaxies are the $n_b=4$ B$+$D galaxies; pseudo-(star-forming) bulge galaxies are the $n_b=1$ B$+$D galaxies. The unclassifiable galaxies are S\'ersic galaxies. The errorbars are the Poisson errors in the number of  galaxies in each bin.}
\label{fig:fracSig5}
\end{figure}
}

\newcommand{\figMorphMass}{
\begin{figure*}
\includegraphics[width=6in]{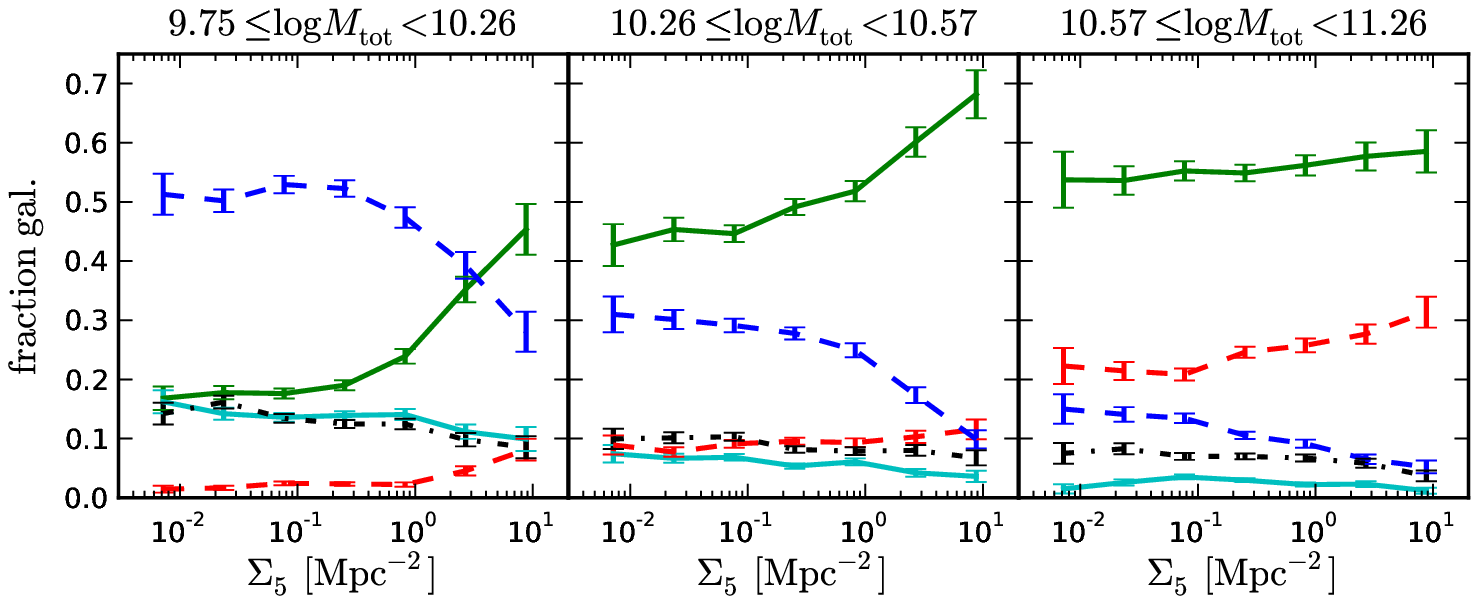}
\caption{The distribution of morphological types as a function of
  local density for three different stellar mass bins. The different
  types are as in Fig.~\ref{fig:fracSig5}. Although the trends with
  morphology at constant mass are the same as those in Fig.
  \ref{fig:fracSig5}, galaxy mass is strongly-correlated with
  morphology at all local densities. }
\label{fig:fracSig5Mass}
\end{figure*}
}

\newcommand{\figredfrac}{
\begin{figure}
\includegraphics[width=3in]{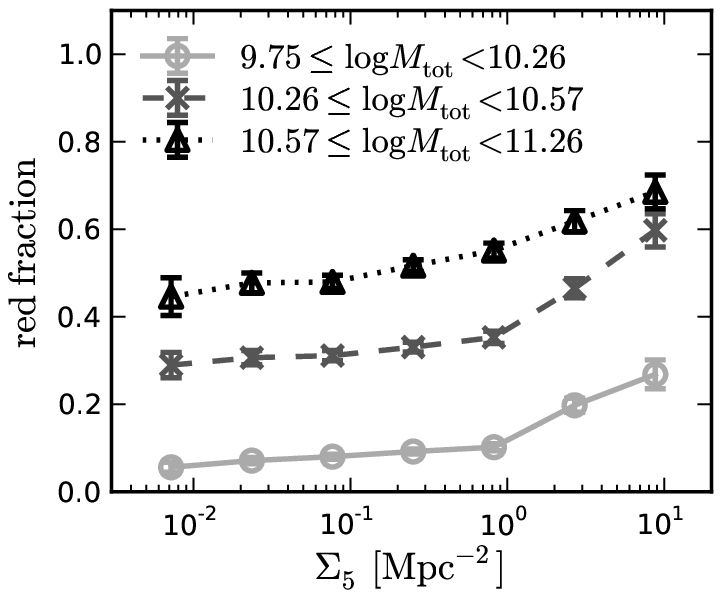}
\caption{The fraction of red galaxies ($u-r>2.22$) as a function of
  $\Sigma_5$ for galaxies in three mass bins. The fraction of red galaxies
is highly dependent on galaxy mass, but turns up sharply at $\Sigma_5
\approx 1$ Mpc$^{-2}$.}
\label{fig:redfrac}
\end{figure}
}

\newcommand{\figRfiveColor}{
\begin{figure*}
\includegraphics[width=6.1in]{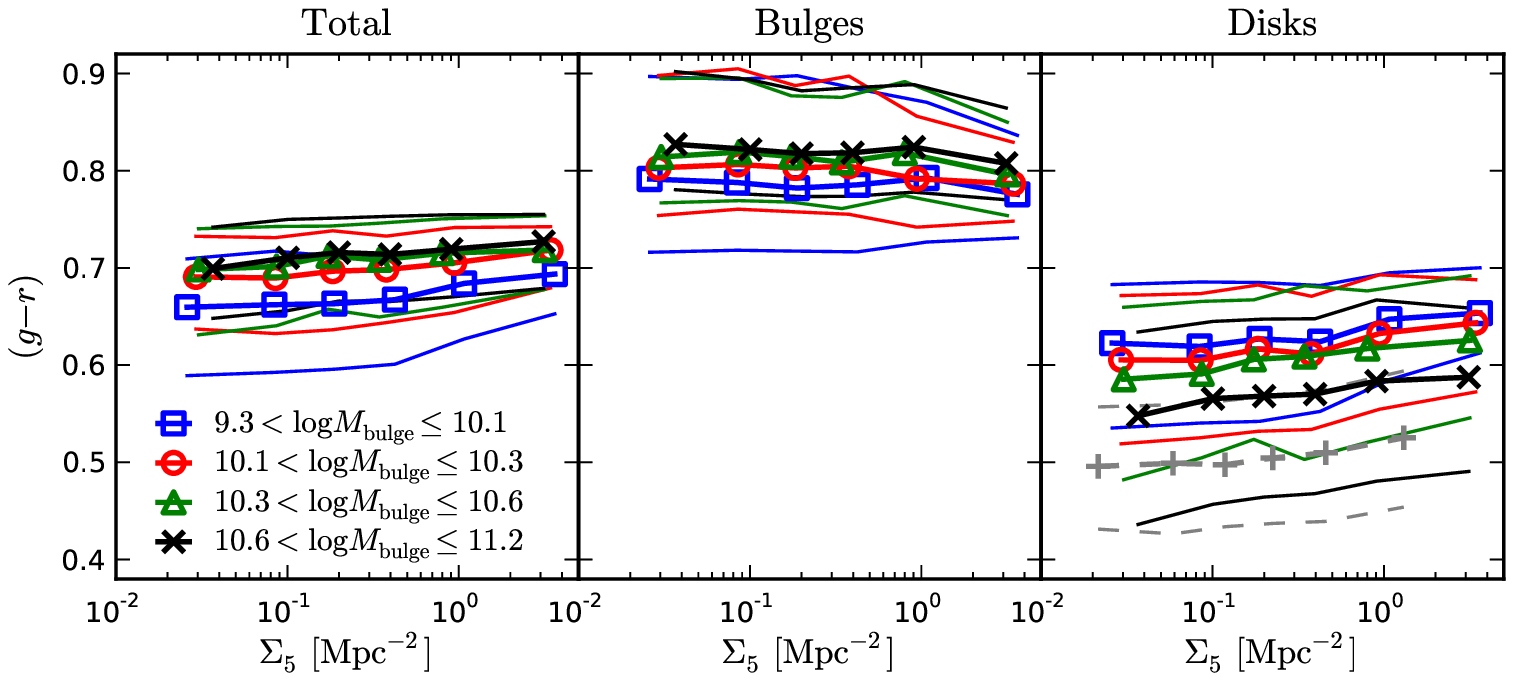}
\caption{Colours of bulge$+$disc galaxies as a function of
  environment. The galaxies are divided into four bins of equal number
  based on bulge stellar mass. The trends in total galaxy $g-r$, bulge
  $g-r$, and disc $g-r$ are shown in the three panels. The points and
  thick lines indicate the median colours in each bulge mass bin, while
  the same colour thin lines indicate the inter-quartile ranges. In the
disc (right) panel, the grey dashed lines and crosses show the
median and inter-quartile range in disc colour for bulge-less
($M_{\mathrm{bulge}} \approx 0$) galaxies, including pseudo-bulge
galaxies and unclassified galaxies.}  
\label{fig:r5Color}
\end{figure*}
}

\newcommand{\figBulgeLdiscColor}{
\begin{figure}
\includegraphics[width=3in]{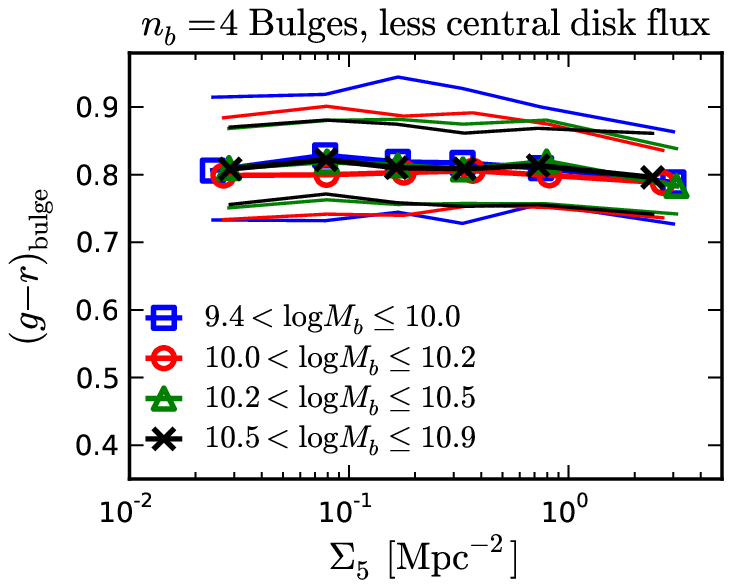}
\caption{Colours of classical bulges as a function of
  $\Sigma_5$ using 
  a model which suppresses the disc flux in the central regions. These
  models essentially eliminate the anti-correlation between bulge colour
  and $\Sigma_5$ seen in the middle panel of Fig.
  \ref{fig:r5Color} The points and thick lines indicate the median 
  colour in each bin, while the same colour thin lines indicate the
  inter-quartile ranges.} 
\label{fig:BcolorLD}
\end{figure}
}
\newcommand{\figbulgedn}{
\begin{figure}
\includegraphics[width=3in]{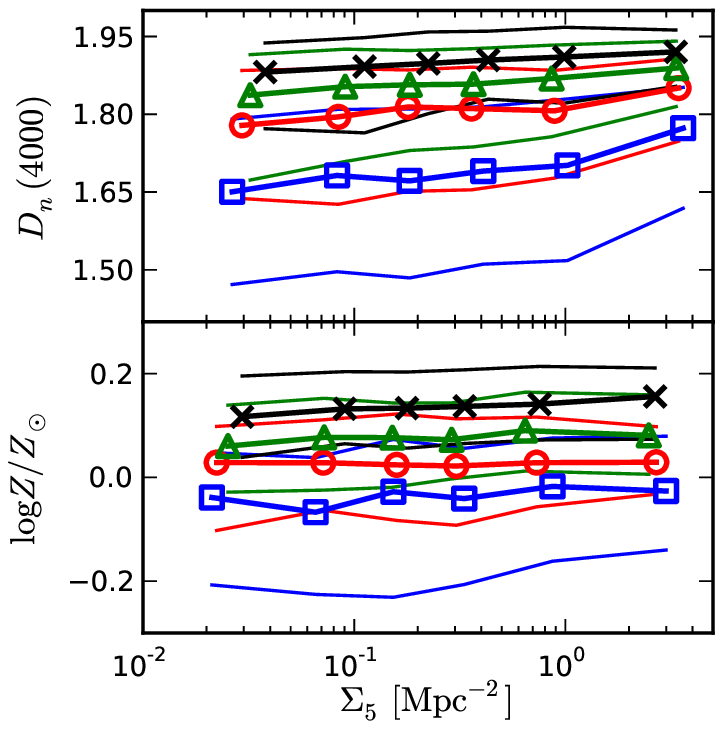}
\caption{{\it Top:} The $4000$\AA{} break measured by SDSS as a
  function of $\Sigma_5$. The sample includes both classical bulges
  and ellipticals. The different colours/symbols denote different
  bulge masses, and are the same as in Fig.~\ref{fig:r5Color}. The
  points denote medians and the thin lines represent the
  inter-quartile ranges. All trends are
  statistically significant. Note that the different
  colours/symbols do not all represent the same number of galaxies (as
  in Fig.~\ref{fig:r5Color}) since
  we include ellipticals in these plots. {\it Bottom:} Same 
  as above for galaxy stellar metallicity computed by
  \citet{Gallazzi2005}. This sample is based on SDSS DR4 and,
  therefore, only contains half the galaxies   plotted in the top
  panel.  }  
\label{fig:dn4000r5}
\end{figure}
}

\newcommand{\figDTTrhoFive}{
\begin{figure}
\includegraphics[width=3in]{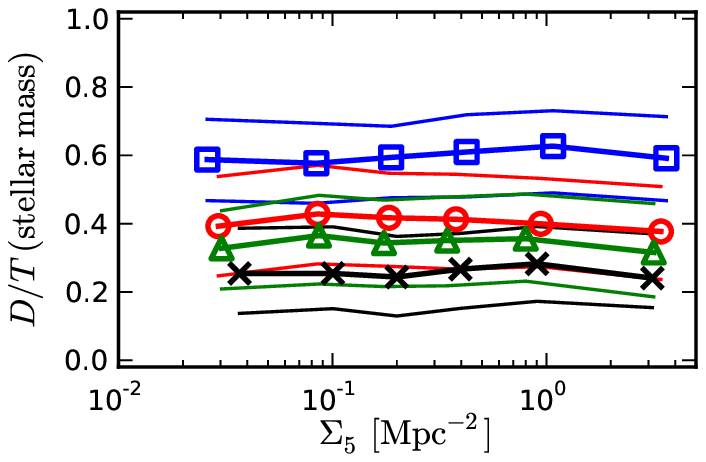}
\caption{The stellar mass disc-to-total ratio for classical bulge$+$disc
  galaxies, at fixed bulge mass, as a function of $\Sigma_5$. The bulge
  mass bins (indicated by colour and symbol) are the same as in Fig.
  \ref{fig:r5Color}. The thick lines and point denote the medians, and
  the thin lines denote the inter-quartile ranges
  for each mass bin. There are no statistically significant trends in
  disc mass at fixed bulge mass as a function of local density. } 
\label{fig:r5DTT}
\end{figure}
}

\newcommand{\figDRerhoFive}{
\begin{figure}
\includegraphics[width=3in]{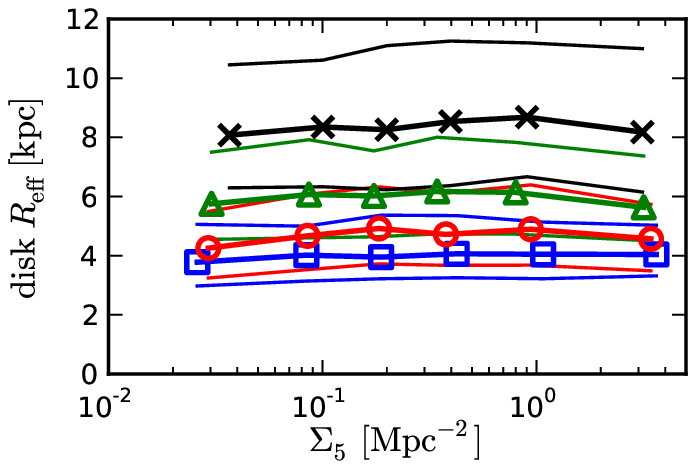}
\caption{Same as Fig.~\ref{fig:r5DTT}, but for the disc half-light
  radius, \Reff{}. There are no statistically significant trends in
  disc \Reff{} at fixed bulge mass as a function of local density. } 
\label{fig:r5DRe}
\end{figure}
}

\newcommand{\figQdvcDTT}{
\begin{figure*}
\includegraphics[width=6in]{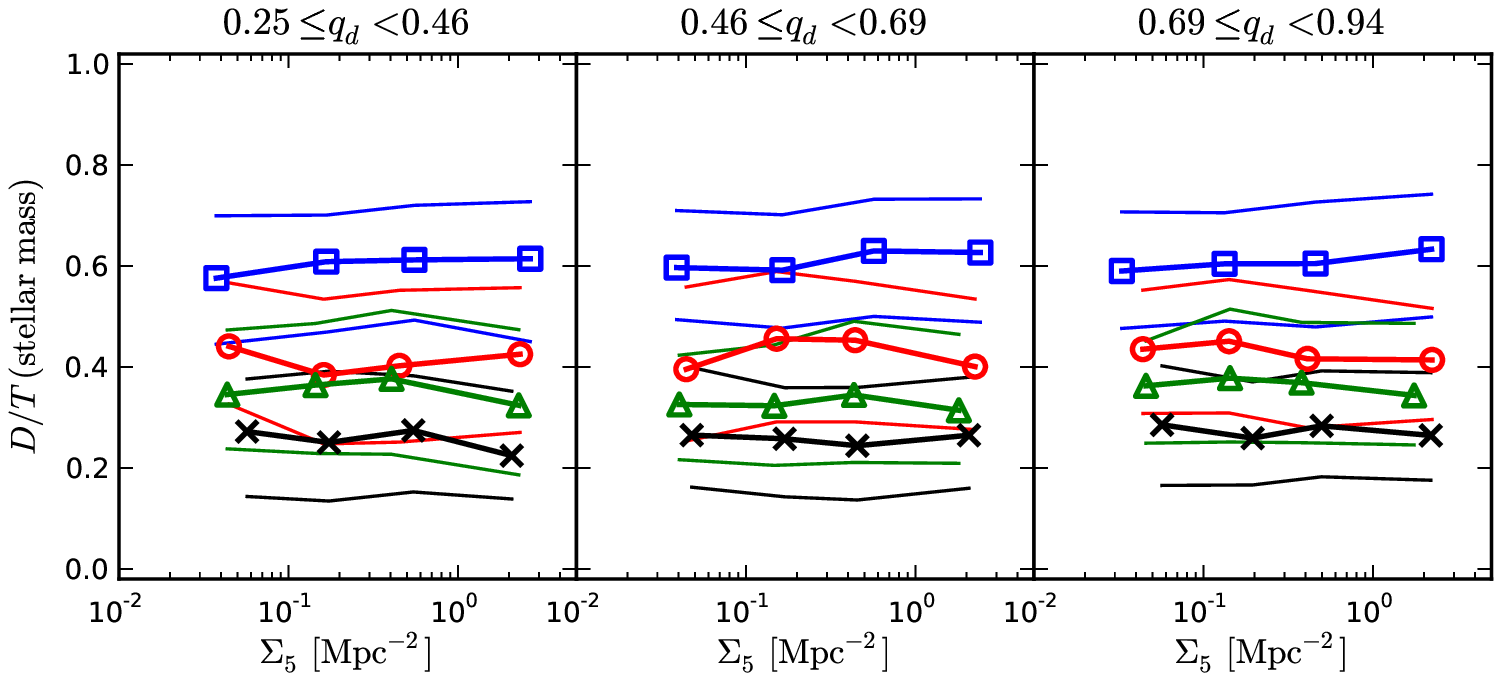}
\caption{Disc-to-total mass ratio as a function of $\Sigma_5$ for
  classical bulge$+$disc  
  galaxies for three bins in disc inclination
  angle (reported  as disc axis ratio, $q_d$). The rightmost panel
  contains face-on galaxies. The different colours and
  symbols represent the same bulge mass bins as in
  Fig.~\ref{fig:r5Color}. Each plot shows the same 
  number of galaxies, $3240$.  The symbols denote the weighted
  medians, while the thin lines represent the inter-quartile ranges. } 
\label{fig:QdttMass}
\end{figure*}
}
\newcommand{\figQdvcDgr}{
\begin{figure*}
\includegraphics[width=6in]{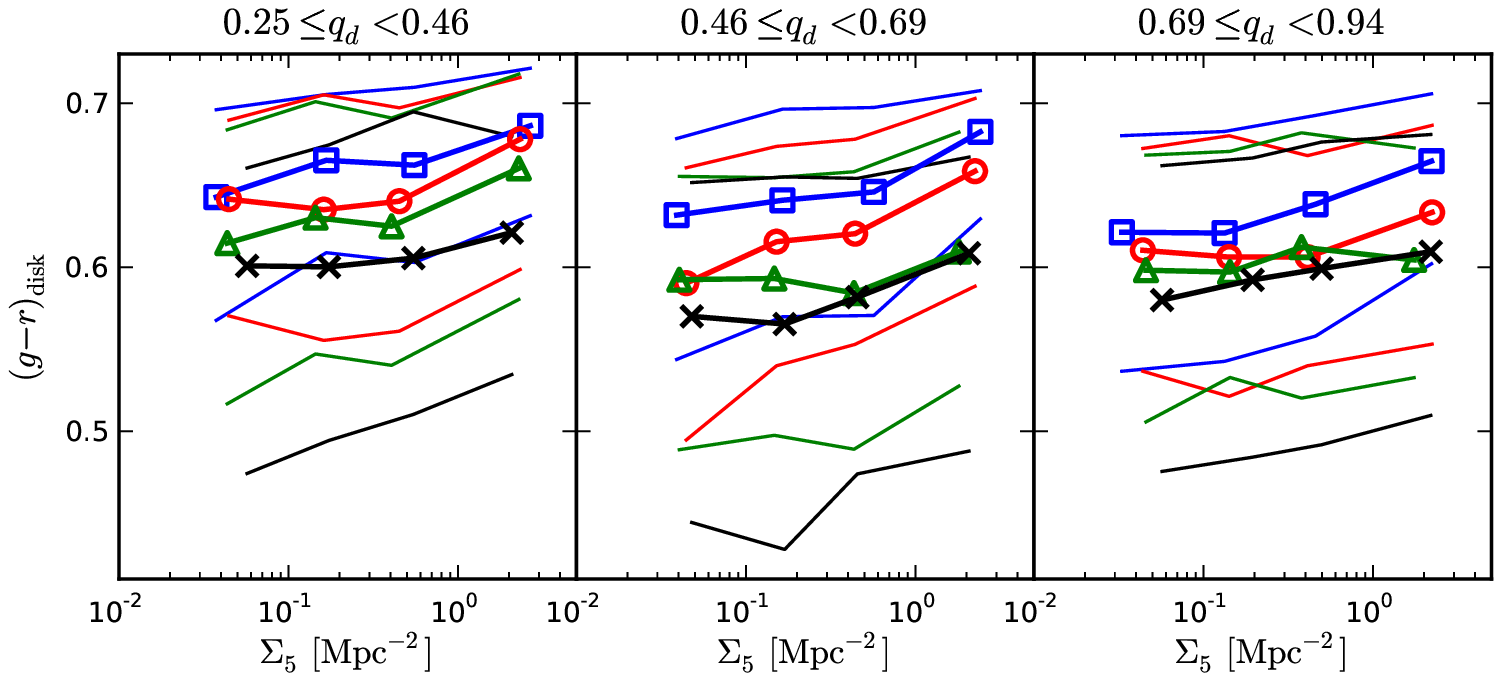}
\caption{Same as Fig.~\ref{fig:QdttMass}, but for the disc colour.
  The disc colours are {\it not} corrected for inclination. The
  correlation coefficients and fitted slopes are reported in Table \ref{tab:spearQ}.}
\label{fig:Qdiscgr}
\end{figure*}
}

\newcommand{\figRichColor}{
\begin{figure*}
\includegraphics[width=6.1in]{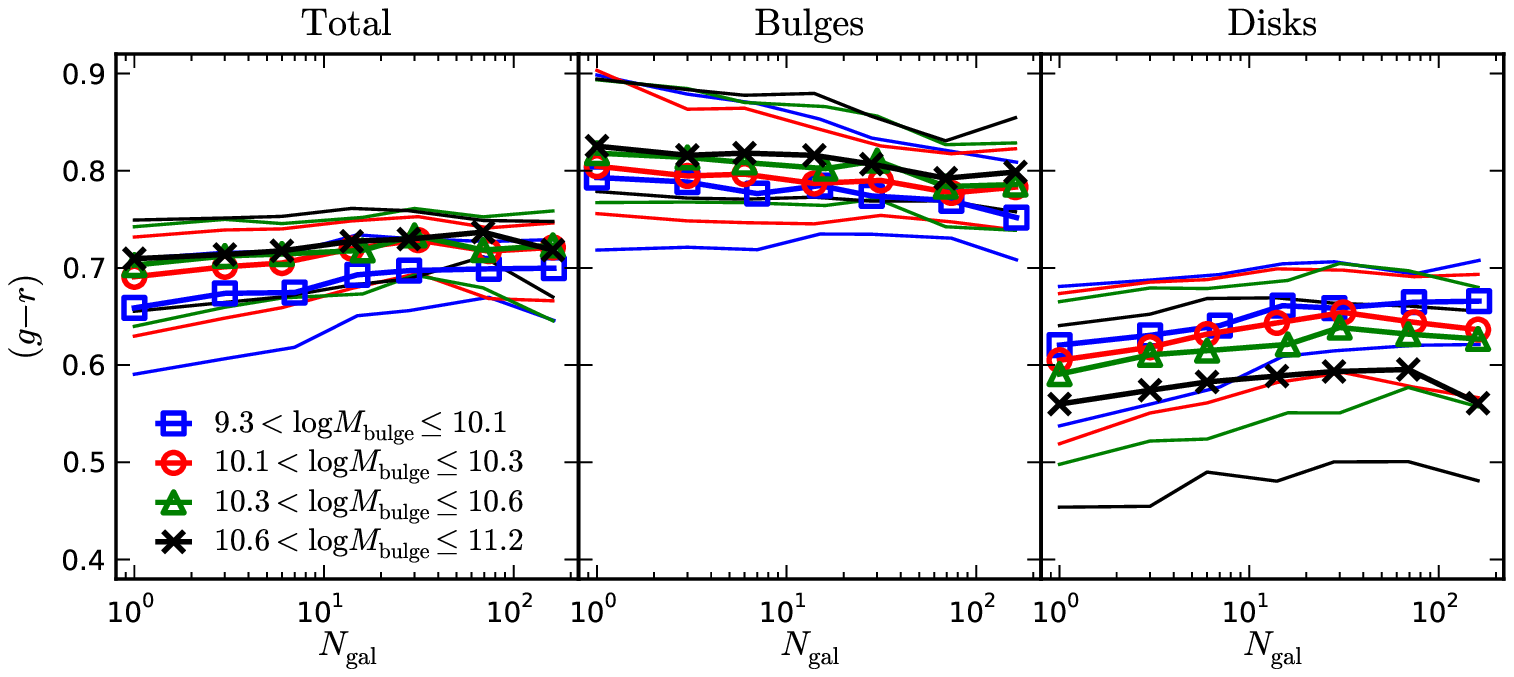}
\caption{Colours of classical bulge$+$disc galaxies as functions of
  group richness ($N_{\mathrm{gal}}$ is the number of galaxies with
  $M_{^{0.1}r} < -19.77$ in a group). Galaxies are binned by bulge
  mass, as in Fig. 
  \ref{fig:r5Color}. The trends 
  in total galaxy $g-r$, bulge   $g-r$, and disc $g-r$ are shown in
  the three panels. The points and thick lines indicate the median
  colours, while the same colour thin lines indicate the
  inter-quartile ranges.} 
\label{fig:richColor}
\end{figure*}
}
\newcommand{\figDgrRichPoor}{
\begin{figure}
\includegraphics[width=3in]{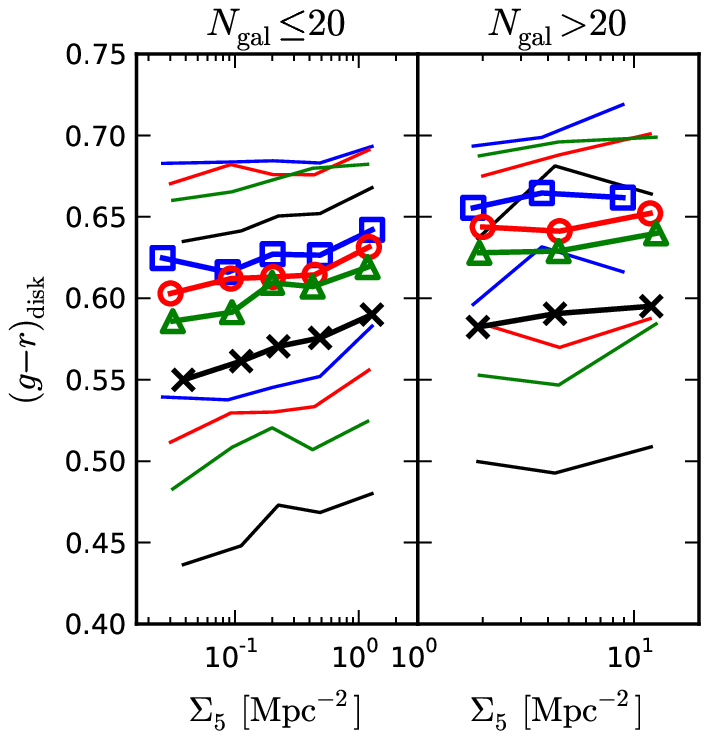}
\caption{Disc colours as a function of $\Sigma_5$. The sample
  is divided by host group richness. Isolated galaxies and galaxies in poor groups ($N_{\mathrm{gal}} \leq 20$) are shown on the left, while galaxies in rich
  groups ($N_{\mathrm{gal}} > 20$) are shown on the right. The bulge
  mass bins (indicated by colour and symbol) are the same as in Fig.
  \ref{fig:richColor}. Each point in the left (right) plot represents
  $\sim 730$ ($150$) galaxies, There is no statistically significant
  trend in disc colour at fixed bulge mass for the larger groups and
  clusters (right panel).  }
\label{fig:r5rpDgr}
\end{figure}
}

\newcommand{\figNMorphfrac}{
\begin{figure}
\includegraphics[width=3in]{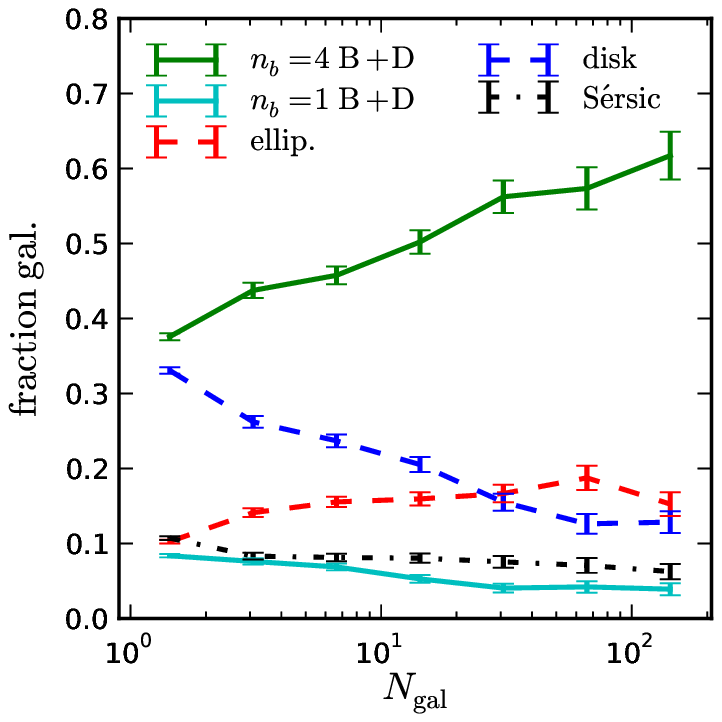}
\caption{The fractions of the various morphological types as functions
  of 
  group richness. The decrease in elliptical galaxies in the highest
  bin may be due to the inclusion unvirialised (and possibly unbound)
  systems in the FoF group catalogue. None the less, there is little
  change in the morphological fractions for groups with more than
  $\sim 20$ members. }
\label{fig:richMorphfrac}
\end{figure}
}

\newcommand{\figDMMorphfrac}{
\begin{figure}
\includegraphics[width=3in]{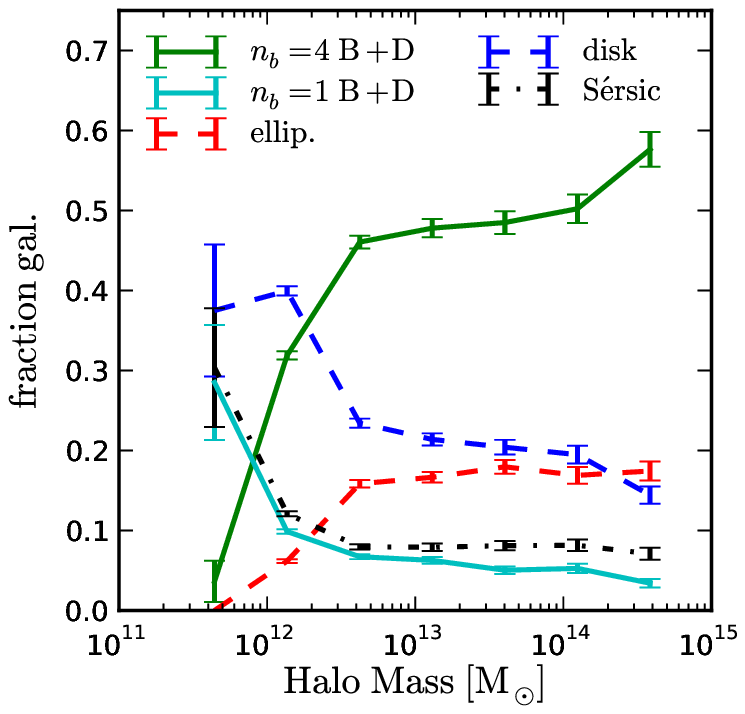}
\caption{The fractions of the various morphological types as functions of
  group dark matter halo mass, $M_{200}$. The constant elliptical
  galaxy fraction agrees with results from 
  \citet{Hoyle2011}. }  
\label{fig:DMMorphfrac}
\end{figure}
}

\newcommand{\figColorTcross}{
\begin{figure*}
\includegraphics[width=6in]{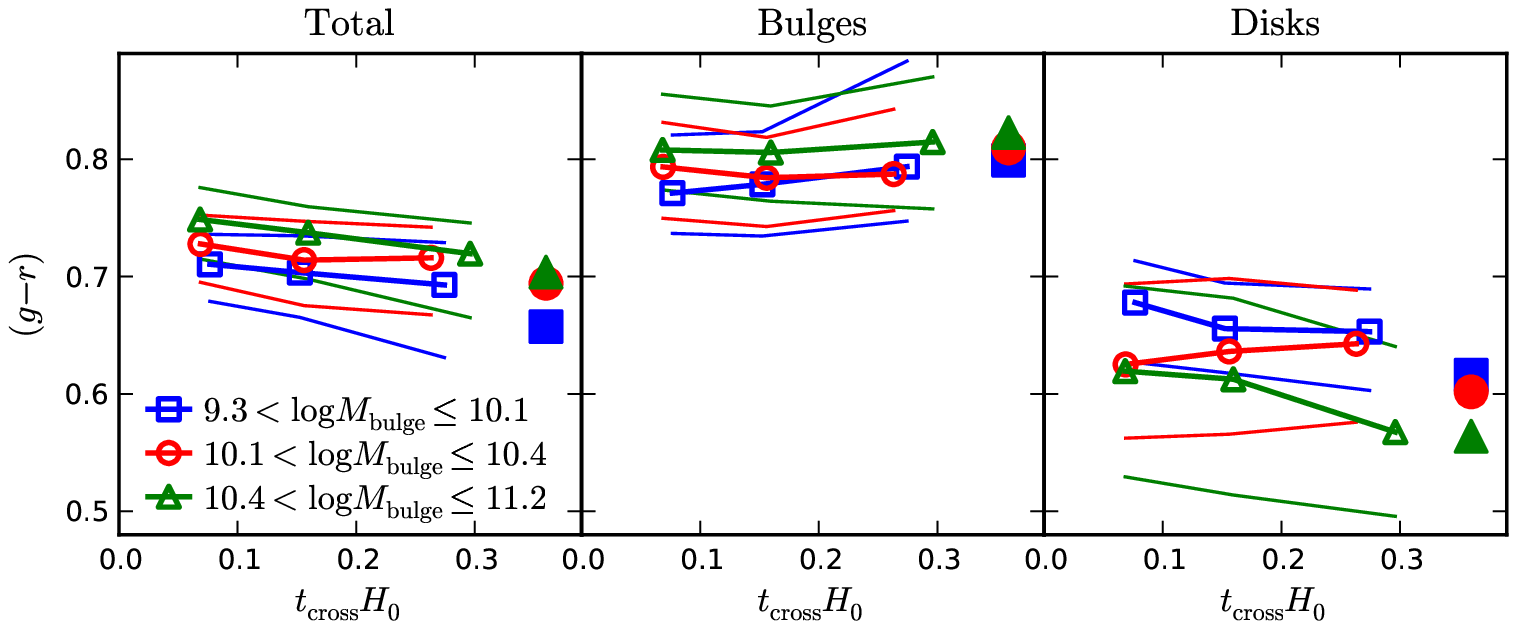}
\caption{Total, bulge, and disc colours as a function of crossing time for
  classical bulge$+$disc galaxies in round groups with at least $20$ members
  (see text for explanation). Crossing time is shown units of the
  Hubble time. The different colours/symbols represent
  different bulge masses. The points show the medians in
  three bins of \tc{} while the thin lines show the
  inter-quartile ranges. The filled points are the median colours
  for isolated galaxies ($N_{\mathrm{gal}} = 1$) of the same bulge mass. }
\label{fig:colorTcross}
\end{figure*}
}

\newcommand{\figDttTcross}{
\begin{figure}
\includegraphics[width=3in]{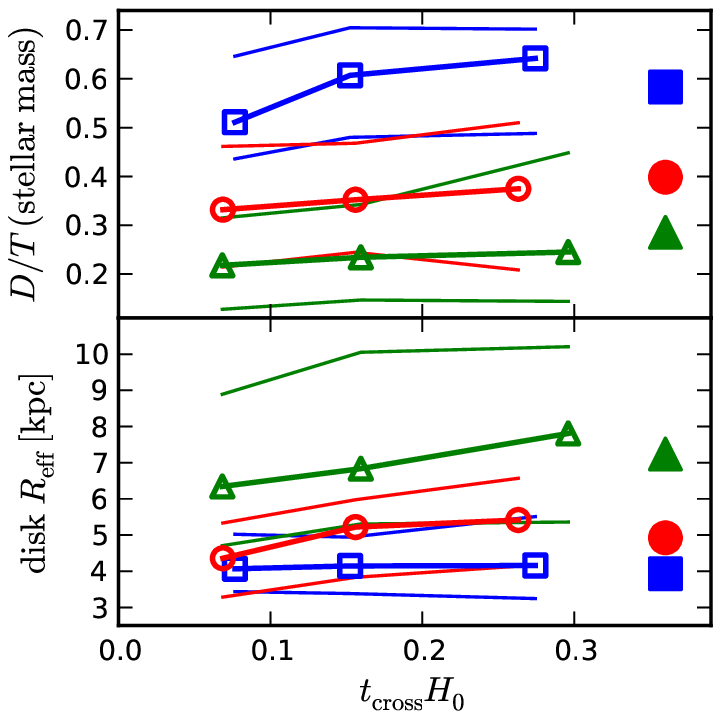}
\caption{Same as Fig.~\ref{fig:colorTcross}, but showing the
  disc-to-total mass ratio (top) and the disc scale length (bottom) as
  a function of crossing time. The top plot shows that disc mass is correlated with crossing time. }
\label{fig:dttTcross}
\end{figure}
}

\newcommand{\figMags}{
\begin{figure}
\includegraphics[width=3in]{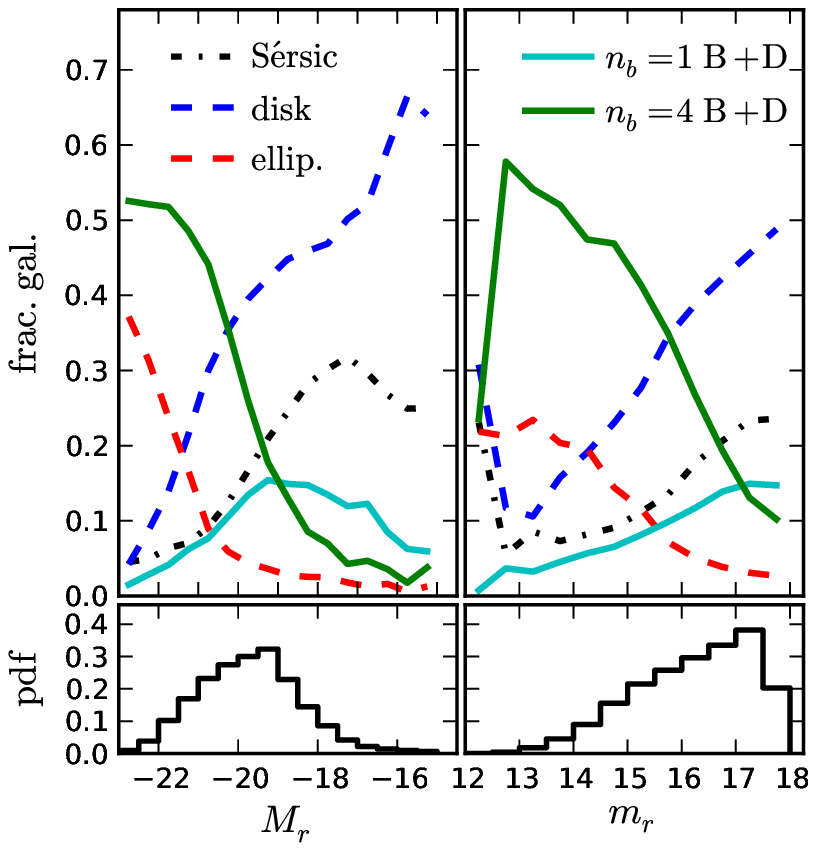}
\caption{ The absolute (left) and apparent (right) magnitude
  distributions for galaxies from the L12 sample classified using the classification scheme described in \S\ref{sssec:classify} and appendix \ref{app:Class}.
  Galaxies fit with a 
  S\'ersic profile are unclassifiable. They make up 17 per cent of the
  sample from L12, and less than 10 per cent of the bright ($M_r \lesssim
  -19.77$) subsample used in this work.
}
\label{fig:mags}
\end{figure}
}

\newcommand{\figSOEllip}{
\begin{figure*}
\includegraphics[width=6in]{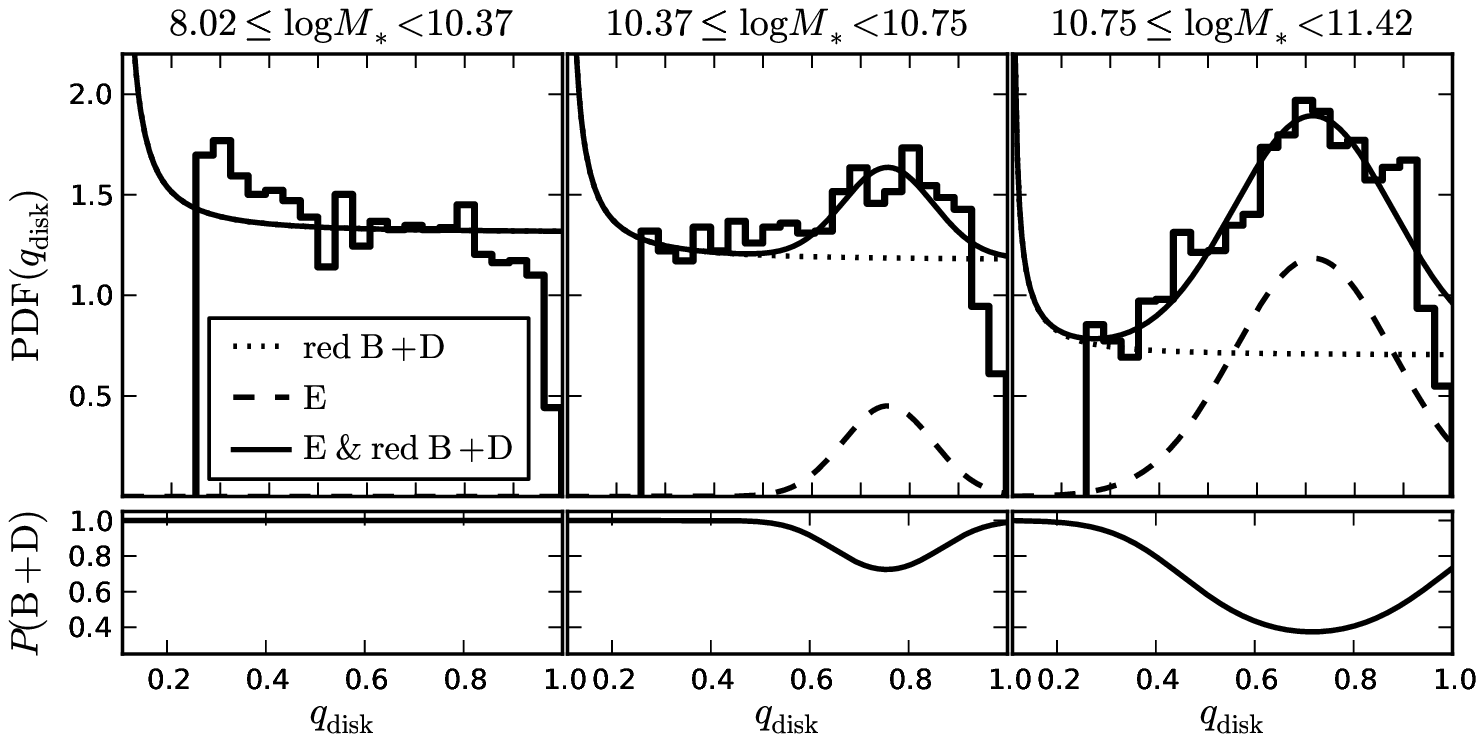}
\caption{ {\it Top row:} The disc axis ratio distribution for red bulge$+$red disc galaxies and
  ellipticals divided into three stellar mass bins. These galaxies are selected to have red
  ($u-r > 2.22$) $n_b=4$ bulges 
  and discs. The dotted curve is the best-fitting disc axis ratio
  distribution for 
  red B$+$D galaxies (flat above $q_d \gtrsim 0.1$) and the dashed curve is a
  Gaussian fit to the distribution of disc axis ratios from
  bulge$+$disc models fit to ellipticals. The sum of these curves is
  shown by the solid line, which is fit to the histogram. {\it
    Bottom row:} The probability that a red galaxy with a given axis
  ratio and mass 
  is a red classical bulge$+$red disc galaxy as opposed to an elliptical. }
\label{fig:S0E}
\end{figure*}
}

\newcommand{\figClassPseudo}{
\begin{figure}
\includegraphics[width=3in]{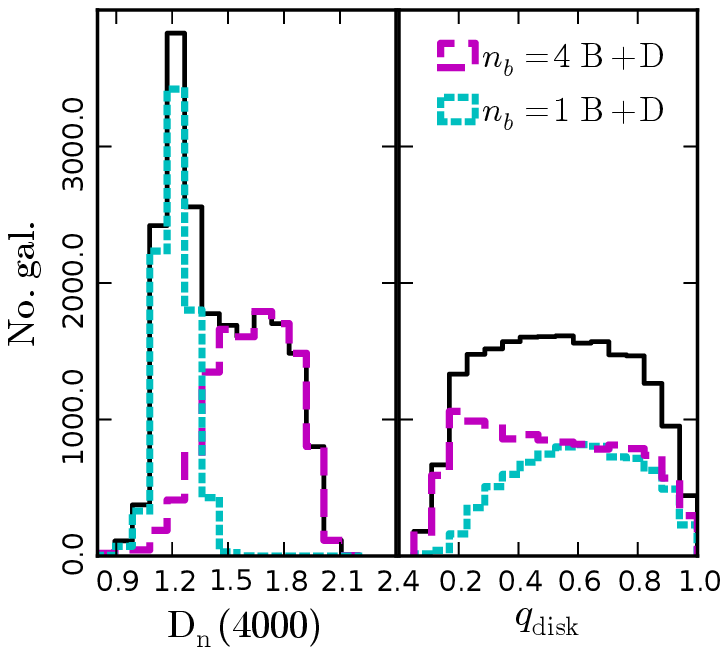}
\caption{ {\it Left:} The distribution of the $4000$\AA{} break
  (D$_n(4000)$) for classical (magenta) and
  pseudo- (cyan) bulges. The total distribution (black line) is fit
  with two Gaussians in order 
  to separate the the two types of bulges. {\it
    Right:} The distribution of the disc axis ratios for classical and
  pseudo-bulges. The excess of edge-on discs around 
  classical bulges is due to the inclusion of edge-on red B$+$D galaxies. The lack of  edge on discs around pseudo-bulges is due to the difficulty of 
  detecting a flat pseudo-bulge in an edge-on disc. }
\label{fig:classpseudo}
\end{figure}
}

\newcommand{\tableClass}{
\begin{table}
  \caption{Classification scheme for galaxies. The number of galaxies refers to the number in the bright, grouped sample. For most of this work, we only use the classical bulge$+$disc galaxies.}
  \label{tab:class}
  \begin{tabular}{llp{4.6cm}}
    \hline
    Type & Num & Classification \\
    \hline

    elliptical & 3681 & either $B/T>0.9$ \emph{or} bulge and disc $u-r > 2.22$ and statistically separated from red B$+$D galaxies based on disc axis ratio (Fig.~\ref{fig:S0E}) \\

    classical B$+$D & 12523 & if bulge and disc $u-r > 2.22$, statistically separated from ellipticals based on disc axis ratio (Fig.~\ref{fig:S0E}); otherwise, selected to be quiescent based on $D_n(4000)$ to distinguish from pseudo-bulge hosts(Fig.~\ref{fig:classpseudo}) \\

    pseudo B$+$D & 5361 & bulge \emph{or} disc $u-r \leq 2.22$ and statistically separated from classical bulge hosts based on $D_n(4000)$; pseudo-bulges are selected to be star-forming  \\

    bulgeless disc & 8505 & $B/T < 0.1$ \\

    unclassifiable & 2857 & none of the above categories, typically blue, disc-like irregulars \\
\hline
  \end{tabular}
\end{table}
}
\newcommand{\tableSpear}{
\begin{table}
\caption{Spearman rank correlation coefficients ($\rho_S$) and linear
  slopes for galaxies shown in Fig.~\ref{fig:r5Color}. The p-value
  for $\rho_S$ is the probability that the colours and $\Sigma_5$ are
  uncorrelated. We only report values larger than $10^{-9}$. The
  errors in the slope are the 1$\sigma$ errors.}
\label{tab:spear}
\begin{tabular}{cccc}
  \hline
 $\log M_{\mathrm{bulge}}/M_\odot$ & $\rho_S$ & p-value, $\rho_S$ & slope \\
 \hline
  \multicolumn{4}{c}{Total $g-r$ vs. $\log \Sigma_5$} \\
  \hline
   9.3--10.1 & 0.17 & 0.0   & $0.024 \pm 0.001$ \\
  10.1--10.3 & 0.14 & 0.0   & $0.013 \pm 0.001$ \\
  10.3--10.6 & 0.13 & 0.0   & $0.012 \pm 0.001$ \\
  10.6--11.2 & 0.11 & $0.0$ & $0.010 \pm 0.001$ \\
  \hline
  \multicolumn{4}{c}{Bulge $g-r$ vs. $\log \Sigma_5$} \\
  \hline
  9.3--10.1  & $-0.04$ & $3.7 \times 10^{-2}$ & $-0.006 \pm 0.002$ \\
  10.1--10.3 & $-0.10$ & $1.1 \times 10^{-6}$ & $-0.012 \pm 0.002$ \\
  10.3--10.6 & $-0.07$ & $2.6 \times 10^{-4}$ & $-0.006 \pm 0.002$ \\
  10.6--11.2 & $-0.07$ & $2.4 \times 10^{-4}$ & $-0.007 \pm 0.002$ \\
  \hline
  \multicolumn{4}{c}{Disc $g-r$ vs. $\log \Sigma_5$} \\
  \hline
  9.3--10.1  & 0.18 & $0.0$              & $0.022 \pm 0.002$ \\
  10.1--10.3 & 0.13 & $0.0$              & $0.017 \pm 0.002$ \\
  10.3--10.6 & 0.12 & $0.0$              & $0.019 \pm 0.002$ \\
  10.6--11.2 & 0.09 & $9.1\times 10^{-9}$ & $0.019 \pm 0.002$ \\
  Bulge-less & 0.11 & $0.0$              & $0.017 \pm 0.001$ \\
  \hline

\end{tabular}
\end{table}
}

\newcommand{\tableSpearQ}{
\begin{table}
  \caption{Spearman rank correlation coefficients ($\rho_S$)
    and linear slopes for galaxies shown in Fig.~\ref{fig:Qdiscgr}. }
   \label{tab:spearQ}
  \begin{tabular}{cccc}
    \hline
    $\log M_{\mathrm{bulge}}/M_\odot$ & $\rho_S$ & p-value, $\rho_S$ & slope \\
    \hline
    \multicolumn{4}{c}{Disc $g-r$ vs. $\log \Sigma_5$, $0.25 \leq
      q_d < 0.46$} \\
    \hline
    {\rm 9.3--10.1}  & {\rm 0.18} & {\rm 0.0} & 
    $\mathrm{0.021 \pm 0.003}$ \\
    {\rm 10.1--10.3} & {\rm 0.12} & $1.4 \times 10^{-4}$ & 
    $0.014 \pm 0.004$ \\
    {\rm 10.3--10.6} & {\rm 0.14} & $1.4 \times 10^{-6}$ & 
    $0.025 \pm 0.005$ \\
    {\rm 10.6--11.2} & {\rm 0.12} & $1.1 \times 10^{-4}$ & 
    $0.022 \pm 0.005$ \\
    \hline
    \multicolumn{4}{c}{Disc $g-r$ vs. $\log \Sigma_5$, $0.46 \leq
      q_d < 0.69$} \\
    \hline
    9.3--10.1  & 0.23 & $0.0$ & $0.028 \pm 0.003$ \\
    10.1--10.3 & 0.20 & $0.0$ & $0.027 \pm 0.004$ \\
    10.3--10.6 & 0.09 & $5.0 \times 10^{-3}$ & $0.015 \pm 0.005$ \\
    10.6--11.2 & 0.07 & $5.2 \times 10^{-2}$ & $0.012 \pm 0.005$ \\
    \hline
    \multicolumn{4}{c}{Disc $g-r$ vs. $\log \Sigma_5$, $0.69 \leq
      q_d < 0.94$} \\
    \hline
    9.3--10.1  & 0.16 & $0.0$               & $0.020 \pm 0.003$ \\
    10.1--10.3 & 0.06 & $0.10$              & $0.010 \pm 0.004$ \\
    10.3--10.6 & 0.06 & $0.10$              & $0.009 \pm 0.004$ \\
    10.6--11.2 & 0.09 & $5.6\times 10^{-3}$ & $0.019 \pm 0.004$ \\
    \hline
  \end{tabular}
\end{table}
}
\ifthenelse{\boolean{astroph} \or \boolean{mn2e}}{
  
  \newcommand{\tailfig}[1]{{}}
}
\title[Environmental effects on discs and bulges]{The Effect of Environment on
  Discs and Bulges}
\author[C.~N.~Lackner]{C.~N.~Lackner$^{1,2}$\thanks{E-mail:
    claire.lackner@ipmu.jp} and J.~E.~Gunn$^{1}$\\
$^{1}$Department of Astrophysical Sciences, Princeton University, Princeton, NJ
08544 \\
$^{2}$Kavli Institute for the Physics and Mathematics of the Universe, Todai Institutes for Advanced Study, \\
the University of Tokyo, Kashiwa, Japan 277-85823 (Kavli IPMU, WPI)}

\begin{document}
\date{\today}
\pagerange{\pageref{firstpage}--\pageref{lastpage}} \pubyear{2012}
\maketitle

\label{firstpage}

%%abstract
\begin{abstract}
 We examine the changes in the properties of galactic bulges and discs with
 environment for a volume-limited
 sample of $12500$ nearby galaxies  from SDSS. We focus on galaxies
 with classical bulges. Classical bulges seem to have the same
 formation history as ellipticals of the same mass, and we test
 if environment determines whether or not a classical bulge {\rm possesses}
 %acquires and retains 
a disc.  Using the projected fifth
 nearest neighbour density as a measure of local 
 environment, we look for correlations with environment at fixed bulge
 stellar mass. {\rm In groups with fewer than 20 members, we find no evidence for changes in disc morphology with local density. At fixed bulge mass, disc mass and disc scale length are independent of local density. However, disc colour does increase ($\Delta(g-r) \sim 0.05$ mag) as a function of local density in relatively poor groups. Therefore, the colour--density relation for classical bulge$+$disc galaxies in the field and in poor groups is due solely to changes in disc colour with density. In contrast, we find no correlations between disc colour and local density for classical bulge$+$disc galaxies in large, relaxed groups and clusters. However, there is a weak correlation between disc mass and group crossing time, suggesting morphological transformation takes places  in rich groups. Our results add to the evidence that star formation is quenched in group environments, instead of clusters, and that star formation quenching and morphological transformation are separate processes. }
%%  We show that the colour--density relation for galaxies
%%  with a classical bulge and disc is due solely to changes in disc colour with
%%  environment, not changes in morphology. At fixed bulge mass, the disc
%%  $g-r$ colour increases by  $\sim 0.05$ {\rm mag} from the lowest to the highest
%%  density regions. We find no statistically significant correlations
%%  between density and disc mass or  disc scale length. 
%%  The change in disc colour occurs entirely in relatively poor  ($N_{\mathrm{gal}}
%%  \lesssim 20$)  groups. 
%% This adds to existing evidence that star
%%  formation is truncated in group environments, instead of
%%  clusters and that star formation quenching and morphological
%%  transformation are separate processes. In groups with at least $20$ members, we find 
%%  no correlations 
%%  between disc colour and group crossing time or local 
%%  density. However, there is a weak correlation between disc mass and
%%  crossing time, suggesting morphological transformation takes places
 %%in rich groups. 
Overall, we show that  environment has two
 effects on galactic discs: relatively  low density environments can
 quench star formation in discs, while  processes occurring in higher
 density environments 
 contribute to the morphological transformation  from disc-dominated
 systems to bulge-dominated systems. %However,  the changes in disc
% properties as functions of environment are not large enough to
% fully explain disc formation around  classical bulges.
\end{abstract}


%  Intro
\section{Introduction}
\label{sec:introduction}

Galaxy morphology, stellar mass, star formation rate, and projected number
density are all known to correlate. Generally, massive galaxies are
bulge-dominated \citep{deVaucouleurs61, Blanton03}, not forming stars
(red) \citep{Strateva01, Kauffmann2004, Baldry2006}, and
reside in high density 
regions \citep{Oemler1974, Dressler80, Postman1984, Goto2003,
  Lewis2002, Gomez2003, Yang2007, Bamford09}. Low mass galaxies are
disc-dominated, 
star-forming (blue), and reside in low density regions. There are, of
course, exceptions: passive discs make up a significant fraction of
the red sequence \citep[e.g.][]{Bernardi03I, Maller09}, blue
ellipticals are still forming 
stars \citep{Schawinski2009}, and there are many early-type,
passively evolving galaxies in relatively low density regions
\citep[e.g.][]{Mulchaey1999}. For galaxies which follow the general
pattern, it is unclear which of the correlations mentioned above are
the result of physical processes, and which, if any, are simply
consequences of other correlations. Morphology, star formation,
and density are   all strongly correlated with stellar mass
\citep[e.g.][]{Hamilton1988,   Brinchmann2000, Blanton03,
  Kauffmann2004, Blanton2005a,   Thomas05}, and correlations between
these properties are partially due to their correlations with
stellar mass. Yet, at 
fixed stellar mass, studies have found correlations between density
and morphology \citep{Bamford09}, density and stellar age
\citep{Thomas05,   Cooper2010}, density and colour  
\citep{Balogh2004, Skibba2009, Cibinel2012}, and density and star formation rate
\citep{Kauffmann2004, Christlein2005}. %However, 
{\rm Furthermore,} at fixed luminosity,
neither blue nor red galaxy colours seem to depend on density, i.e. blue galaxies do not 
get redder as a function of density, only their number fraction
decreases \citep{Balogh2004, Hogg2003}. Furthermore, correlations
with morphology and density {\rm seem to} disappear for high mass  galaxies
\citep[e.g.][]{Tasca2009,  Grutzbauch2011}.   

The local correlations between galaxy properties and density extend to
higher redshift; the morphology--density and colour--density relations
are in place by $z\approx1$ \citep{Dressler1997, Postman05, Treu2003,
  Smith2005}, but the relations do evolve with redshift.  The fraction
of blue galaxies in high density regions increases with redshift
\citep[e.g.][]{ButcherOemler1978}, while the fraction of S0s and red
discs galaxies decreases with increasing redshift \citep{Smith2005,
  Moran2007, Bundy2010, Bruce2012}.  In addition, the
hierarchical growth of structure implies that galaxies generally move
from low density regions to high density regions as a function of
time. Therefore, any environmental effects and trends will be more
pronounced today than in the past \citep{Tasca2009}.

From these observations, a general outline of  galaxy evolution has
been 
developed. At early times, galaxies are blue, intensely star-forming, disc
systems. As galaxies become more massive, star formation is quenched
(by internal feedback mechanisms), creating a
population of red, massive galaxies. This mass quenching
\citep{Peng2010} is compounded by environmental effects. At fixed
mass, galaxies in  higher density regions become red and
bulge-dominated earlier, creating the colour--density and
morphology--density relations. Star formation quenching is thought to
occur before morphological transformation, which leads to an increase in
the fraction of S0s at intermediate densities
\citep[e.g.][]{Dressler80, McIntosh2004,
  Cooper2006,  Moran2007,   Bundy2006}. In order to separate the
effects of environment from the effects of stellar mass, we examine
correlations between galaxy morphologies, colours, and local  density
at {\it fixed} stellar mass.  

The physical processes responsible for the environment-driven
transformations are unknown, although there are many
candidates. Processes can be divided into those which truncate 
star formation, and those which also cause morphological
transformations \citep[see][for a review of these
processes]{Boselli2006}. Ram-pressure stripping of ISM from galaxies
entering clusters \citep{GunnGott72} and the removal of hot halo gas
(strangulation) \citep{Larson80, Balogh2000b} both act to truncate
star formation in discs and can transform spiral discs into S0s,
but do not drastically alter a galaxy's stellar disc.  Tidal stripping by
the cluster potential \citep{Merritt1984} and high speed encounters
with other cluster galaxies (harassment) \citep{Moore96, Moore98,
  Moore99} both act to transform disc-dominated galaxies into
bulge-dominated galaxies. All these processes act on a variety of
timescales and require different minimum local densities in order to
be effective. It is likely that more than one process is responsible
for the morphology--density relation. %, especially since star formation
%quenching and morphology changes happen on different
%timescales. %Furthermore, these processes act on star-forming disc 
%galaxies; suggesting that the impact of local environment is different
%for different galaxy morphologies.

In this work, we examine low redshift galaxies with both a bulge and
disc, and study the correlations of the separate bulge and disc
properties with local density. These galaxies, which include S0s, may
represent a transition from disc-dominated to
bulge-dominated, and the environmental processes enumerated above
should have observable effects on the discs and possibly the bulges of
these transitional 
galaxies. For the bulge and disc properties, we use the bulge$+$disc
decompositions from our earlier work 
\citep[][hereafter L12]{Lackner2012}. L12 presents bulge$+$disc
decompositions for nearly $72,000$ low redshift ($0.002<z<0.05$)
galaxies from the Sloan Digital Sky Survey (SDSS). The galaxies we use
for in this work are a luminous subsample from L12. 

We focus on galaxies which host classical bulges. These bulges
have properties and, presumably, formation histories identical to
elliptical galaxies {\it of the same  mass}. Classical bulges are
concentrated, pressure-supported systems \citep{FalconBarroso02,
  MacArthur08}, with old stellar populations \citep[][but see
\citealp{Gadotti09}]{Peletier99,   Moorthy06,  MacArthur2010}. In L12, 
we model the light profiles of both classical bulges and ellipticals using
a de Vaucouleurs profile \citep[but see][]{Caon93}.  We show in L12
that classical bulges and ellipticals follow the same size-density relation
\citep{Kormendy77}. Taken together, these studies support the assertion
that classical bulges and ellipticals of the same stellar mass are
indistinguishable. Therefore,  by exploring  how local density affects
the discs around classical bulges, we can determine if changes in
environment correspond to a transition from 
classical bulges with discs to disc-less elliptical galaxies.
% and pseudo-bulges, which are modelled
%with an exponential (disc) profile \citep[e.g.][]{Kormendy77,
%  Kormendy93, Fisher08} and are thought to arise from secular processes
%within discs \citep{Kormendy04, Athanassoula05, Weinzirl09}. In this
%work, we focus on classical bulges and their relation to
%ellipticals. 

To date, there have been a handful of studies which explore the effects 
of environment on 
bulges and discs separately. Both \citet{McIntosh2004} and
\citet{Hudson2010} show that disc colour is a function of cluster
radius. \citet{McIntosh2004} also show that the amount of substructure
in discs declines with increasing density, further demonstrating that star
formation is quenched in dense environments. The sample we use in this
work is considerably larger than the samples in previous studies, and
it covers the entire spectrum of local densities, not just rich
clusters and the field. Furthermore, our large sample can easily be
divided into 
subsamples of constant bulge mass, eliminating trends with stellar
mass and environment, and still yield statistically significant
results.

Below, we use our sample of classical 
bulge$+$disc galaxies to determine whether the colour--density relation
for these galaxies is due to changes in bulge or disc colour, indicating star
formation truncation, or changes in bulge-to-total ratio, indicating
morphological transformation. {\rm We find that the colour--density relation for these galaxies \emph{at fixed bulge mass} is due entirely to changes in disc colour, not changes in disc mass or size. Next, we divide the sample into rich and poor groups to determine if the trends in disc properties with environment depend on group size or halo mass.  We find that while disc colour is a function of local density in relatively poor groups, disc colour is independent of local density in larger, relaxed groups and clusters. However, Disc mass decreases slightly with increasing local density 
%(parametrised by crossing time) 
in large groups and clusters, while disc mass is independent of local density for galaxies in the field and in poor groups. From these results, we conclude that environment-driven star formation quenching occurs in relatively low density environments, while structural changes to discs only occur in higher density environments. }
%there is a local density threshold above which star formation in discs is truncated. 
%Finally, we follow
%\citet{McIntosh2004} and \citet{Hudson2010} and study intra-cluster
%trends in bulge and disc properties.  These analyses are all done at
%fixed bulge mass, eliminating known trends with environment and
%mass. %Our analysis shows that the colour--density relation for
%classical bulge galaxies is entirely due to changes in disc colour;
%the disc mass and disc scale length are independt of local density, at
%fixed bulge mass. 

{\rm In order to perform the studies detailed above, we require a robust sample of classical bulge$+$disc galaxies. It is especially important to distinguish classical bulge$+$disc galaxies from disc-less ellipticals. } We present a probabilistic
method for separating classical bulge hosts and ellipticals using bulge$+$disc
decompositions in Section \ref{sssec:classify}. The remainder of
Section \ref{sec:Sample} details the environment metrics employed. In
order to determine group membership and local density,  we use an
updated group catalogue from A. Berlind, 
(priv. comm.), which is based on the group catalogues presented in
\citet{Berlind2006}. Section \ref{sec:wholeGalEnv} presents the
morphology--density relation and colour--density relation for our
sample. The results of this section help to confirm the
assignment of galaxy morphologies in Section
\ref{sssec:classify}. Sections \ref{sec:bdGalEnv} and \ref{sec:richness}
are devoted to the correlations between environment and bulge and disc
properties. Section \ref{sec:bdGalEnv} focuses on correlations with
projected local density, while Section \ref{sec:richness} discusses classical bulge$+$disc galaxies in relatively rich groups.

Throughout this paper we use the
$\Lambda\mathrm{CDM}$ cosmology: $\Omega_m = 0.3$, $H_0 = 70\ \kmps \mathrm{Mpc^{-1}}$,
and $\Omega_\lambda = 0.7$.

%%Sample section
%%
\section{Sample}
\label{sec:Sample}

The sample used in the work consists of a luminous subsample of galaxies from our earlier bulge$+$disc decompositions matched to an updated group catalogue from A. Berlind (priv. comm.). This group catalogue employs the method presented in \citet{Berlind2006}, but uses data from SDSS data
release 7. The sample contains $29781$ galaxies and is complete 
to an absolute $^{0.1}r-$band magnitude of $-19.77$  
and covers the redshift range $0.02 \leq z < 0.05$. Below, we describe
the bulge$+$disc decompositions and the environmental information
extracted from the group catalogue.

\subsection{Bulge-Disc Decompositions}
\label{ssec:bdecomp}
For the bulge and disc properties, we use the results of our earlier
work (L12).  L12 presents bulge$+$disc decompositions for
$72,000$ galaxies from SDSS data release 8 
(the data is the same as that in data release 7, but the reductions
have been improved). The galaxies have redshifts between $0.003$ and
$0.05$. All of the galaxies are in the SDSS spectroscopic sample,
which implies a limiting magnitude of $m_r  < 17.77$. Two dimensional
bulge-disc decompositions are performed for the $r-$band images. The results are then linearly scaled to fit the galaxy images in the $u$, $g$,
$i$, and $z$ 
bands, yielding colours for the bulge and disc components. {\rm Because we only linearly scale the fits in each band, our bulge$+$disc models do not take into account colour gradients within each component.} Each galaxy
is fit with $5$ different models: a de Vaucouleurs bulge and an
exponential disc ($n_b=4$ B+D), an exponential bulge and exponential
disc ($n_b=1$ B+D), a single de Vaucouleurs profile, a single
exponential profile, and a single S\'ersic profile. 

The two
bulge$+$disc models allow us to fit both elliptical-like,
pressure-supported  classical
bulges (de Vaucouleurs profile) and disc-like, rotationally-supported
pseudo-bulges (exponential profile) \citep{Kormendy77, Kormendy93,
  Fisher08}. Pseudo-bulges are thought to arise from secular
processes within discs, such as bar-driven instabilities
\citep[e.g.][]{Kormendy04, Athanassoula05,Weinzirl09}, and, as such, have 
very different formation histories than classical bulges \citep[but
see][]{Elmegreen09}. Often, pseudo-bulges are still forming stars today
\citep{Kormendy04, Fisher06}. Since pseudo-bulges are a
disc phenomenon, we do not include them in our sample of bulge$+$disc
galaxies. %separate pseudo-bulge galaxies from our sample of
%classical bulge$+$disc galaxies based on spectroscopic signatures of
%recent star formation.  %Details of the  
%fitting and tests of the systematics with signal-to-noise ratio and
%resolution can be found in L12. 

For each galaxy, we tabulate the bulge and disc magnitudes and
colours in all $5$ SDSS bands. These values are  Galactic extinction corrected
\citep{SFD1998} and k-corrected to $z=0$ using the \verb+IDL+ package
\verb+kcorrect+ \verb+v4_2+ \citep{Blanton07}. In addition, we
correct the colours and magnitudes of galaxies with discs for
intrinsic extinction using corrections from \citet{Maller09} and
L12. These corrections remove trends in colours with disc inclination,
but they do not correct for extinction due to dust in face-on
discs. Finally, we 
calculate the stellar masses for the bulge and disc using the relation
from \citet{Bell2003}:
\begin{align}
\label{eqn:bell}
\log M/\Msun =& -0.22 + 0.66(g-i) - 0.15 \\\nonumber
&- 0.4(M_r - M_{\odot,r} + 1.3z) \ ,
\end{align}
where $0.15$ accounts for the difference between the diet Salpeter
initial mass function (IMF) used by \citet{Bell2003} and
the Kroupa IMF \citep{Kroupa2002} we employ.  The
colours and magnitudes used for the stellar mass are {\it not}
corrected for intrinsic extinction, in keeping with the derivation of
the relation in \citet{Bell2003}. Because the mass-to-light ratio is a convex function of 
galaxy colour, the sum of the masses of the bulge and disc is always
slightly larger than the mass measured using the total galaxy
colour and magnitude. For most of the galaxies, this difference is
small; the median $M_{\mathrm{total}}/(M_{\mathrm{disc}}+
M_{\mathrm{bulge}})$ is $0.9996$, and for $95$ per cent of the
galaxies, this ratio is between $0.85$ and $1.0$. 
%{\rm 
%The accuracy of the
%bulge and disc masses helps verify our bulge and disc colour
%measurements. 
%The close agreement between the sum of the bulge and
%disc mass and the total galaxy mass is only possible if the
%bulge-to-total ratios in each band agree reasonably well. Thus, the
%accuracy of the bulge and disc masses verifies that the linear-scaling
%procedure to obtain bulge and disc colours is robust. }

\subsubsection{Classifying Galaxies}
\label{sssec:classify}

Since we fit each galaxy with five different models, we require a
method for selecting the best-fitting model.  In L12, we show
that the $\chi^2$ values of the various model fits are indistinguishable at
the resolution of SDSS \citep[but see][]{Simard2011}. Instead, we
develop an algorithm that relies on the sizes, shapes, and colours of the
bulges and discs in order to select the best-fitting, physically-sensible
model for each galaxy. In this work, we present a simplified
classification algorithm that attempts to classify most of the
galaxies in our sample and emphasises the distinction between
classical bulge galaxies and ellipticals. Additionally, instead of
assigning each  
galaxy a best-fitting model, we assign each galaxy a probability of
being fit by each model (often, the  probability is unity for one of
the models).  Although this does not allow us to accurately classify a
given galaxy, it does allow us to study the properties of a large
sample of galaxies. When we examine properties of bulges and discs, we weight each galaxy by its probability of
having a bulge and a disc.  

\tableClass

{\rm We separate galaxies into {\rm five} different categories: bulge-less disc galaxies, disc-less ellipticals, classical bulge$+$disc galaxies, pseudo-bulge$+$disc galaxies, and unclassifiable galaxies. Our goal is to assemble a sample of galaxies which are accurately modelled by a classical bulge plus a disc. %Thus, we identify and set aside pure disk galaxies, ellipticals, %irregular galaxies, and poorly modelled galaxies in the sample. In addition, we %distinguish pseudo-bulges from classical bulges based on recent star formation %in the bulge. 
A brief outline of the classification is given below and summarised in Table \ref{tab:class}. Details can be found in Appendix \ref{app:Class}. 

First, we identify bulge-less and disc-less galaxies. Bulge-less disc galaxies are defined to have $B/T < 10$ per cent. %%In our earlier work, we show that the %%typical uncertainty in $B/T$ is $10$ per cent.  
Disc-less galaxies (ellipticals) are more difficult to identify. The few galaxies with $B/T > 90$ per cent are considered ellipticals. These make up only $6$ per cent of the our sample. As shown in L12 \citep[see also][]{Allen06}, elliptical galaxies are often best fit by a de Vaucouleurs component along with a low surface brightness exponential component. This ``disc'' is not a physical disc and has several possible origins, i.e. the outer halo of ellipticals, a S\'ersic index larger than four, and/or inadequate sky subtraction around bright galaxies in SDSS.  These model ``discs'' make ellipticals indistinguishable from face-on bulge$+$disc galaxies based on the 2-dimensional bulge$+$disc decomposition alone.

However, the distributions of inclination angles for real and spurious discs will be different; the former will be randomly oriented, while the latter will be preferentially face-on. We use this fact to statistically separate ellipticals from face-on classical bulge host galaxies. Galaxies with a small measured disc axis ratio have a high probability of being a bulge$+$disc galaxy, while galaxies with a large disc axis ratio (face-on) might be either a bulge$+$disc galaxy or an elliptical with an exponential halo. For this statistical separation, we examine galaxies with $0.1 < B/T < 0.9$ and  $(u-r) > 2.22$ for \emph{both} the bulge and disc component. This means all the ellipticals we find will be red. Restricting ourselves to red galaxies will enhance the fraction of ellipticals relative to bulge$+$disc galaxies, making the two inclination angle distributions easier to fit. Blue ellipticals are relatively rare \citep{Schawinski2009}, and,  therefore, are  a small contamination in our sample of classical bulge host galaxies.

After setting aside ellipticals, we distinguish between galaxies with quiescent bulges and those with young, star-forming bulges based on the 4000\AA{} break strength ($D_n(4000)$) measured by SDSS\footnote{http://www.mpa-garching.mpg.de/SDSS/DR7/raw\textunderscore
  data.html}. Based on results from morphology studies at higher resolution \citep[e.g.][]{Fisher06}, we associate star-forming bulges with pseudo-bulges and quiescent bulges with classical bulges.  As above, we assign each galaxy a probability of having a classical bulge or pseudo-bulge based on its $D_n(4000)$ (see Fig.~\ref{fig:classpseudo}). %Pseudo-bulges are thought to be %currently forming via secular processes in discs (disc instabilities, bars, %etc.) \citep{Kormendy04}. 
Since pseudo-bulges are a disc phenomenon, we consider galaxies with pseudo-bulges to be bulge-less and exclude them from our analysis. Because $90$ per cent of the star-forming bulges are less massive than the median classical, quiescent bulge in our sample, excluding pseudo-bulge hosts does not significantly affect correlations in galaxy properties with local density \emph{at fixed bulge mass}. However, the separation of quiescent and star-forming bulges will eliminate any star-forming classical bulges from our sample.

Finally, we exclude galaxies which are not well fit by any of the above models. These galaxies are modelled by a single S\'ersic profile. Unclassifiable galaxies make up less than 10\% of our sample. Seventy-five percent of unclassifiable galaxies have a S\'ersic  index less than $2.3$ and the same fraction  lie in the blue cloud ($u-r<2.22$). These galaxies are probably disc-like irregulars, which are unlikely to have a well-defined bulge and disc. The remaining $25$ per cent of unclassifiable galaxies are mostly merger remnants, starbursts, and other complicated morphologies. None the less, because the majority of unclassifiable galaxies exhibit  disc-like properties, we group them with other bulge-less galaxies.}

Below, we use
a bright ($M_{^{0.1}r} < -19.77$) subsample of $29781$ galaxies from
the L12 sample. This sample contains  $12523$ classical bulge$+$disc
galaxies, $3681$ ellipticals, $2214$ pseudo-bulge galaxies, $8505$
bulge-less galaxies, and $2857$ unclassifiable galaxies. These numbers
are inexact, since we only assign each galaxy a probability of being a
certain type.

\subsection{Group Catalogues}
\label{ssec:group_catalogs}
We study the environmental properties of bulges and discs using the
group catalogue from A. Berlind (priv. comm.). This catalogue is
built using the methods in \citet{Berlind2006}, but is based on SDSS
data release 7 instead of data release 4.  The group catalogue allows
us to relate galaxy properties to 
their host group (and dark matter halo) properties as well as to study the
properties of galaxies as a function of intra-group environment. We
select this group catalogue from the many group catalogues available for
SDSS data because it extends to low redshift, and overlaps
significantly with our
sample of bulge$+$disc galaxies.  The group catalogue is
volume-limited and includes all galaxies with  absolute magnitudes
$M_{^{0.1}r} \leq -19.77$ in the redshift range $0.02<z<0.067$. {\rm We include isolated galaxies in the group catalogue (groups of richness one).} The
catalogue is created   
 using a friends-of-friends algorithm to determine group membership
 \citep[see][for details]{Berlind2006}. Two
galaxies are linked if the projected and transverse distances between
them are smaller than $b_{\perp}\bar{n}_g^{-1/3}$ and
$b_{\parallel}\bar{n}_g^{-1/3}$, respectively, where
$b_{\parallel,\perp}$ are the linking lengths and $\bar{n}_g$ is the
average galaxy number density in the sample. Here, 
$\bar{n}_g = 5.275 \times 10^{-3}\, \mathrm{Mpc^{-3}}$, $b_\parallel = 0.14$, and $b_\perp
=0.75$, which corresponds to physical linking lengths of $0.8$~\Mpc{}
in the transverse direction and $300$~\kmps{} along the line of sight.  

%Two environment metrics available from the group catalogue include group
%richness (number of members brighter than $M_{^{0.1}r}-5\log\h_{100}=-19$)
%and projected  distance from group centre. We also use the galaxies in
%the group catalogue to calculate the projected density ($\Sigma_5\
%\mathrm{[Mpc^{-2}]}$) to the fifth nearest  neighbour in a small
%redshift slice for each galaxy. The width of the redshift slice used
%is based on the the velocity dispersion of the galaxy's parent group,
%with a lower limit set by the line-of-sight linking length
%($300$~$\mathrm{km\ s}^{-1}$) used in creating the group catalogue.

In the group catalogue, there are $90893$ galaxies, of which $29781$
have a bulge$+$disc model fit from L12. This subsample includes $83$ per cent
of the galaxies below 
$z=0.05$ with spectroscopic redshifts in the group catalogues.  Galaxies
without spectroscopic redshifts (due to fibre collisions in the SDSS
spectrograph) make up $4$ per cent of the sample below $z=0.05$, and are
therefore a small omission. The remaining $6124$ galaxies are missing from the L12 bulge$+$disc sample because cuts in the galaxy axis ratio ($1289$), model fits with surface brightness consistent with zero ($2830$), and galaxies
which did not make our quality cuts due to problems with deblending
and cosmic ray removal. 
\figComplete

Despite the missing galaxies,  the bulge$+$disc matched sample
is a representative subsample of the group catalogue for $z \leq
0.05$.  This is demonstrated by Fig.~\ref{fig:complete}, which shows
the distributions of group sizes
and projected fifth nearest neighbour densities ($\Sigma_5$, see
\S\ref{ssec:qenv} for explanation) for the full group 
catalogue and for the subsample which overlaps with the bulge$+$disc
decomposition sample. The bulge$+$disc sample follows essentially the
same distributions in $\Sigma_5$ and $N_{\mathrm{gal}}$ as the full
group catalogue. A Kolmogorov--Smirnov (KS) test shows that the
distributions of $\Sigma_5$ (lower panel) are indistinguishable. The
distributions of group sizes are not identical; a KS test yields a
probability of $3\times10^{-5}$ that the distributions are the
same. The bulge$+$disc matched sample is missing galaxies in mid-sized
groups, but the number of missing galaxies is small.
%Fig.~\ref{fig:complete} shows that the smaller, matched sample is missing
%galaxies in mid-sized groups, but that the number of missing galaxies
%is small.  These two 
%tests demonstrate that we are not missing galaxies from a specific
%environment. 
%In addition, we ascertain that 
%the fraction of missing galaxies in a group is not a function of
%group size. 
Furthermore, for the larger groups, the missing galaxies are not a
function of position in the group or of local density. This is not
surprising, since the number of galaxies missing a spectrum due to
fibre collisions is a small fraction ($4$ to $6$ per cent) of the
sample. If fibre collisions were more prevalent, we would expect to be
missing more galaxies in high density regions than in low density regions.

\figMagBerlind
We also check that matching the L12 sample to the group catalogue does
not change the distributions 
of galaxy morphology as a function of magnitude. The top panel of
Fig.~\ref{fig:magBerlind} shows the absolute magnitude distribution of
the total L12 sample, down to a magnitude of $M_r=-19.77$. The lower
panel shows the same for the galaxies from the group catalogue. Although
there are small differences at the faint end, the distributions of
galaxy morphology as a function of magnitude are essentially the same
for the both samples. The fraction of galaxies of each type in the two
samples is the same to within $450$ galaxies, or $1.5$ per cent of
the grouped sample. 

In this work, we address the effects of environment on galaxy
properties at fixed bulge
mass. Therefore, although it is important that certain types of
galaxies are not systematically excluded as a function of environment,
we are not concerned with the overall completeness of the sample. We
are only concerned that the galaxies in the different environments are
a representative sample, which is demonstrated by Figs.
\ref{fig:complete} and \ref{fig:magBerlind}.

%%Environment measures
\subsection{Measuring environment}
\label{ssec:qenv}
In order to examine the correlations between galaxy properties and
environment, we
use several measures of environment, including both group halo
properties and more local measures of environment. Directly from the
group catalogue, we obtain the group richness, $N_{\mathrm{gal}}$, the
group line-of-sight velocity dispersion, and the 
total stellar mass in galaxies brighter than  $M_{^{0.1}r} = -19.77$. Using
the relation  between total group stellar mass and halo mass from
\citet{Leauthaud2012}, we calculate the group dark matter mass, $M_{200}$,
defined as the mass enclosed in a region $200$ times denser than the
critical density\footnote{\citet{Leauthaud2012} report halo masses as
    $500$ times the critical density. We convert to $M_{200}$ using
    the concentration-mass relation from \citet{Maccio2008} for a 
    WMAP5 cosmology.}.
 We do not take into account the difference in stellar mass
 completeness between our sample and that used in
 \citet{Leauthaud2012}, nor do we correct the stellar masses for contamination from  non-group galaxies. However, we expect the corrections to the
 stellar masses to be small \citep[see][]{Leauthaud2012}. 

We can compute the
 line-of-sight velocity dispersion ($\sigma$) for a halo of a given
 mass using  the $M_{\mathrm{halo}}-\sigma$ relation from
 \citet{Yang2007}. For groups with more than $10$ galaxies, the
 $\sigma$ obtained from the dark matter halo mass is a factor of $1.4$
 larger than the $\sigma$ measured directly from the galaxies. 
   Half of this discrepancy is due to the small value 
 for the line-of-sight linking length used to build the group
 catalogue. The small $b_\parallel$  biases the measured velocity dispersion down by $\sim 20$ per
 cent \citep{Berlind2006}. In the following analysis, we do not make
 extensive use of the  dark matter halo mass. We do use the group
 velocity dispersion, but 
 since we are only interested in making comparisons between different
 environments, the absolute values of $\sigma$ are not relevant. 

The group catalogue also contains the the projected distance, $R_p$,
from each galaxy to its host group centre (the number-weighted mean angular
position). We use this distance, along with the velocity dispersion of
the group to define the crossing time $\tc =
R_p/\sigma$, where $\sigma$ is the the group velocity dispersion,
measured directly from the galaxy redshifts
\citep[see][]{Berlind2006}. %The crossing time is similar to the
%position of the galaxy in a group normalised by the the virial radius
%of the group. 
Clearly, \tc{} is only a sensible measure of environment for
relatively large groups and clusters, which have a definite centre.
%Both of 
%these measures are only sensible proxies for environment in large
%groups and clusters ($N_{\mathrm{gal}} \gtrsim 20$). Additionally,
While the number-weighted (or mass-weighted) centre of a group always is
well-defined, the FoF grouping algorithm does not guarantee it is physically
meaningful. This is especially true for non-relaxed groups. We will
address this problem in our investigation of intra-group trends by limiting our sample to galaxies from relaxed
groups (\S \ref{sssec:ingrouptrends}). 

Finally, in addition to the halo properties, we measure
the local surface density around each galaxy. We use the projected
fifth nearest neighbour density ($\Sigma_5\ [\mathrm{Mpc}^{-2}]$). Neighbours
are selected from the volume-limited group catalogue in a redshift
slice which has a width equal to the velocity dispersion of a galaxy's
host group. The minimum width is 300 \kmps{}, the line-of-sight
linking length used in the group catalogue. For galaxies in large
clusters, $\Sigma_5$ is a measure of their immediate vicinity, not the
underlying large-scale dark matter density field
\citep{Muldrew2012}.

In the sections below, we rely mostly on $\Sigma_5$ as a metric
for environment. We will demonstrate that our results
are essentially unchanged if group richness, $N_{\mathrm{gal}}$,  is
substituted for $\Sigma_5$. In \S \ref{sssec:ingrouptrends} we
explore trends in galaxy properties within massive, relaxed groups and
use the crossing time, \tc{}, as a measure of galaxy environment. 

%%Whole galaxy properties with environment
%%
\section{Whole galaxy properties with $\Sigma_5$}
\label{sec:wholeGalEnv}
\figGoto

Before examining trends in bulge and disc properties with density,
we confirm trends in whole galaxy properties with density. The
low redshift morphology--density and colour--density relations have been
measured numerous times using various data sets
\citep[e.g.][]{Dressler80, Goto2003, Balogh2004, DePropris2004,
  Kauffmann2004, Tanaka2004, Weinmann2006,   Blanton2005a, Hansen2009,
  Skibba2009, Bamford09}. First, we confirm that the morphological 
classifications described in \S\ref{sssec:classify} follow the
expected morphology--density relation. Fig.~\ref{fig:fracSig5} shows
the distribution of the five morphological types   as a
function of $\Sigma_5$. The 
trends with density are in agreement with the morphological trends
from other studies \citep[e.g.][]{Dressler80, Goto2003,
  Bamford09}. Our results are in close agreement with those from
\citet{Goto2003} who find that early and intermediate disc galaxies
(Sa and S0) dominate in almost all environments, while late-type discs
drop off quickly at high densities. In our sample, classical
  bulge galaxies dominate 
  at all but the  lowest densities. The fraction of classical bulge host galaxies rises  quickly above $\Sigma_5\approx1$ Mpc$^{-2}$. 
%We find more S0s than
%elliptical galaxies at all densities. This is probably due to our
%separation of ellipticals and S0s, which finds many face-on
%S0s. 
Figure \ref{fig:fracSig5} is missing roughly $1300$ galaxies which
have axis ratios below $0.2$. These galaxies are excluded from the
bulge$+$disc catalogue in L12 because our exponential disc model is
unsuitable for edge-on discs. Including these galaxies slightly
decreases the  fraction of elliptical galaxies, but the changes are
small and the overall trends remain the same. 

Note that although both \citet{Goto2003} and this work use $\Sigma_5$
to measure local environment, the numerical values cannot be directly
compared since $\Sigma_5$ depends on the lower flux limit of the
galaxy catalogue used and the width of the redshift interval
used. \citet{Goto2003} measure the distance to the fifth nearest
neighbour within a redshift slice of $\pm 1000$ \kmps{}, while the
redshift slice we use depends on the galaxy's host group velocity
dispersion and is typically smaller.  

\figMorphMass
Fig.~\ref{fig:fracSig5Mass} shows the morphology trends for galaxies
divided into three
stellar mass bins. The stellar mass is computed using the most 
likely model for each galaxy (although the difference in total
stellar mass between two reasonable model fits is always small). The trends
with environment are the same for all three mass bins, but the fraction of
early/late type galaxies is a strong function of mass in all
environments, in agreement with previous results
\citep[e.g.][]{Kauffmann2004, Blanton2009, Bamford09}. Classical
bulges dominate over all other types of galaxies in
the two highest mass 
bins. %As expected from Fig.~\ref{fig:S0E}, the ratio of
%ellipticals to S0s increases dramatically with galaxy mass, but this
%ratio is not a function local density at a given mass. This suggests
%that, among red galaxies, there is little mass segregation as a
%function of environment \citep[e.g.][]{Rood1972}. 

Since colour
and morphology are generally correlated, the morphology--density
relation implies a relation between local density and galaxy colour
such that red galaxies dominate in high density regions while blue
galaxies are dominant in the field \citep{Gomez2003,
  Hansen2009,Balogh2004, Baldry2006, Skibba2009}. There is evidence
that the correlation between colour and density is more fundamental
than the correlation between morphology and density
\citep{Kauffmann2004, Christlein2005, Skibba2009}. We show the fraction of
red ($u-r>2.22$) galaxies as a function of environment and mass in
Fig.~\ref{fig:redfrac}. In 
agreement with previous work, we find the fraction of red galaxies is
an increasing function of galaxy mass and
environment. For all masses, the fraction of red galaxies increases
sharply near $\Sigma_5 \approx 1$ Mpc$^{-2}$, in agreement with results from
\citet{Gomez2003}, who find a break in the star formation rate--density
relation at approximately the same local density. 
%\citet{Balogh2004}
%show that this increase in red 
%galaxies is not due to changes in the colours of red or blue galaxies,
%but rather a shift in the morphological fractions at a given local
%density. 
Figs.~\ref{fig:fracSig5} and \ref{fig:redfrac}
show that both the colour--density and morphology--density relations hold for galaxies of constant stellar
mass \citep{Bamford09, Skibba2009}. Additionally, comparing the two
figures demonstrates that the red fraction of galaxies increases
sharply at the same  local density as the classical bulge host and elliptical galaxy fractions. {\rm We note ellipticals are not
the majority of red galaxies, but only contribute $25$ per cent by number;
the remainder of red galaxies are classical bulge hosts, with a small
contribution ($5$ per cent) from unclassifiable galaxies.}
\figredfrac

In the following section, we explore the colour--density and
morphology--density relations for classical bulge host
galaxies. The colours of these galaxies follow the same trends as
the colours of the galaxy population as a whole. Using bulge$+$disc
decompositions, 
we can study the colour--density relations for bulges and discs
separately, and determine whether changes in galaxy morphology (e.g. $B/T$ ratio) or changes in the component colours drive the increase in the fraction of galaxies with red integrated colour as a function of $\Sigma_5$. 

%In this work, we focus on galaxies with both a disc and a bulge, and
%mainly on galaxies with a prominent classical, de Vaucouleurs
%bulge. This includes S0s and ellipticals (although the latter lack
%discs). We exclude galaxies with pseudo-bulges, counting them among
%bulgeless galaxies. Since pseudo-bulges
%arise from disc instabilities \needref, we do not expect their properties
%to follow the same relations with environment as classical bulges and
%their discs. However, since the distinction between classical bulges
%and pseudo-bulges is uncertain, we will show their inclusion does not
%significantly alter our results. Additionally, galaxies which are not
%well-fit by either bulge$+$disc model (S\'ersic galaxies) are placed
%in the same category as bulgeless galaxies. These galaxies make up
%less than ten per cent of the sample and seventy-five per cent of these
%galaxies have $n < 2.6$ and $u-r < 2.22$, suggesting most of these
%galaxies are small, disc-like starforming galaxies.

\section{Bulge and disc properties with $\Sigma_5$}
\label{sec:bdGalEnv}

Many studies show that a galaxy's properties are largely
determined by a its stellar mass \citep[e.g.][]{Brinchmann2000,
  Kauffmann2004} and dark matter host halo 
mass \citep[e.g.][]{Blanton2006, Weinmann2006}. Since the most massive haloes are
strongly clustered, galaxy mass and environment are strongly
correlated. However, there are correlations with environment at fixed
stellar mass \citep[e.g.][]{Balogh2004, Bamford09, Peng2011, Grutzbauch2011,
  Cooper2010, Cibinel2012}, 
and these are the trends we explore. Below, we show trends with
environment at fixed bulge mass; however, our results are essentially
unchanged if we choose to hold total galaxy stellar mass fixed. In
this section, we focus on our sample of $12500$ galaxies which have
both a classical bulge and a disc. When calculating medians and
Spearman rank correlation coefficients, we weight each galaxy by
its probability of having a classical bulge and a disc, as explained
in \S\ref{sssec:classify}.

%As discussed above, we choose to
%exclude pseudo-bulges from the sample since the physical origin of
%pseudo-bulges is different from that of classical
%bulges. Pseudo-bulges arise from disc phenomena and are therefore,
%more closely associated with bulge-less disc galaxies. Since pseudo-bulges
%are typically low mass, including them
%would only contribute significantly to the lowest bulge mass bin in
%the next sections. Furthermore, since pseudo-bulges 
%are bluer and reside in lower density environments than classical
%bulges (see Fig.~\ref{fig:fracSig5Mass}), including pseudo-bulges in
%our bulge$+$disc sample would steepen any existing trends with
%$\Sigma_5$.  We also place unclassifiable galaxies  (S\'ersic
%galaxies) in the same category as bulgeless galaxies. These galaxies
%make up  less than ten per cent of the sample and are have properties
%similar to bulge-less disc galaxies; seventy-five per cent of
%unclassifiable galaxies have $n < 2.6$ and $u-r < 2.22$. 

%\subsection{Colors}
%\label{ssec:bdcolors}
\figRfiveColor

The left panel in Fig.~\ref{fig:r5Color} shows the colours of galaxies
with classical 
bulges and discs as a function of $\Sigma_5$. We divide the galaxies
into quartiles based on bulge mass. We remove galaxies with the most and least
massive $0.5$ per cent of bulges in order to eliminate galaxies with
large errors in modelled colours and masses. The mass bin divisions
are at $\log M_{\mathrm{bulge}}/M_\odot =[9.3,10.1,10.3,10.6,11.2]$. These mass
bins are then divided into six bins of equal galaxy number at
different $\Sigma_5$. Thus, each point in Fig.~\ref{fig:r5Color}
represents $1/24$ of the sample ($\sim 520$ galaxies). These points are the weighted
median colour in each bin. We define the weighted median as the value
for which the sum of weights for values smaller than the median is
equal to the sum of weights for values larger than the median. The
thin lines in Fig.~\ref{fig:r5Color} denote the similarly-defined
weighted inter-quartile ranges. Although there is significant scatter
in the colour, 
the trends with environment are statistically
significant\footnote{By statistically significant, we mean that the probability
  that the abscissa and ordinate are uncorrelated is less than $0.3$
  per cent, corresponding to a $3\sigma$ result for Gaussian
  distributions.}.  
The trend in total galaxy colour is simply the colour--density relation for classical bulge$+$disc
galaxies. The weighted Spearman rank coefficients are given in
Table \ref{tab:spear}. We also calculate the best-fitting linear
slopes of the relation (see Table \ref{tab:spear}). In order to reduce
the errors on the slope, we remove points offset  from the fitted relation by more than $3$ standard
deviations and then re-fit the
relation using the slightly smaller sample. 
 Despite being statistically significant, {\rm the change in 
median integrated colour is small, $\sim 0.01-0.03$ per dex in $\Sigma_5$. The change is largest for galaxies which host low mass bulges.} This is in agreement
with the conclusions of \citet{Bamford09}, who show that the
colour--density relation is most significant for lower mass galaxies. 

The
middle and right panels of Fig.~\ref{fig:r5Color} show the colour
density relation for the bulge and disc components of these galaxies
separately. It is immediately evident that the change in disc colour
with $\Sigma_5$ must contribute significantly to the change in total
galaxy colour. In fact, we demonstrate below that the change in disc
colour is the 
{\it only} contribution to the change in total colour. In the
following subsections, we will discuss the  
changes in bulge properties (\S \ref{sssec:bcolors}),
and the changes in disc properties (\S \ref{sssec:dcolors}) as functions of
$\Sigma_5$. 
 \tableSpear

\subsection{Bulges}
\label{sssec:bcolors}
The classical bulges in Fig.~\ref{fig:r5Color} were selected to
have large
$4000$\AA{} breaks, similar to ellipticals.   Elliptical 
galaxy colours depend on galaxy mass \citep[e.g.][]{deVaucouleurs61,
  Brinchmann2000}, but only weakly, if at all, on environment
\citep[e.g.][]{Dressler87, Bernardi2003III, Balogh2004, Trager2008,
  Hansen2009}. The 
central panel in Fig.~\ref{fig:r5Color} shows that the median bulge 
colours lie on the red sequence ($g-r\sim0.8$), and that there is a
small increase ($\Delta(g-r) \approx 0.03$) in median colour from the
lowest mass bulges to the highest mass bulges. The weakness of this
trend  may be due in part to large scatter in the model bulge colours
compared to the scatter in model elliptical colours (see fig. 32 in
L12).  

In addition to trends in colour with bulge mass, there is a weak but
statistically significant anti-correlation between environment and
bulge colour; bulges in low density regions are
redder than those in higher density regions. The Spearman rank
correlation coefficients for this trend are listed in Table
\ref{tab:spear}. This result is in contrast to the results of
\citet{Hudson2010} who find no 
variation in median bulge colour with increasing local
density. This trend is strongest for the second lowest bulge mass bin (red
circles).  Since pseudo-bulges are
typically blue and are not as strongly clustered as classical bulge
galaxies (see Figs.~\ref{fig:fracSig5} and
\ref{fig:fracSig5Mass}), contamination from pseudo-bulges will tend to
flatten
the trend in bulge colour for the lowest mass bulges, possibly explaining the weaker trend for the lowest mass bulges. 

Although the
bulge colours decrease as a 
function of environment,  there is  no significant correlation between
$\Sigma_5$ and SDSS fibre colours, and the correlation between
$\Sigma_5$ and $D_n(4000)$ is in the opposite sense (see
Fig.~\ref{fig:dn4000r5}). However, if we replace the L12 bulge$+$disc
decompositions with those from \citet{Simard2011}, the same trend in
bulge colour with density is recovered. 
 
We propose that the decrease in bulge colour with environment
is an artefact of 2-dimensional bulge$+$disc decompositions. Specifically, the
colours of small bulges in blue discs are biased redwards by
the bulge$+$disc decomposition. The disc model used in L12 (and
\citet{Simard2011}) is an exponential profile, which is centrally
peaked. Therefore, the disc can contribute significantly to the
central flux of the galaxy. Moreover, since the disc is usually bluer
than the bulge (especially in low density regions), the bulge$+$disc
decomposition will attribute blue light from the central region to the
disc, thus making the bulge appear redder. This effect will be largest for
the smallest bulges in the bluest discs. Since the bulge-to-total flux
ratio and disc colour are both increasing functions of environment, 
the bias in bulge colour will be largest for small bulges in low density
environments.

\figBulgeLdiscColor
We test this explanation by refitting classical bulge galaxies with
models that suppress the disc flux in the central
region. In the new models, the disc flux goes rapidly to zero within
one bulge scale radius, $I_{\mathrm{disc}}\propto
(r/r_{\mathrm{bulge}})/[1+(r/r_{\mathrm{bulge}})^4]^{0.25}$. Keeping all the other model
parameters fixed, we linearly scale the bulge and disc flux of this
new model to fit each galaxy. By design, the new fits have larger
bulge-to-total flux ratios, and the differences are largest for
galaxies with small bulges. The resulting trends in bulge colour with
environment for these disc-suppressed models  are shown in
Fig.~\ref{fig:BcolorLD}. There is only a 
statistically significant trend with mass for galaxies with $10.2 <
\log M_* \leq 10.5$ (note that because the bulge and disc masses have
been recomputed, the mass quartiles in
Fig.~\ref{fig:BcolorLD} are different from those in
Fig.~\ref{fig:r5Color}). The choice of 
parametrisation for the central region of the disc plays an important
role in the bulge colours. 

In general, it is unknown whether the
stellar disc continues unchanged through the bulge or if the disc only
exists outside the central bulge. In the Sombrero galaxy,
photometry and spectroscopy show an inner cutoff for the stellar
disc, suggesting the disc-suppressed models may be a better choice for
early-type disc galaxies \citep{Emsellem1996}. In this
work, we will continue to use the centrally-peaked model for the disc,
knowing that the colours of small bulges will be biased
redwards. 

This bias will also occur in the opposite case, when a
small disc surrounds a large, red bulge. However, we do not expect the
bias in disc colours of galaxies with large 
bulges to be as severe. 
%First, the measured disc
%colours are more robust than the bulge colours (L12;
%\citealp{Simard2011}), so the errors in disc colour are typically
%smaller than the errors in the bulge colours.  Second, 
Since the
majority of the bulge flux comes from regions above and below the disc, there is
no physical basis for the bias in disc colour, as in the case for
small bulges.

\subsubsection{Spectroscopic Properties}
\label{sssec:spectro}

\figbulgedn

Recent work has shown statistically
significant positive correlations  between stellar age and local density \citep[e.g.][]{Thomas05,
  Bernardi2006, Clemens2006, Smith2006, Cooper2010} and between metallicity and local density \citep{Cooper2010} for early type, or red sequence,
galaxies. Our sample of classical bulge$+$disc galaxies contains both early
type, passively evolving, galaxies and galaxies with ongoing star
formation in their  discs. However, we expect  all classical bulges
and ellipticals to follow the same relations with stellar mass and
environment.  

In order to examine the stellar population properties of classical
bulges more closely, we use the SDSS fibre spectra and the line
indices and metallicities 
reported in the MPA/JHU SDSS spectroscopic catalogue.  At low
redshift, the fibre spectra are dominated by bulge stellar light, but
the mixture of bulge and disc light in the fibre is a function of
redshift. These aperture effects may influence correlations between
measured line indices and $\Sigma_5$.
%For galaxies at
%$z\approx 
%0.05$, the median classical bulge scale length is 2.9 arcsec, while
%the median disc scale length is 8.9 arcsec. In the inner $3$ arcsec, the
%median $B/T$ flux ratio is $0.74$ and eighty per cent of the galaxies
%have $B/T$ larger than $0.5$.
For classical bulge$+$disc galaxies in our sample, the median $(B/T)_r$ in the inner 3 arcsec is $0.74$, and eighty per cent of the galaxies have $(B/T)_r$ within $3$ arcsec 
larger than $0.5$. For galaxies at $z>0.04$, the median ($3$ arcsec) $(B/T)_r$ is only
slightly smaller at $0.72$.  Therefore, the 3 arcsec SDSS spectroscopic
fibres are dominated by bulge stellar light, even at the highest
redshift ($z=0.05$) in our  sample. 

Since our sample is in the regime
where angular size scales linearly with distance, the 
spectroscopic measurements will include more disc flux at higher
redshift. This effect is partially countered by a slight bias
toward larger bulges at higher redshifts; in this sample, the average {\it physical} bulge size increases by 8 per cent from $z=0.02$ to $z=0.05$, due to the fact that small bulges at high redshift are more difficult to
accurately fit. However, there is no statistically significant
trend in $\Sigma_5$ with redshift. Therefore, even though the
the spectroscopic properties do change with redshift, the 
correlations between spectroscopic properties and $\Sigma_5$ will be
unaffected, since $\Sigma_5$ and $z$ are uncorrelated.  

The top panel of Fig.~\ref{fig:dn4000r5} shows the $4000$\AA{}
break (D$_n(4000)$) \citep{Balogh1999} as a function of $\Sigma_5$ for
four different bulge masses.  In this figure, we include both
classical bulges and ellipticals, since we expect the stellar
populations of classical bulges and ellipticals to be the same at a
given mass \citep[][]{MacArthur2010}. Indeed, excluding
ellipticals from Fig.~\ref{fig:dn4000r5} 
does not noticeably alter the results. The four mass bins used are
the same as in Fig.~\ref{fig:r5Color}. The increase in D$_n(4000)$ as
a function of density is statistically significant for all bulge
masses, but the 
trend is strongest for the lowest mass bulges. 
%This is consistent with
%the trend in bulge $u-r$ colour, which increases noticeably for the
%lowest mass bulges as a function of $\Sigma_5$. 
These results imply
that bulges in
high density environments have had less recent star formation than
those in lower density environments. The same results are obtained if we plot
the line index H$\delta_A$ as a proxy for stellar age (the equivalent
width of H$\delta$ is anti-correlated with environment)
\citep{Kauffmann2003b}. The increase in stellar age as a function of
local density at fixed bulge mass is in agreement with previous results
\citep[][but see \citealp{Thomas2007}]{Trager2000, Kauffmann2004, Thomas05,
  Clemens2006, Bernardi2006, Smith2006,
  Cooper2010}. \citet{Cooper2010} note that the trend in age of red
sequence galaxies with density is evidence for galaxy assembly bias
\citep{Croton2007}; namely, older galaxies are more strongly clustered
than younger ones at fixed stellar mass.

The lower panel of Fig.~\ref{fig:dn4000r5} shows stellar metallicity
as a function of $\Sigma_5$. The values for $\log Z/Z_\odot$ are taken
from \citet{Gallazzi2005,
  Gallazzi2006}\footnote{http://www.mpa-garching.mpg.de/SDSS/DR4/
Data/stellarmet.html}. The 
metallicities are only computed for galaxies from SDSS data release 4,
which includes approximately half of the sample used in this
paper. \citet{Gallazzi2005} compute the stellar ages and metallicities
by fitting model spectra from \citet{Bruzual2003} to a combination of
iron and magnesium line indices, the 
$4000$\AA{} break, and three Balmer lines. The trend in stellar
metallicity at fixed bulge mass is statistically significant for all
but the second lowest mass bin (red circles). The trends are strongest for
the highest mass bin, where the metallicity changes by $0.01$ dex per
decade in $\Sigma_5$. Despite being statistically significant, the
change in metallicity is quite small.%, especially compared to the
%change in $D_n(4000)$ as a function of bulge stellar mass. 

The
increasing trend in stellar ages {\it and} metallicities is orthogonal
to the age-metallicity degeneracy \citep{Worthey1994, Gallazzi2005},
which makes the results 
more robust. The small increase in metallicity is in agreement with
results from \citet{Cooper2010} \citep[but see][]{Thomas05,
  Smith2006}, as well as studies of the gas-phase metallicity which
show an increase in metallicity in star-forming galaxies as a function
of local density \citep{Cooper2008}. These results demonstrate that bulges
in high density regions formed earlier and with higher star formation
rates \citep[see][]{Cooper2010}. {\rm The increases in stellar age and metallicity with increasing density suggest that classical bulges should be redder in higher density regions. This disagrees with the results from \S \ref{sssec:bcolors}, where we find that bulge colours may even be slightly bluer in high density regions. This disagreement is likely due to biases in our model bulge colours. %It may also indicate a decrease in dust extinction as a function of local density.
}
%Since the majority
%of galaxies shown in Fig.~\ref{fig:dn4000r5} are bulge$+$disc
%galaxies, and not necessarily red sequence galaxies, the agreement
%between our results and those of \citet{Cooper2010} show that bulge
%stellar populations follow the same relations with local density as ellipticals,
%regardless of whether the galaxy has a disc or not. 

\subsection{Discs}
\label{sssec:dcolors}

In the previous section, we show that the change in bulge colours with
local density is small, and any statistically significant change is
probably due to our choice of disc profile for galaxies with small
bulges. On the 
other hand, the trends in disc colour with density are
statistically significant. These are shown in the right panel of
Fig.~\ref{fig:r5Color}. The Spearman rank
coefficients and slopes of the linear relation between $\log \Sigma_5$
and $(g-r)_{\mathrm{disc}}$ are given in Table
\ref{tab:spear}. Although the correlations are statistically
significant, the changes in disc colour are small; the colour only
increases by $\sim 0.015$ mag per dex in $\Sigma_5$. However, this change is $2-3$
times larger than the change in colour measured for bulges. For all
four bulge mass bins, the slope
of the relation between $\log \Sigma_5$ and disc colour is comparable or
larger than the slope in total galaxy  colour, which suggests the
change in integrated galaxy colour can be fully explained by the change in disc colour. We will
return to this conclusion  in \S\ref{ssec:discSize}. 

As with the bulge
colours, there is significant 
scatter in disc colour at fixed bulge mass; the inter-quartile range
of $(g-r)_{\mathrm{disc}}$ colours is typically $0.1-0.2$ mag. This scatter is due to
  the scatter in disc mass at fixed
  bulge mass and fixed  $\Sigma_5$. Disc mass and disc colour are
  strongly correlated \citep[e.g.][]{deJong1994}, so a range of disc
  masses will yield a range of 
  disc colours.  We show below that the correlation with disc mass and
environment does {\it not} contribute to the trend in disc colour with
environment.  In Fig.~\ref{fig:r5Color}, we plot the relation between $\Sigma_5$ and disc
colour for bulge-less discs, including pseudo-bulges and unclassifiable
galaxies (grey crosses and dashed
lines). This relation is offset to lower values of $\Sigma_5$ since
bulge-less discs are typically found in lower density
environments. None the less, discs in galaxies without prominent bulges
follow the same colour--density relation as discs with large
classical bulges.  

Like the change in total galaxy colour, the change
in disc colour is largest for discs around the smallest bulges.  This 
does not extend to bulge-less disc galaxies, for which the relation
between $\log \Sigma_5$ and $(g-r)_{\mathrm{disc}}$ is not as steep as
the relation for galaxies with low, but non-zero, mass bulges (see
Fig. \ref{fig:r5Color} and Table \ref{tab:spear}). However, since
each bulge mass bin includes a large range of disc masses, this does
not contradict the observation that the colour--density relation is
strongest for low- (total) mass galaxies \citep{Bamford09,
  Tasca2009}. 
%morphology--density relations are strongest for lower mass galaxies
%\citep{Bamford09, Tasca2009}. 

The trend of
increasing disc colour with environment is a signature of star
formation being halted in denser environments. Our
results are in agreement with earlier studies of discs becoming redder in
cluster environments \citep[e.g.][]{Hashimoto1998, McIntosh2004, Hudson2010}
and  the rise of red disc  galaxies (anaemic spirals) with increasing
local density \citep[e.g.][]{Dressler80, Goto2003, Gomez2003,
  Bamford09}.
% The
%physical explanation for this effect is elimination of a disc's cold
%gas supply, which will end star formation.  
Two models for
removing gas from galaxies in dense environments include ram-pressure
stripping by the intra-group medium \citep{GunnGott72}, and the
removal of the hot halo gas supply around disc galaxies (strangulation)
\citep{Larson80}. These two mechanisms have different 
timescales for shutting off star formation; ram-pressure stripping
almost immediately ends star formation, while galaxies losing their
halo gas reservoirs undergo an exponential decay in star formation
rate \citep{Balogh2000, vandenBosch2008}. From the disc fading shown
here, these two processes are indistinguishable. We return to the
differences between ram-pressure stripping and strangulation in
\S\ref{sec:richness} when discussing the correlation between disc colour
and $\Sigma_5$ for rich and poor groups, separately.  

Fig.~\ref{fig:r5Color} shows that the most massive 
bulges in our sample also have the bluest median disc colours (black
crosses in 
Fig.~\ref{fig:r5Color}). This seemingly contradicts observations that find
redder, and presumably more massive, bulges have redder discs
\citep{deJong1994,Peletier96,Wyse1997,Cameron09}. However,
Fig.~\ref{fig:r5Color} and \ref{fig:BcolorLD}, show little change in
bulge colour as a function of bulge mass. Furthermore, while we separate the
sample based on bulge mass, the disc mass does not monotonically
increase with bulge mass.  The trend of decreasing disc
colour with increasing bulge mass exists even if the inclination 
correction is removed (see Fig.~\ref{fig:Qdiscgr}). This trend
  may be partially explained by the same modelling artefact which
  affects bulge colours (\S \ref{sssec:bcolors}).  In this case, the
  subtraction of a 
  large, red bulge from 
an image will leave behind an abnormally blue disc. This modelling effect will
be strongest for galaxies with large red bulges, and relatively small
discs. 
%However, the
%colours of discs shown in Fig.~\ref{fig:r5Color} are not exceptional,
%as is the case for bulge colours (the bulge colours are unusually
%red). 
%The second effect is due to the 
%algorithm for distinguishing ellipticals from bulge$+$disc
%galaxies. Galaxies with massive bulges and red discs have a
%significant probability of being an elliptical. If we overestimate the
%number of ellipticals, then the average disc colour around classical
%bulges will be shifted bluewards. 

\subsubsection{Disc mass and size}
\label{ssec:discSize}

In \S\ref{sssec:dcolors}, we show that disc colour and environment are
significantly correlated, thus disc fading is a significant
contribution to
the colour--density relation. However, morphological transformation
could also play a role. If discs are being stripped at the same time
they are fading, we expect that the mass contribution from the disc to
be a decreasing function of $\Sigma_5$. Since bulges are typically 
redder than discs, decreases in disc mass will lead to increases in
total galaxy colour.  
\figDTTrhoFive
In Fig.~\ref{fig:r5DTT} we show the disc-to-total
stellar mass ratio ($D/T$) for galaxies in four bulge mass bins. The
mass bins are the same as in Fig.~\ref{fig:r5Color}. We find no
statistically significant trend in $D/T$ with $\Sigma_5$. For the
highest bulge mass bin (black $\times$s), the median $D/T$ is above 10
per cent, the cutoff $D/T$ below which 
we classify galaxies as ellipticals. Therefore, it is unlikely that
the constant $D/T$ at high bulge mass is due to the minimum detectable
$D/T$. Although we
have separated the galaxies by bulge mass, changes in median bulge
mass {\it within} each bin could affect the relation between $D/T$ and
$\Sigma_5$. We confirm there is no statistically significant change in
median bulge mass in each of the four mass bins as a function of
$\Sigma_5$. Therefore, at fixed bulge mass, the disc mass is independent of
local density, and the changes in total galaxy colour
as a function of $\Sigma_5$ are due solely to the changes in disc
colour. 

\figDRerhoFive
Fig.~\ref{fig:r5DRe} strengthens the argument against morphological
transformations by demonstrating there are no statistically
significant correlations between disc half-light radius (measured in
the 
$r$ band) and $\Sigma_5$ at fixed bulge mass.  Neither the disc size nor
mass changes significantly with 
increased density, as would be expected if galaxy harassment or 
mergers play a major role in galaxy
evolution in high density environments. These results are in agreement
with results from \citet{McIntosh2004}. They find large differences in
star formation 
rates between cluster galaxies and field galaxies, as evidenced by 
differences in disc colours and disc structures (e.g. spiral arms) between
cluster and field galaxies, but they do not find changes in the
bulge-total-ratio distributions for field and cluster galaxies at
fixed galaxy colour.

It is important to keep in mind that Figs.~\ref{fig:r5DTT} and
\ref{fig:r5DRe}  are based entirely on galaxies with classical bulges and
discs. If we include ellipticals ($D/T=0$) in the sample, the median
$D/T$ is a very weakly decreasing function of 
$\Sigma_5$, but only for the highest mass bulges. This is expected
since ellipticals are more  strongly clustered than typical 
bulge$+$disc galaxies (see Fig.~\ref{fig:fracSig5Mass}). 
%However, the
%differences in $D/T$ as a function of bulge mass are substantially
%larger than the changes in 
%$D/T$ as a function of $\Sigma_5$, suggesting that the median $D/T$ is
%not correlated with environment beyond its correlation with galaxy
%mass. 
Similarly, if we include bulge-less 
galaxies in Fig.~\ref{fig:r5DTT} (and separate galaxies into bins of
constant total stellar mass instead of constant bulge mass), the
median $D/T$ is a decreasing function of $\Sigma_5$.  By focusing on
galaxies with a bulge and a disc, we are investigating whether there
are morphological transformations within this population that help
explain the transition from disc-dominated galaxies in the field to
bulge-dominated galaxies in large groups and clusters. The lack of
evolution in $D/T$ with increasing $\Sigma_5$ indicates that either
higher density environments do not lead to disc destruction or the
transition from a disc-dominated galaxy to a bulge-dominated galaxy
occurs on very short timescales, and we do not observe any galaxies in
this transitional phase. 

\subsubsection{Inclination Angle}
\label{ssec:discq}

The previous section demonstrates that the changes in total galaxy
colour for bulge$+$disc galaxies are due to changes in disc
colour. However, these results could be biased by the inclusion of
ellipticals in high density regions. In this section, we address this
concern by dividing the sample into three bins of disc inclination
angle.  If we mistakenly model the
elliptical outskirts and halo as a disc, we would expect no colour
change with environment for face-on discs, where contamination from
ellipticals is most significant. Conversely, if contamination from
ellipticals is zero and our sample does not suffer from any biases
based on disc inclination, we  expect the same trends in bulge and
disc properties with environment, irrespective of disc inclination.   

\figQdvcDTT
Figures \ref{fig:QdttMass} and
\ref{fig:Qdiscgr} show the disc 
colour and disc-to-total stellar mass ratio as a function of
$\Sigma_5$ for three bins in disc inclination angle. The galaxies plotted are the same as in previous figures
(classical bulge$+$disc galaxies).  The bins in
inclination are chosen such that each panel shows the same number of
galaxies. Assuming a flat disc, the inclination angle limits for each
panel are ($76\degree$, $63\degree$, $46\degree$,
$20\degree$), where $0\degree$ is face-on. From Fig.~\ref{fig:S0E}, it
is clear that galaxies in the 
leftmost panels of Figs.~\ref{fig:QdttMass} and \ref{fig:Qdiscgr}, with
$q_d < 0.46$,  have a very low probability of being
ellipticals, while the galaxies in
the rightmost panel are the most likely to include ellipticals. The galaxies are divided into four bulge mass bins, using the same
divisions as in Fig.~\ref{fig:r5Color}.  Note that the bulge mass bins
do not all have the same number of galaxies, as they did above; at
low $q_d$ (highly inclined galaxies), there are more massive bulges,
while at high $q_d$ (face-on galaxies), there are more low mass
bulges. However, the number of galaxies in each bin differs by at
most $\sim 200$, and each bin has $800$ galaxies on
average. Therefore, any bias introduced by changes in average bulge
mass with disc inclination will be small.  

\figQdvcDgr
Figure \ref{fig:QdttMass} shows no statistically significant trend in
$D/T$  as a function of density. Furthermore, $D/T$ is independent of
disc inclination 
angle. This supports our previous conclusions. Namely,
at constant bulge mass, disc mass is not a function of
environment. 

The dependence of disc colour on environment and disc
inclination is less straightforward. Although the rank correlation
coefficients are always positive, the trends are only
statistically significant for the low mass bulges or highly inclined
discs (see Table \ref{tab:spearQ}). This hints at contamination from
elliptical galaxies in the highest $q_d$ and highest bulge mass bins. For
the most inclined galaxies, where contamination from ellipticals is
minimal, the slopes in disc colour reported in Table \ref{tab:spearQ} agree with
those reported for disc colours in Table \ref{tab:spear}. Additionally, for the
lowest mass bulges (blue squares), 
the slope in $(g-r)_{\mathrm{disc}}$ with $\Sigma_5$ is essentially
independent of  $q_d$. Although
the trends for the higher mass bulges are not always statistically
significant, the slopes measured are also in reasonable agreement with
those for the whole sample. Together with the lack of correlation
between $D/T$ and $\Sigma_5$ at fixed $q_d$, the trends in disc colour
with $\Sigma_5$ at fixed $q_d$ reinforce our conclusion that the changes in
bulge$+$disc galaxy colour, while small, are entirely due to changes in
disc colour with $\Sigma_5$.
\tableSpearQ

The colours plotted in
Fig.~\ref{fig:Qdiscgr} are {\it not} corrected for disc inclination,
which accounts for the differences in the median disc colour across the three
panels. Without the intrinsic inclination correction, the redder
colours for inclined discs are expected. However, using the
uncorrected colours will allow us to verify the disc inclination
correction used in the previous sections. Fig.~\ref{fig:Qdiscgr} shows that the 
change in median colour as a function of $q_d$  is largest for 
the the two middle mass bins (red circles and green triangles). In this mass range, the median disc colour decreases by $\sim 0.04$ mag from the highest
inclination to the lowest inclination galaxies. For the high and low
mass bins, the median disc colours only decrease $0.02$ mag. %If we include
%pseudo-bulges in this sample, only the low mass bulges are affected,
%and the change in disc colour for low mass bulges increases
%(pseudo-bulge galaxies add blue, face-on discs to the sample). 
This demonstrates that the intrinsic inclination correction should be a
function of bulge mass; galaxies with very low mass bulges and very
high mass bulges seem to suffer less intrinsic extinction in their
discs.  Following \citet{Maller09}, our extinction
correction addresses the low bulge mass effect. The correction
includes a term that depends on $K-$band magnitude such that 
higher mass galaxies have a larger extinction correction. Thus,
our extinction correction is typically too large for galaxies with
massive bulges. Finally, the lack of extinction due to ``discs'' around high
mass bulges may be due to contamination from ellipticals, where we
expect little extinction. %\todo{is disc thickness also and issue?} 

Reddening of discs is both a function of disc inclination and a
function of the amount of dust in the disc. Above, we argue that discs
in 
galaxies with different bulge masses suffer from different amounts of
extinction. We can also use the differences in disc colour across the
three 
panels in Fig.~\ref{fig:Qdiscgr} to explore changes in the amount of
extinction as a function of local density. In principle, the change in the
correlation between the uncorrected disc colour and $\Sigma_5$ as a
function of $q_d$ could be used to measure the change in extinction as
a function of local density. However, this requires a detailed
understanding of how 
extinction depends on both the amount of dust and the disc
inclination. Reddening is a complicated, possibly non-monotonic,
function of these variables \citep[see][for one
parametrisation]{Tuffs2004}. Therefore, we limit ourselves to a simple
test; if the extinction in high density environments is negligible,
then the disc colour in the highest density
environments should be independent of disc inclination. 
Fig.~\ref{fig:Qdiscgr} shows this is not true, with the possible
exception of the highest bulge mass galaxies (black crosses). As
explained above, this mass bin is likely to contain misrepresented
ellipticals, which would counter the change in disc colour with disc
inclination. Therefore, any 
  contamination from ellipticals strengthens the argument that discs
  in high density regions are not free from extinction.

\section{Other environment measures}
\label{sec:richness}
\figRichColor
As discussed in \S\ref{ssec:qenv}, we compute environment metrics in
addition to the local density measure, $\Sigma_5$. Below, we use
  the group richness, $N_{\mathrm{gal}}$, as our environment
  measure. We define richness as the number of galaxies in a group
  above the group catalogue magnitude limit, $M_{^{0.1}r} =
  -19.77$. We do not correct $N_{\mathrm{gal}}$ for contamination, nor
  do we attempt to include lower luminosity galaxies in
  $N_{\mathrm{gal}}$. For rich groups, 
  $N_{\mathrm{gal}}$ is closely related to a galaxy's host dark matter
  halo mass, and is notably different from $\Sigma_5$, which
  measures the local (intra-group) density around a galaxy. Since many
  of the effects of environment depend, at least indirectly, on the
  group potential, correlations between bulge and disc properties and
  $N_{\mathrm{gal}}$ may help determine which environmental
  effects are most relevant. In Section \ref{sssec:ingrouptrends2},
  we use crossing time, \tc{} as a proxy for environment {\it within}
  relaxed, relatively rich groups. Unlike $N_{\mathrm{gal}}$, \tc{}
  is different for each galaxy in a group. It measures
  how long a  galaxy has been affected by its host group.

In Fig.~\ref{fig:richColor} we plot the trends in total, bulge, and disc colour
as a function of group richness, $N_{\mathrm{gal}}$, instead of
$\Sigma_5$. Unsurprisingly, 
the trends in galaxy colours with host group richness are similar to
the trends with $\Sigma_5$; total galaxy colour and disc colour are
increasing functions of richness, while the bulge colour decreases
slightly with increasing $N_{\mathrm{gal}}$. The latter effect is explained by the
modelling bias we discuss in \S\ref{sssec:bcolors}. Additionally, we find no
significant correlation between $D/T$ and $N_{\mathrm{gal}}$, in
agreement with the lack of correlation in 
Fig.~\ref{fig:r5DTT}. However, the plots in Fig.
\ref{fig:richColor} do indicate that galaxy colours (and disc colours)
do not redden above $N_{\mathrm{gal}} \sim 20$.  This is in
agreement with results from previous studies
\citep[e.g.][]{Balogh2004, vandenBosch2008}. \citet{Balogh2004} show
that the fraction of red galaxies is independent of cluster velocity
dispersion for $\sigma > 250$~\kmps, which corresponds to $N_{\mathrm{gal}}\approx15-25$. Thus, the majority of the colour
evolution of bulge$+$disc galaxies occurs in smaller groups.  

\figDgrRichPoor

We can test this hypothesis by examining the colour--density relation
in relatively rich groups and poor groups separately. This test is shown in Fig.
\ref{fig:r5rpDgr}; the left panel shows the relation between disc
colour and $\Sigma_5$ for groups with at most $20$ members, while
the right panel shows the same for larger groups. There is no
statistically significant correlation 
between disc colour and $\Sigma_5$ in the large groups, but the trends are
statistically significant for galaxies in the field and in smaller
groups. Therefore, the transformation of disc colour does not require rich
clusters, and star formation in discs is effectively halted by
group density environments. This is in agreement with other studies of
colour and environment \citep[e.g.][]{Zabludoff1998, Gomez2003,
  Cooper2006, BlantonBerlind2007}. However, 
Fig.~\ref{fig:r5rpDgr} does show a significant offset ($\gtrsim 0.01$)
in disc colour between galaxies in poor groups just below $\Sigma_5
\approx 1\, \mathrm{Mpc}^{-2}$ and galaxies in rich groups at $\Sigma_5 \gtrsim 1\, \mathrm{Mpc}^{-2}$. This density
threshold is close to the density of the inflection point in the red
fraction seen in Fig.~\ref{fig:redfrac}.

\figNMorphfrac
\figDMMorphfrac

In addition to examining the changes in colour as a function of
environment, we show the changes in morphological types as a function
of $N_{\mathrm{gal}}$ and dark matter halo mass in
Figs.~\ref{fig:richMorphfrac} and \ref{fig:DMMorphfrac}. In general,
the  trends in morphological type fraction are in agreement with those
in Fig.~\ref{fig:fracSig5}. For the largest groups, there is 
little change in the elliptical galaxy fraction as a function of halo
mass, in agreement with 
\citet{Hoyle2011}, who find that the fraction of early type galaxies
is $\approx 0.2$ and independent of halo mass for masses above  $\sim 10^{13}
\Msun$. \citet{Hoyle2011} use morphologies from the Galaxy Zoo project
\citep{Lintott2011}, which are qualitative classifications, and cannot
distinguish between face-on S0s and elliptical galaxies.  As such,
their definition of early type galaxies only corresponds approximately
to our definition of ellipticals. There are certainly some early type
galaxies in our classical bulge$+$disc category. 
%, and, unlike ellipticals, the fraction of
%S0s increases with increasing halo mass. 
In Fig.~\ref{fig:richMorphfrac}, the
downturn in the elliptical fraction in rich groups is partially due to the inclusion of large, but not necessarily bound structures, in the FoF group
catalogue. We work to eliminate these non-virialised groupings the
next section. 

\subsection{Round groups}
\label{sssec:ingrouptrends}

Although we show that the change in disc colour  occurs in
relatively small groups, the original morphology--density relation was
constructed using  cluster-centric distance as a proxy for
environment \citep{Oemler1974}. Indeed, many studies compare galaxies
on the outskirts of massive clusters to those at 
the centres
\citep[e.g.][]{Dressler80, Postman1984, Whitmore1991, Gomez2003,
  Treu2003, Smith2006,  Weinmann2006, Trager2008, Hudson2010}. Although   
our sample does not extend to significant redshifts, there are several
massive groups included in the group catalogue which we use to study trends in
bulge and disc properties with local density within groups. These trends
may help reduce the scatter in the relations with $\Sigma_5$ and
$N_{\mathrm{gal}}$ shown in Figs.~\ref{fig:r5Color} and \ref{fig:richColor}.

In order to study trends within
clusters, we first limit our sample to groups with at least $20$
members. However, as discussed in \S \ref{ssec:qenv}, the FoF
grouping algorithm does not necessarily identify relaxed or even
bound systems; filaments and merging galaxy groups are often
identified by the FoF algorithm as galaxy groups. Furthermore, many
large galaxy groups do not have well-defined centres
\citep[e.g.][]{Zabludoff1998, Skibba2010}, making trends in galaxy
properties with distance from the group centre almost meaningless. In order
to use environment measures 
such as the crossing time or distance from the group centre, we
select a sample of groups 
%with at least twenty members and 
that have
a symmetric projected distribution of galaxies. This eliminates groups
in the process of forming or merging, for 
which the distance from the cluster centre has little physical meaning. 

If a group is relaxed and approximately spherical, we expect that
galaxies should have a rotationally symmetric distribution on the
sky. We generalise the requirement of rotational symmetry by
  allowing galaxies to be evenly distributed along elliptical
  contours, instead of just circular ones. In practise, we accomplish this
  by compressing the coordinates along the minor axis of the group,
  making an elliptical group appear spherical. Allowing groups to be
  elliptical in projection does not accurately represent oblate,
  prolate, or triaxial groups in projection (expect from special
  viewing angles), but it does account for some of the asymmetries real groups.

We 
test whether a group is symmetric by comparing the distribution 
of galaxies observed for a given group, to a distribution drawn from a
rotationally symmetric radial profile. The functional form of the
radial profile, however, is uncertain
\citep[e.g.][]{Adami2001}. Instead of relying on an 
analytic function for the profile, we create separate comparison
radial profiles for each galaxy group by fitting a smooth curve to the 
binned radial distribution of galaxies in each group. To limit the
Poisson noise,  this fit is smoothed using a Gaussian filter with a
width ($\sigma$) equal to half the root-mean-squared (rms) radius,
$R_{\mathrm{rms}}$, of the group.  

We then compare the actual
distribution of galaxies to the smoothed radial distribution using a
$\chi^2$ test: $\chi^2 = \Sigma_{i,j}[(N_{\mathrm{obs.},i,j} -
\rho_{i,j}N_{\mathrm{gal}})^2/(\rho_{i,j}N_{\mathrm{gal}})]$, where
$N_{\mathrm{obs.},i,j}$ is the number of galaxies observed in a region
($i,j$) and $\rho_{i,j}N_{\mathrm{gal}}$ is the expected number of
galaxies from the rotationally symmetric distribution. The denominator
is the Poisson uncertainty in the number of galaxies.  The size of the
($i,j$) region used in the sum will affect the $\chi^2$ value; if the
region is too large, the two distributions will trivially agree. We use
a square region which is $0.5 R_{\mathrm{rms}}$ on a side. This is
large enough to ensure most regions in the sum have at least one
galaxy, but small  
enough to distinguish between rotationally symmetric and asymmetric
groups. If the galaxy distribution is rotationally symmetric, it will
be statistically indistinguishable from the smoothed radial
distribution. On the other hand, if the galaxy distribution is markedly non-symmetric 
(e.g. bimodal), the
$\chi^2$ value will be large. We consider a group to be
round if the $\chi^2$ value from this test has a probability between
$0.1$ and $0.9$. This selects $95$ of the $197$ groups with
at least 20 members in the group catalogue. Below, we designate
these groups round groups. Of these groups, $46$
  are represented in the bulge$+$disc catalogue; the remainder are at
  redshifts greater than $0.05$. There are $\sim 720$
classical bulge$+$disc galaxies in round groups, compared to $1470$
classical bulge$+$disc galaxies in groups with $N_{\mathrm{gal}} > 20$.

\subsection{Intra-group trends}
\label{sssec:ingrouptrends2}
\figColorTcross

In order to examine the effect of intra-group environment on bulges
and discs, we define the crossing time
(\tc) for each galaxy as the distance from the galaxy
to the
group centre divided by the line-of-sight velocity dispersion of the
group. For a single galaxy, \tc{} is unphysical, since
the galaxy is typically not moving radially with the group velocity
dispersion. However, 
for a large sample of galaxies, \tc{} is a 
measure of how long galaxies have been affected by the group
environment.  Crossing time is 
anti-correlated with local density $\Sigma_5$, although there is significant scatter. We have not divided our sample
  into central galaxies and satellite galaxies as previous studies
  have done 
  \citep[e.g.][]{vandenBosch2008, SkibbaSheth2009, Skibba2009A,
    Peng2011}. However, the number of groups is much smaller than the
  number of galaxies in this sample, so the contribution from central
  galaxies is small. %The correlations with galaxy properties and \tc{} are almost entirely based on satellite galaxies.

Fig.~\ref{fig:colorTcross} shows the
relation between galaxy colours and \tc{} for classical
bulge$+$disc galaxies in round groups with $N_{\mathrm{gal}} >
20$. Since these constraints greatly reduce the sample size ($\sim
720$ classical bulge$+$disc galaxies), we 
only divide the galaxies into three mass bins and only show three
median points in \tc{}. The largest
crossing times in our sample are less than 40 per cent of a Hubble
time, in agreement with those found at higher redshift
\citep{Grutzbauch2011} and with the timescale for relaxation
\citep{GunnGott72, Ferguson1990}. Crossing times shorter than
the Hubble time indicate that the galaxies are not falling into a
group for the first time, and have probably orbited the group centre
at least once. 

The crossing time gives an indication of how long a
  galaxy has been affected by the group environment; galaxies with the
  shortest crossing times have presumably passed through the centre of
  the group most often and have experienced the highest average
  local density. In addition, a small \tc{} implies a high mass
  density within the group ($\tc{} \propto
  \rho^{-1/2}$); galaxies in compact groups will have smaller
 crossing times. Therefore, any trends in galaxy properties with
 \tc{} will  
  be sensitive to how the environmental processes causing the trends
  depend on local density and time, although the exact dependencies are
  not straightforward. 
%Therefore, a galaxy's current
%position in the group is not typically indicative of its historically
%averaged 
%position in the group, and we expect any correlations with crossing time
%and galaxy properties to be weak. 
In addition, the median distance to
the group centre in our sample of round groups is two thirds of the group virial
radius. The galaxies in our sample are all much closer to their host
group centre than $3-4$ virial radii, the distance at which
\citet{Gomez2003} find a large change in the star formation
rate.  Samples that extend to much larger radii are probably needed
in order to see large trends in galaxy colours with group-centric distance.

The left panel in Fig.~\ref{fig:colorTcross} shows that the total
galaxy colour is  
anti-correlated with \tc{}; galaxies with short group
crossing times typically have redder colours. The correlation is only
statistically significant for the highest mass bulges. If we do not
limit the sample to galaxies from round groups, the statistically
significance of all the trends decreases. The middle panel in
Fig.~\ref{fig:colorTcross} shows no statistically significant trend in
bulge colour with crossing time. For the smaller bulges, there is a 
weak correlation with \tc{}, but, as shown in \S\ref{sssec:bcolors},
these trends are likely due to uncertainties in the disc model. The
final panel in 
Fig.~\ref{fig:colorTcross} shows the disc colours as a function of
\tc{}. In this case, none of the correlations are
statistically significant (for the highest mass bulges,
\tc{} and disc colour are anti-correlated with
$2\sigma$ significance). These weak trends in disc colour are in
agreement with the lack of correlation between $(g-r)_{\mathrm{disc}}$
and $\Sigma_5$ for all groups with more than 20 members (see
Fig.~\ref{fig:r5rpDgr}). There are
significant colour differences between discs in large groups and discs
in the field. Isolated classical bulge$+$disc galaxy colours are denoted by solid points in
Fig.~\ref{fig:colorTcross}, and, at least for low mass bulges, discs
in the field are significantly bluer than discs in groups. This agrees
with our earlier conclusion that most disc colour evolution occurs in
smaller groups \citep[see also][]{vandenBosch2008, Cooper2006}.  

The lack of
correlation between disc colour and \tc{} is in 
contrast with the results of 
\citet{Hudson2010}, who find a statistically significant correlation
between disc colour and a galaxy's distance to the
group centre. This is
probably due to the selection of the samples; \citet{Hudson2010} do
not divide their sample into galaxies with bulges and discs only. If we
include bulge-less galaxies in our sample, there is a statistically
significant correlation with disc colour and crossing time.

\figDttTcross

%Since the  colours of galaxies with massive bulges (green
%triangles in Fig.~\ref{fig:colorTcross}) are correlated  with
%\tc{}, but the bulge and disc colours are not, we must examine
%other properties of bulges and discs to account for the change in
%colour with \tc{}. 
Fig.~\ref{fig:dttTcross} shows the change in disc-to-total
mass ratio and disc scale length as a function of \tc{}. Unlike the
case of the relations with $\Sigma_5$, there are statistically
significant correlations between disc mass and \tc{} and between disc size
and \tc{}.  The positive correlation between $D/T$ and \tc{} is
statistically significant for the highest and lowest mass bins, and
significant at the $2\sigma$ level for the middle mass bin. For the
highest mass bulges, the disc-to-total mass 
ratio increases by 6 per cent from $\tc = 0$ to $\tc = 0.3/H_0$. There
is no change in average bulge mass as a function of \tc{}, so the
change in $D/T$ is due entirely to changes in the disc mass. In
addition, the lower panel of Fig.~\ref{fig:dttTcross} shows a
correlation between \tc{} and the disc \Reff{} for the most massive
bulges, which is significant at the $2.3\sigma$ level. Although these
trends with \tc{} are weak, they are also present if we substitute
$\Sigma_5$ for \tc{} (correlations become anti-correlations). These trends suggest that processes active in
the highest density regions in groups either suppress disc formation
or destroy discs around infalling galaxies. We can expand our sample
to include ellipticals ($D/T=0$), to determine if the trends in $D/T$ and
\tc{} continue until the disc is negligible. In this case, the
correlation between $D/T$ and \tc{} for the highest bulge mass bin becomes
insignificant.  Thus, at least for high mass bulges, the trend in
$D/T$ with  \tc{} does not extend to disc-less ellipticals. %Furthermore,
%the lack of correlation when including ellipticals, suggests that our
%earlier result may be sensitive to the separation of ellipticals and
%S0s.

Together, Figs.~\ref{fig:colorTcross} and \ref{fig:dttTcross}
tentatively suggest that morphological changes play some role in the colour--density relation for galaxies in rich groups. This
seems to conflict with 
the conclusions in \S\ref{sssec:dcolors}, which show no significant
trends in disc mass with density. Taken together, these results
suggest that star formation quenching and morphological transformation
are separate physical processes and that these transitions take place 
in different environments. This agrees with previous studies that find
an increase in the number of passive, red discs as a function of
time \citep{Dressler1997, Moran2007, Bundy2010} and environment \citep{Goto2003,
  McIntosh2004, Bundy2006, Bamford09}. The existence of these non-star
forming disc galaxies demonstrates that star formation  quenching
occurs before morphological transformation. In addition,
\citet{Skibba2009} note that there is a weak correlation between red galaxy
morphology and density at small scales, in agreement with our
findings. Namely, we
  demonstrate that morphological transformation may be taking place in
  rich groups and clusters, and that these transformations are not
  associated with star formation quenching since disc colours are not correlated with \tc{}. Galaxy harassment, i.e. high-speed encounters
  between galaxies, is a plausible explanation for the decrease in disc
  mass we observe. The number of encounters a galaxy experiences will 
  increase with galaxy number density, and galaxies with short
  crossing  times  are exposed to the highest average densities. Thus
  galaxy harassment would yield correlations between \tc{} and disc
  mass and disc size similar to those in Fig.~\ref{fig:dttTcross}.

%%Summary Section
\section{Summary}
\label{sec:sumup}

In this work, we examine the changes in bulge and disc properties for
a sample of $12500$ galaxies as functions of local projected
density. Since galaxy mass and environment are strongly correlated, we
divide the sample into bins of equal bulge mass in order to study
any residual trends in bulge and disc properties with local density. Using 2-dimensional bulge$+$disc decompositions, we are able to study the colour--density and morphology--density relations for bulges and discs,
separately. Our sample consists of galaxies with a classical,
elliptical-like bulge surrounded by a disc. Classical bulges are
observed to have the
same characteristics as ellipticals of 
the same mass; the only difference is the
encompassing disc. 
%We distinguish red, dust-free,
%face-on classical bulge$+$disc galaxies from ellipticals 
%based on the distribution of disc axis ratios from the 2-dimensional
%bulge$+$disc fits. Real discs should be randomly oriented, while
%``discs'' around ellipticals will be preferentially face-on. This
%separation is probabilistic; galaxies are assigned a probability of
%being an elliptical or a classical bulge host based on their disc inclination. 
By studying the properties of classical bulge hosts in different environments, we can deduce if local density is the determining factor in whether or not a classical bulge acquires and retains a disc. 
%The
%separation of ellipticals and S0s allows us to study the correlations
%of both red and blue discs with local density. 
%By looking at trends in disc properties with
%environment, we can determine if environment controls whether or not a 
%classical bulge acquires and retains a disc. 
In addition, we study
whether the population of classical bulge$+$disc galaxies is undergoing a
transition from disc-dominated and star-forming 
in low-density regions to bulge-dominated and passive in high-density
regions, as is expected from the morphology--density and colour--density
relations. 

Based on our results, we can draw two conclusions about the effects of
environment on discs around classical bulges. First, both the
colour and the mass of these discs changes with increasing density, but
these changes occur at different densities and, presumably, on
different time-scales. This suggests that star formation quenching and
morphological transformation are caused by different physical
processes, in agreement with many previous studies of galactic 
environment \citep[e.g.][]{Goto2003, Gomez2003, McIntosh2004,
  Kauffmann2004, Christlein2005, Cooper2006, Bundy2006,
  vandenBosch2008, Bamford09, Skibba2009, Hudson2010}. Second, although disc
properties are a function of environment, the changes in disc colour
and disc mass are insufficient to explain why classical bulges have
discs and ellipticals do not have discs. While environment clearly affects disc formation and evolution around classical bulges, there must be
other parameters
%, which do not depend directly on present-day local density, 
that determine whether or not a classical bulge is surrounded by a disc.

The first piece of evidence for our conclusions comes from
  separating the colour--density relation for bulge$+$disc galaxies
  into relations for bulges and discs. We find that the correlation
between total galaxy $g-r$ and local density is due {\it entirely} to changes
in disc colour with density. The small
change in bulge colour with increasing density depends on our choice of disc
model, but this choice does not noticeably affect disc colours. At
fixed bulge mass, disc mass and disc 
scale length are independent of local density.  
Thus, there is no evidence for morphological transformation as a function of 
local density, but there is evidence for star formation quenching in
discs with increased local density. 
%This supports the claim that star formation quenching and morphological
%transformation  are distinct physical processes that probably do not
%occur simultaneously. 
Morphological transformation and star
formation quenching must occur separately, and any process responsible for
star formation quenching cannot dramatically alter stellar discs. 

Even for low bulge masses, where the colour--density relation is
strongest, the change in disc colour with increasing density is
small; $(g-r)_{\mathrm{disc}}$ increases by $\sim 0.05$ mag over two
decades in projected galaxy number density. 
This colour change is not enough to explain the colour--density
relation for the entire galaxy population, which is twice as steep. 
%The median total galaxy
%$g-r$ colour increases by $0.1$ from the lowest to the highest density
%regions. 
Therefore, the correlations
between stellar mass and density and between stellar mass and galaxy
colour contribute significantly to the colour--density
relation for classical bulge$+$disc galaxies. We expect processes internal  to galaxies, which may 
depend on stellar mass, are at least as important in moderating
star formation as external, environment-related, processes. 

The colour--density relation
for discs is not linear. By dividing the sample into
groups with more or fewer than $20$ members, we show that the relation
between density and disc colour disappears for galaxies in rich groups, regardless of bulge mass. This is in agreement with results that star formation is
quenched well before galaxies enter massive clusters \citep{Gomez2003,
  Cooper2006, McIntosh2004}. It also puts limits on the physical
processes responsible for stopping star formation. For example,
ram-pressure stripping requires cluster-scale velocities and is
unlikely to account for star formation cessation in smaller
groups. Strangulation, on the other hand, is effective at relatively
low densities and is, therefore, a possible cause of star formation
quenching in discs. {\rm Although our analysis is not sensitive to colour  gradients in discs, changes in these gradients as a function of environment may put additional constraints on the physical processes responsible for disc fading as a function of local density \citep[see][]{Roediger2011}.}

After determining that star formation quenching occurs outside
  rich groups, we examine trends in bulge and disc properties {\it
  within} rich groups. Instead of local projected density, we use the group
crossing time, \tc{},  as a measure of environment. Since \tc{}
depends on distance to the group centre, we restrict our sample to
``round'' groups, which have a smooth angular distribution of
galaxies. Even for these groups, the luminosity-weighted group centre
may not be physically relevant; it is not unusual for the most massive
galaxy to be offset from the group centre
\citep{Skibba2010, George2012}. The average crossing time for galaxies in
  this restricted sample is roughly one third of a Hubble time, suggesting most
  of these galaxies have already orbited their host group at least
  once. %The trends in galaxy properties with \tc{} will be sensitive
  %to how the environmental processes causing the trends depend on
  %density and time. 
We find no statistically significant correlations with
disc colour and \tc{}, but we do find moderately significant ($2-3\sigma$)
correlations between disc-to-total mass ratio ($D/T$) and \tc{}, and between disc
scale length and \tc{}, at fixed bulge mass. The trends in $D/T$ are
not a result of correlations between group richness and \tc{}, or
between richness and $D/T$. In addition, the correlations are also
present at the same 
significance if we replace \tc{} with local projected density. Like the changes  in disc colour as a function of projected density, the changes in
  disc mass as a function of \tc{} are small; the median $D/T$
  decreases by less than $10$ per cent from the largest to the
  smallest crossing times. 
  None the less, we infer that bulge$+$disc galaxies
do undergo morphological transformations in large groups, but they do
not undergo star formation quenching at the same time. This is
direct evidence for the separation of morphological transformation and
star formation quenching. %In rich groups, while 
%we find no evidence for star formation truncation, {\rm we do find
%tentative evidence for morphological transformation as stellar discs
%are smaller and less massive in the central regions of massive groups.}

The above results demonstrate that environment has two 
distinct effects on the discs surrounding classical bulges. First,
disc star formation is truncated in (relatively) poor groups, leading
to the colour--density relation for discs (Fig.~\ref{fig:r5Color}).
Although gas-stripping requires relatively high 
velocities, tidal interactions and heating by the intra-group median 
may remove a disc's outer halo gas supply (strangulation), thus
quenching star formation over several gigayears
\citep{Larson80}. This preprocessing of galaxies in small groups has
been suggested before as the origin of S0 galaxies
\citep[e.g.][]{Dressler1997}. The second effect is a morphological
transformation. As these quenched galaxies enter higher density
environments in larger groups, the stellar disc is
disrupted over several orbits, leading to the observed $D/T$--\tc{}
correlation. This disruption may be caused by galaxy  harassment
\citep{Moore96}, which is most effective for galaxies in high density regions with short crossing times.

Although environment does affect discs around
classical bulges,  these effects are small. %The
%median disc
%$g-r$ colour reddens by $0.05$ from the lowest to the highest density
%regions, and the average $D/T$ ratio  decreases by less than $10$ per
%cent as a function of \tc{}. 
Therefore, while 
statistically significant, the changes in disc colour with density and
disc mass with crossing time are insufficient to explain the full
range of the colour--density relation and morphology--density
relation. 
%As a function of projected density, the medain total galaxy $g-r$
%reddens by $\sim 0.1$ and the $D/T$ decreases by $30$ per cent. These
%trends are in large part due to the correlations between galaxy mass
%and density, colour, and $B/T$. 
Furthermore, since the changes are small, present-day local density cannot be the determining factor in
whether or not a classical bulge has a disc. 
%Although we find evidence
%for disc fading and destruction, the changes in disc
%mass  are small and do not extend to negligible disc
%masses. 
There must be other processes, unrelated to
present-day environment, which regulate disc formation around
classical bulges. None the less, because classical bulges and
ellipticals of the same mass seem to have the same formation
history, the processes that regulate disc formation
are probably external to the galaxy. {\rm By only examining trends with density at fixed bulge stellar mass, we cannot explore trends in bulge mass with density. Environmental processes, such as ram pressure stripping, may drive bulge growth \citep[e.g.][]{Tonnesen2009}. We plan to explore the effects of environment on bulges at fixed \emph{total} stellar mass in a later paper. }

One way to further explore disc formation
around classical bulges would be to look for evolution in discs around
classical bulges as a
function of redshift. 
%Together the effects of environment on
%discs 
%help explain why some 
%classical bulges have discs and others do not. However, the changes in
%disc mass in rich clusters is small, and  Fig.~\ref{fig:fracSig5}
%shows ellipticals exist even in relatively low density regions 
%\citep[e.g.][]{Mulchaey1999}. Therefore, although local density plays
%some role in disc formation and acquisition, it cannot be the only
%relevant variable. 
In this paper, we present correlations between disc
properties with environment in the local universe. In drawing
conclusions, we have assumed these correlations are signatures of the
evolution of bulges and discs as galaxies move from low density
environments to high density environments. However, these
conclusions need to be supplemented with observations of bulges and
discs at higher redshifts. The comparison data needed for such
studies is easily available from space-based optical and near-infrared
surveys. For example, using data from the CANDELS survey,
\citet{Bruce2012} presents bulge$+$disc decompositions for massive
galaxies beyond $z=1$. Careful comparisons of bulges and discs at
different redshifts and in different environments will better constrain what
effect environment has on the evolution of galaxies and their
component bulges and discs.

\section*{Acknowledgments}
CNL is supported by NSF grant AST0908368. 

This work makes extensive use of data from SDSS-III. Funding for
SDSS-III has been provided by the Alfred P. Sloan 
Foundation, the Participating Institutions, the National Science
Foundation, and the U.S. Department of Energy Office of Science. The
SDSS-III web site is http://www.sdss3.org/. 

SDSS-III is managed by the Astrophysical Research Consortium for the
Participating Institutions of the SDSS-III Collaboration including the
University of Arizona, the Brazilian Participation Group, Brookhaven
National Laboratory, University of Cambridge, University of Florida,
the French Participation Group, the German Participation Group, the
Instituto de Astrofisica de Canarias, the Michigan State/Notre
Dame/JINA Participation Group, Johns Hopkins University, Lawrence
Berkeley National Laboratory, Max Planck Institute for Astrophysics,
New Mexico State University, New York University, Ohio State
University, Pennsylvania State University, University of Portsmouth,
Princeton University, the Spanish Participation Group, University of
Tokyo, University of Utah, Vanderbilt University, University of
Virginia, University of Washington, and Yale University.

\bibliographystyle{mn2e}
\bibliography{lackner_short}

\begin{thebibliography}{}

\bibitem[\protect\citeauthoryear{{Adami}, {Mazure}, {Ulmer} \&
  {Savine}}{{Adami} et~al.}{2001}]{Adami2001}
{Adami} C.,  {Mazure} A.,  {Ulmer} M.~P.,    {Savine} C.,  2001, \aap, 371, 11

\bibitem[\protect\citeauthoryear{{Aihara}, {Allende Prieto}, {An}, {Anderson},
  {Aubourg}, {Balbinot}, {Beers}, {Berlind}, {Bickerton}, {Bizyaev}, {Blanton},
  {Bochanski}, {Bolton}, {Bovy}, {Brandt}, {Brinkmann}, {Brown}
  et~al.,}{{Aihara} et~al.}{2011}]{SDSSDR82011}
{Aihara} H.,  {Allende Prieto} C.,  {An} D.,  {Anderson} S.~F.,  {Aubourg}
  {\'E}.,  {Balbinot} E.,  {Beers} T.~C.,  {Berlind} A.~A.,  {Bickerton} S.~J.,
   {Bizyaev} D.,  {Blanton} M.~R.,  {Bochanski} J.~J.,  {Bolton} A.~S.,  {Bovy}
  J.,  {Brandt} W.~N.,  {Brinkmann} J.,  {Brown} P.~J.,    et~al., 2011, \apjs,
  193, 29

\bibitem[\protect\citeauthoryear{{Allen}, {Driver}, {Graham}, {Cameron},
  {Liske} \& {de Propris}}{{Allen} et~al.}{2006}]{Allen06}
{Allen} P.~D.,  {Driver} S.~P.,  {Graham} A.~W.,  {Cameron} E.,  {Liske} J.,
  {de Propris} R.,  2006, \mnras, 371, 2

\bibitem[\protect\citeauthoryear{{Athanassoula}}{{Athanassoula}}{2005}]{Athana%
ssoula05}
{Athanassoula} E.,  2005, \mnras, 358, 1477

\bibitem[\protect\citeauthoryear{{Baldry}, {Balogh}, {Bower}, {Glazebrook},
  {Nichol}, {Bamford} \& {Budavari}}{{Baldry} et~al.}{2006}]{Baldry2006}
{Baldry} I.~K.,  {Balogh} M.~L.,  {Bower} R.~G.,  {Glazebrook} K.,  {Nichol}
  R.~C.,  {Bamford} S.~P.,    {Budavari} T.,  2006, \mnras, 373, 469

\bibitem[\protect\citeauthoryear{{Balogh}, {Baldry}, {Nichol}, {Miller},
  {Bower} \& {Glazebrook}}{{Balogh} et~al.}{2004}]{Balogh2004}
{Balogh} M.~L.,  {Baldry} I.~K.,  {Nichol} R.,  {Miller} C.,  {Bower} R.,
  {Glazebrook} K.,  2004, \apjl, 615, L101

\bibitem[\protect\citeauthoryear{{Balogh} \& {Morris}}{{Balogh} \&
  {Morris}}{2000}]{Balogh2000b}
{Balogh} M.~L.,  {Morris} S.~L.,  2000, \mnras, 318, 703

\bibitem[\protect\citeauthoryear{{Balogh}, {Morris}, {Yee}, {Carlberg} \&
  {Ellingson}}{{Balogh} et~al.}{1999}]{Balogh1999}
{Balogh} M.~L.,  {Morris} S.~L.,  {Yee} H.~K.~C.,  {Carlberg} R.~G.,
  {Ellingson} E.,  1999, \apj, 527, 54

\bibitem[\protect\citeauthoryear{{Balogh}, {Navarro} \& {Morris}}{{Balogh}
  et~al.}{2000}]{Balogh2000}
{Balogh} M.~L.,  {Navarro} J.~F.,  {Morris} S.~L.,  2000, \apj, 540, 113

\bibitem[\protect\citeauthoryear{{Bamford}, {Nichol}, {Baldry}, {Land},
  {Lintott}, {Schawinski}, {Slosar}, {Szalay}, {Thomas}, {Torki}, {Andreescu},
  {Edmondson}, {Miller}, {Murray}, {Raddick} \& {Vandenberg}}{{Bamford}
  et~al.}{2009}]{Bamford09}
{Bamford} S.~P.,  {Nichol} R.~C.,  {Baldry} I.~K.,  {Land} K.,  {Lintott}
  C.~J.,  {Schawinski} K.,  {Slosar} A.,  {Szalay} A.~S.,  {Thomas} D.,
  {Torki} M.,  {Andreescu} D.,  {Edmondson} E.~M.,  {Miller} C.~J.,  {Murray}
  P.,  {Raddick} M.~J.,    {Vandenberg} J.,  2009, \mnras, 393, 1324

\bibitem[\protect\citeauthoryear{{Bell}, {McIntosh}, {Katz} \&
  {Weinberg}}{{Bell} et~al.}{2003}]{Bell2003}
{Bell} E.~F.,  {McIntosh} D.~H.,  {Katz} N.,    {Weinberg} M.~D.,  2003, \apjs,
  149, 289

\bibitem[\protect\citeauthoryear{{Berlind}, {Frieman}, {Weinberg}, {Blanton},
  {Warren}, {Abazajian}, {Scranton}, {Hogg}, {Scoccimarro}, {Bahcall},
  {Brinkmann}, {Gott} III et~al.,}{{Berlind} et~al.}{2006}]{Berlind2006}
{Berlind} A.~A.,  {Frieman} J.,  {Weinberg} D.~H.,  {Blanton} M.~R.,  {Warren}
  M.~S.,  {Abazajian} K.,  {Scranton} R.,  {Hogg} D.~W.,  {Scoccimarro} R.,
  {Bahcall} N.~A.,  {Brinkmann} J.,  {Gott} III J.~R.,    et~al., 2006, \apjs,
  167, 1

\bibitem[\protect\citeauthoryear{{Bernardi}, {Nichol}, {Sheth}, {Miller} \&
  {Brinkmann}}{{Bernardi} et~al.}{2006}]{Bernardi2006}
{Bernardi} M.,  {Nichol} R.~C.,  {Sheth} R.~K.,  {Miller} C.~J.,    {Brinkmann}
  J.,  2006, \aj, 131, 1288

\bibitem[\protect\citeauthoryear{{Bernardi}, {Sheth}, {Annis}, {Burles},
  {Eisenstein}, {Finkbeiner}, {Hogg}, {Lupton}, {Schlegel}, {SubbaRao},
  {Bahcall}, {Blakeslee}, {Brinkmann}, {Castander}, {Connolly}, {Csabai}
  et~al.,}{{Bernardi} et~al.}{2003a}]{Bernardi03I}
{Bernardi} M.,  {Sheth} R.~K.,  {Annis} J.,  {Burles} S.,  {Eisenstein} D.~J.,
  {Finkbeiner} D.~P.,  {Hogg} D.~W.,  {Lupton} R.~H.,  {Schlegel} D.~J.,
  {SubbaRao} M.,  {Bahcall} N.~A.,  {Blakeslee} J.~P.,  {Brinkmann} J.,
  {Castander} F.~J.,  {Connolly} A.~J.,  {Csabai} I.,    et~al., 2003a, \aj,
  125, 1817

\bibitem[\protect\citeauthoryear{{Bernardi}, {Sheth}, {Annis}, {Burles},
  {Eisenstein}, {Finkbeiner}, {Hogg}, {Lupton}, {Schlegel}, {SubbaRao},
  {Bahcall}, {Blakeslee}, {Brinkmann}, {Castander}, {Connolly}, {Csabai}
  et~al.,}{{Bernardi} et~al.}{2003b}]{Bernardi2003III}
{Bernardi} M.,  {Sheth} R.~K.,  {Annis} J.,  {Burles} S.,  {Eisenstein} D.~J.,
  {Finkbeiner} D.~P.,  {Hogg} D.~W.,  {Lupton} R.~H.,  {Schlegel} D.~J.,
  {SubbaRao} M.,  {Bahcall} N.~A.,  {Blakeslee} J.~P.,  {Brinkmann} J.,
  {Castander} F.~J.,  {Connolly} A.~J.,  {Csabai} I.,    et~al., 2003b, \aj,
  125, 1866

\bibitem[\protect\citeauthoryear{{Blanton} \& {Berlind}}{{Blanton} \&
  {Berlind}}{2007}]{BlantonBerlind2007}
{Blanton} M.~R.,  {Berlind} A.~A.,  2007, \apj, 664, 791

\bibitem[\protect\citeauthoryear{{Blanton}, {Eisenstein}, {Hogg}, {Schlegel} \&
  {Brinkmann}}{{Blanton} et~al.}{2005}]{Blanton2005a}
{Blanton} M.~R.,  {Eisenstein} D.,  {Hogg} D.~W.,  {Schlegel} D.~J.,
  {Brinkmann} J.,  2005, \apj, 629, 143

\bibitem[\protect\citeauthoryear{{Blanton}, {Eisenstein}, {Hogg} \&
  {Zehavi}}{{Blanton} et~al.}{2006}]{Blanton2006}
{Blanton} M.~R.,  {Eisenstein} D.,  {Hogg} D.~W.,    {Zehavi} I.,  2006, \apj,
  645, 977

\bibitem[\protect\citeauthoryear{{Blanton}, {Hogg}, {Bahcall}, {Baldry},
  {Brinkmann}, {Csabai}, {Eisenstein}, {Fukugita}, {Gunn}, {Ivezi{\'c}},
  {Lamb}, {Lupton}, {Loveday} et~al.,}{{Blanton} et~al.}{2003}]{Blanton03}
{Blanton} M.~R.,  {Hogg} D.~W.,  {Bahcall} N.~A.,  {Baldry} I.~K.,  {Brinkmann}
  J.,  {Csabai} I.,  {Eisenstein} D.,  {Fukugita} M.,  {Gunn} J.~E.,
  {Ivezi{\'c}} {\v Z}.,  {Lamb} D.~Q.,  {Lupton} R.~H.,  {Loveday} J.,
  et~al., 2003, \apj, 594, 186

\bibitem[\protect\citeauthoryear{{Blanton} \& {Moustakas}}{{Blanton} \&
  {Moustakas}}{2009}]{Blanton2009}
{Blanton} M.~R.,  {Moustakas} J.,  2009, \araa, 47, 159

\bibitem[\protect\citeauthoryear{{Blanton} \& {Roweis}}{{Blanton} \&
  {Roweis}}{2007}]{Blanton07}
{Blanton} M.~R.,  {Roweis} S.,  2007, \aj, 133, 734

\bibitem[\protect\citeauthoryear{{Boselli} \& {Gavazzi}}{{Boselli} \&
  {Gavazzi}}{2006}]{Boselli2006}
{Boselli} A.,  {Gavazzi} G.,  2006, \pasp, 118, 517

\bibitem[\protect\citeauthoryear{{Brinchmann} \& {Ellis}}{{Brinchmann} \&
  {Ellis}}{2000}]{Brinchmann2000}
{Brinchmann} J.,  {Ellis} R.~S.,  2000, \apjl, 536, L77

\bibitem[\protect\citeauthoryear{{Bruce}, {Dunlop}, {Cirasuolo}, {McLure},
  {Targett}, {Bell}, {Croton}, {Dekel}, {Faber}, {Ferguson}, {Grogin},
  {Kocevski}, {Koekemoer}, {Koo}, {Lai}, {Lotz}, {McGrath}, {Newman} \& {van
  der Wel}}{{Bruce} et~al.}{2012}]{Bruce2012}
{Bruce} V.~A.,  {Dunlop} J.~S.,  {Cirasuolo} M.,  {McLure} R.~J.,  {Targett}
  T.~A.,  {Bell} E.~F.,  {Croton} D.~J.,  {Dekel} A.,  {Faber} S.~M.,
  {Ferguson} H.~C.,  {Grogin} N.~A.,  {Kocevski} D.~D.,  {Koekemoer} A.~M.,
  {Koo} D.~C.,  {Lai} K.,  {Lotz} J.~M.,  {McGrath} E.~J.,  {Newman} J.~A.,
  {van der Wel} A.,  2012, preprint, (arXiv:1206.4322)

\bibitem[\protect\citeauthoryear{{Bruzual} \& {Charlot}}{{Bruzual} \&
  {Charlot}}{2003}]{Bruzual2003}
{Bruzual} G.,  {Charlot} S.,  2003, \mnras, 344, 1000

\bibitem[\protect\citeauthoryear{{Bundy}, {Ellis}, {Conselice}, {Taylor},
  {Cooper}, {Willmer}, {Weiner}, {Coil}, {Noeske} \& {Eisenhardt}}{{Bundy}
  et~al.}{2006}]{Bundy2006}
{Bundy} K.,  {Ellis} R.~S.,  {Conselice} C.~J.,  {Taylor} J.~E.,  {Cooper}
  M.~C.,  {Willmer} C.~N.~A.,  {Weiner} B.~J.,  {Coil} A.~L.,  {Noeske} K.~G.,
    {Eisenhardt} P.~R.~M.,  2006, \apj, 651, 120

\bibitem[\protect\citeauthoryear{{Bundy}, {Scarlata}, {Carollo}, {Ellis},
  {Drory}, {Hopkins}, {Salvato}, {Leauthaud}, {Koekemoer}, {Murray}, {Ilbert},
  {Oesch}, {Ma}, {Capak}, {Pozzetti} \& {Scoville}}{{Bundy}
  et~al.}{2010}]{Bundy2010}
{Bundy} K.,  {Scarlata} C.,  {Carollo} C.~M.,  {Ellis} R.~S.,  {Drory} N.,
  {Hopkins} P.,  {Salvato} M.,  {Leauthaud} A.,  {Koekemoer} A.~M.,  {Murray}
  N.,  {Ilbert} O.,  {Oesch} P.,  {Ma} C.-P.,  {Capak} P.,  {Pozzetti} L.,
  {Scoville} N.,  2010, \apj, 719, 1969

\bibitem[\protect\citeauthoryear{{Butcher} \& {Oemler} Jr.}{{Butcher} \&
  {Oemler}}{1978}]{ButcherOemler1978}
{Butcher} H.,  {Oemler} Jr. A.,  1978, \apj, 219, 18

\bibitem[\protect\citeauthoryear{{Cameron}, {Driver}, {Graham} \&
  {Liske}}{{Cameron} et~al.}{2009}]{Cameron09}
{Cameron} E.,  {Driver} S.~P.,  {Graham} A.~W.,    {Liske} J.,  2009, \apj,
  699, 105

\bibitem[\protect\citeauthoryear{{Caon}, {Capaccioli} \& {D'Onofrio}}{{Caon}
  et~al.}{1993}]{Caon93}
{Caon} N.,  {Capaccioli} M.,  {D'Onofrio} M.,  1993, \mnras, 265, 1013

\bibitem[\protect\citeauthoryear{{Christlein} \& {Zabludoff}}{{Christlein} \&
  {Zabludoff}}{2005}]{Christlein2005}
{Christlein} D.,  {Zabludoff} A.~I.,  2005, \apj, 621, 201

\bibitem[\protect\citeauthoryear{{Cibinel}, {Carollo}, {Lilly}, {Bonoli},
  {Miniati}, {Pipino}, {Silverman}, {van Gorkom}, {Cameron}, {Finoguenov},
  {Norberg}, {Rudick}, {Lu} \& {Peng}}{{Cibinel} et~al.}{2012}]{Cibinel2012}
{Cibinel} A.,  {Carollo} C.~M.,  {Lilly} S.~J.,  {Bonoli} S.,  {Miniati} F.,
  {Pipino} A.,  {Silverman} J.~D.,  {van Gorkom} J.~H.,  {Cameron} E.,
  {Finoguenov} A.,  {Norberg} P.,  {Rudick} C.~S.,  {Lu} T.,    {Peng} Y.,
  2012, preprint, (arXiv:1206.6496)

\bibitem[\protect\citeauthoryear{{Clemens}, {Bressan}, {Nikolic}, {Alexander},
  {Annibali} \& {Rampazzo}}{{Clemens} et~al.}{2006}]{Clemens2006}
{Clemens} M.~S.,  {Bressan} A.,  {Nikolic} B.,  {Alexander} P.,  {Annibali} F.,
     {Rampazzo} R.,  2006, \mnras, 370, 702

\bibitem[\protect\citeauthoryear{{Cooper}, {Gallazzi}, {Newman} \&
  {Yan}}{{Cooper} et~al.}{2010}]{Cooper2010}
{Cooper} M.~C.,  {Gallazzi} A.,  {Newman} J.~A.,    {Yan} R.,  2010, \mnras,
  402, 1942

\bibitem[\protect\citeauthoryear{{Cooper}, {Newman}, {Croton}, {Weiner},
  {Willmer}, {Gerke}, {Madgwick}, {Faber}, {Davis}, {Coil}, {Finkbeiner},
  {Guhathakurta} \& {Koo}}{{Cooper} et~al.}{2006}]{Cooper2006}
{Cooper} M.~C.,  {Newman} J.~A.,  {Croton} D.~J.,  {Weiner} B.~J.,  {Willmer}
  C.~N.~A.,  {Gerke} B.~F.,  {Madgwick} D.~S.,  {Faber} S.~M.,  {Davis} M.,
  {Coil} A.~L.,  {Finkbeiner} D.~P.,  {Guhathakurta} P.,    {Koo} D.~C.,  2006,
  \mnras, 370, 198

\bibitem[\protect\citeauthoryear{{Cooper}, {Tremonti}, {Newman} \&
  {Zabludoff}}{{Cooper} et~al.}{2008}]{Cooper2008}
{Cooper} M.~C.,  {Tremonti} C.~A.,  {Newman} J.~A.,    {Zabludoff} A.~I.,
  2008, \mnras, 390, 245

\bibitem[\protect\citeauthoryear{{Croton}, {Gao} \& {White}}{{Croton}
  et~al.}{2007}]{Croton2007}
{Croton} D.~J.,  {Gao} L.,  {White} S.~D.~M.,  2007, \mnras, 374, 1303

\bibitem[\protect\citeauthoryear{{de Jong}}{{de Jong}}{1994}]{deJong1994}
{de Jong} R.~S.,  1994, PhD thesis, Kapteyn Astronomical Inst.

\bibitem[\protect\citeauthoryear{{De Propris}, {Colless}, {Peacock}, {Couch},
  {Driver}, {Balogh}, {Baldry}, {Baugh}, {Bland-Hawthorn}, {Bridges}, {Cannon},
  {Cole}, {Collins}, {Cross}, {Dalton}, {Efstathiou}, {Ellis} et~al.,}{{De
  Propris} et~al.}{2004}]{DePropris2004}
{De Propris} R.,  {Colless} M.,  {Peacock} J.~A.,  {Couch} W.~J.,  {Driver}
  S.~P.,  {Balogh} M.~L.,  {Baldry} I.~K.,  {Baugh} C.~M.,  {Bland-Hawthorn}
  J.,  {Bridges} T.,  {Cannon} R.,  {Cole} S.,  {Collins} C.,  {Cross} N.,
  {Dalton} G.,  {Efstathiou} G.,  {Ellis} R.~S.,    et~al., 2004, \mnras, 351,
  125

\bibitem[\protect\citeauthoryear{{de Vaucouleurs}}{{de
  Vaucouleurs}}{1961}]{deVaucouleurs61}
{de Vaucouleurs} G.,  1961, \apjs, 5, 233

\bibitem[\protect\citeauthoryear{{Dressler}}{{Dressler}}{1980}]{Dressler80}
{Dressler} A.,  1980, \apj, 236, 351

\bibitem[\protect\citeauthoryear{{Dressler}, {Lynden-Bell}, {Burstein},
  {Davies}, {Faber}, {Terlevich} \& {Wegner}}{{Dressler}
  et~al.}{1987}]{Dressler87}
{Dressler} A.,  {Lynden-Bell} D.,  {Burstein} D.,  {Davies} R.~L.,  {Faber}
  S.~M.,  {Terlevich} R.,    {Wegner} G.,  1987, \apj, 313, 42

\bibitem[\protect\citeauthoryear{{Dressler}, {Oemler} Jr., {Couch}, {Smail},
  {Ellis}, {Barger}, {Butcher}, {Poggianti} \& {Sharples}}{{Dressler}
  et~al.}{1997}]{Dressler1997}
{Dressler} A.,  {Oemler} Jr. A.,  {Couch} W.~J.,  {Smail} I.,  {Ellis} R.~S.,
  {Barger} A.,  {Butcher} H.,  {Poggianti} B.~M.,    {Sharples} R.~M.,  1997,
  \apj, 490, 577

\bibitem[\protect\citeauthoryear{{Elmegreen}, {Elmegreen}, {Fernandez} \&
  {Lemonias}}{{Elmegreen} et~al.}{2009}]{Elmegreen09}
{Elmegreen} B.~G.,  {Elmegreen} D.~M.,  {Fernandez} M.~X.,    {Lemonias} J.~J.,
   2009, \apj, 692, 12

\bibitem[\protect\citeauthoryear{{Emsellem}, {Bacon}, {Monnet} \&
  {Poulain}}{{Emsellem} et~al.}{1996}]{Emsellem1996}
{Emsellem} E.,  {Bacon} R.,  {Monnet} G.,    {Poulain} P.,  1996, \aap, 312,
  777

\bibitem[\protect\citeauthoryear{{Falc{\'o}n-Barroso}, {Peletier} \&
  {Balcells}}{{Falc{\'o}n-Barroso} et~al.}{2002}]{FalconBarroso02}
{Falc{\'o}n-Barroso} J.,  {Peletier} R.~F.,  {Balcells} M.,  2002, \mnras, 335,
  741

\bibitem[\protect\citeauthoryear{{Ferguson} \& {Sandage}}{{Ferguson} \&
  {Sandage}}{1990}]{Ferguson1990}
{Ferguson} H.~C.,  {Sandage} A.,  1990, \aj, 100, 1

\bibitem[\protect\citeauthoryear{{Fisher}}{{Fisher}}{2006}]{Fisher06}
{Fisher} D.~B.,  2006, \apjl, 642, L17

\bibitem[\protect\citeauthoryear{{Fisher} \& {Drory}}{{Fisher} \&
  {Drory}}{2008}]{Fisher08}
{Fisher} D.~B.,  {Drory} N.,  2008, \aj, 136, 773

\bibitem[\protect\citeauthoryear{{Gadotti}}{{Gadotti}}{2009}]{Gadotti09}
{Gadotti} D.~A.,  2009, \mnras, 393, 1531

\bibitem[\protect\citeauthoryear{{Gallazzi}, {Charlot}, {Brinchmann} \&
  {White}}{{Gallazzi} et~al.}{2006}]{Gallazzi2006}
{Gallazzi} A.,  {Charlot} S.,  {Brinchmann} J.,    {White} S.~D.~M.,  2006,
  \mnras, 370, 1106

\bibitem[\protect\citeauthoryear{{Gallazzi}, {Charlot}, {Brinchmann}, {White}
  \& {Tremonti}}{{Gallazzi} et~al.}{2005}]{Gallazzi2005}
{Gallazzi} A.,  {Charlot} S.,  {Brinchmann} J.,  {White} S.~D.~M.,
  {Tremonti} C.~A.,  2005, \mnras, 362, 41

\bibitem[\protect\citeauthoryear{{George}, {Leauthaud}, {Bundy}, {Finoguenov},
  {Ma}, {Rykoff}, {Tinker}, {Wechsler}, {Massey} \& {Mei}}{{George}
  et~al.}{2012}]{George2012}
{George} M.~R.,  {Leauthaud} A.,  {Bundy} K.,  {Finoguenov} A.,  {Ma} C.-P.,
  {Rykoff} E.~S.,  {Tinker} J.~L.,  {Wechsler} R.~H.,  {Massey} R.,    {Mei}
  S.,  2012, \apj, 757, 2

\bibitem[\protect\citeauthoryear{{G{\'o}mez}, {Nichol}, {Miller}, {Balogh},
  {Goto}, {Zabludoff}, {Romer}, {Bernardi}, {Sheth}, {Hopkins}, {Castander},
  {Connolly}, {Schneider}, {Brinkmann}, {Lamb}, {SubbaRao} \&
  {York}}{{G{\'o}mez} et~al.}{2003}]{Gomez2003}
{G{\'o}mez} P.~L.,  {Nichol} R.~C.,  {Miller} C.~J.,  {Balogh} M.~L.,  {Goto}
  T.,  {Zabludoff} A.~I.,  {Romer} A.~K.,  {Bernardi} M.,  {Sheth} R.,
  {Hopkins} A.~M.,  {Castander} F.~J.,  {Connolly} A.~J.,  {Schneider} D.~P.,
  {Brinkmann} J.,  {Lamb} D.~Q.,  {SubbaRao} M.,    {York} D.~G.,  2003, \apj,
  584, 210

\bibitem[\protect\citeauthoryear{{Goto}, {Yamauchi}, {Fujita}, {Okamura},
  {Sekiguchi}, {Smail}, {Bernardi} \& {Gomez}}{{Goto} et~al.}{2003}]{Goto2003}
{Goto} T.,  {Yamauchi} C.,  {Fujita} Y.,  {Okamura} S.,  {Sekiguchi} M.,
  {Smail} I.,  {Bernardi} M.,    {Gomez} P.~L.,  2003, \mnras, 346, 601

\bibitem[\protect\citeauthoryear{{Gr{\"u}tzbauch}, {Conselice}, {Varela},
  {Bundy}, {Cooper}, {Skibba} \& {Willmer}}{{Gr{\"u}tzbauch}
  et~al.}{2011}]{Grutzbauch2011}
{Gr{\"u}tzbauch} R.,  {Conselice} C.~J.,  {Varela} J.,  {Bundy} K.,  {Cooper}
  M.~C.,  {Skibba} R.,    {Willmer} C.~N.~A.,  2011, \mnras, 411, 929

\bibitem[\protect\citeauthoryear{{Gunn} \& {Gott}}{{Gunn} \&
  {Gott}}{1972}]{GunnGott72}
{Gunn} J.~E.,  {Gott} J.~R.~I.,  1972, \apj, 176, 1

\bibitem[\protect\citeauthoryear{{Hamilton}}{{Hamilton}}{1988}]{Hamilton1988}
{Hamilton} A.~J.~S.,  1988, \apjl, 331, L59

\bibitem[\protect\citeauthoryear{{Hansen}, {Sheldon}, {Wechsler} \&
  {Koester}}{{Hansen} et~al.}{2009}]{Hansen2009}
{Hansen} S.~M.,  {Sheldon} E.~S.,  {Wechsler} R.~H.,    {Koester} B.~P.,  2009,
  \apj, 699, 1333

\bibitem[\protect\citeauthoryear{{Hashimoto}, {Oemler} Jr., {Lin} \&
  {Tucker}}{{Hashimoto} et~al.}{1998}]{Hashimoto1998}
{Hashimoto} Y.,  {Oemler} Jr. A.,  {Lin} H.,    {Tucker} D.~L.,  1998, \apj,
  499, 589

\bibitem[\protect\citeauthoryear{{Hogg}, {Blanton}, {Eisenstein}, {Gunn},
  {Schlegel}, {Zehavi}, {Bahcall}, {Brinkmann}, {Csabai}, {Schneider},
  {Weinberg} \& {York}}{{Hogg} et~al.}{2003}]{Hogg2003}
{Hogg} D.~W.,  {Blanton} M.~R.,  {Eisenstein} D.~J.,  {Gunn} J.~E.,  {Schlegel}
  D.~J.,  {Zehavi} I.,  {Bahcall} N.~A.,  {Brinkmann} J.,  {Csabai} I.,
  {Schneider} D.~P.,  {Weinberg} D.~H.,    {York} D.~G.,  2003, \apjl, 585, L5

\bibitem[\protect\citeauthoryear{{Hoyle}, {Masters}, {Nichol}, {Jimenez} \&
  {Bamford}}{{Hoyle} et~al.}{2012}]{Hoyle2011}
{Hoyle} B.,  {Masters} K.~L.,  {Nichol} R.~C.,  {Jimenez} R.,    {Bamford}
  S.~P.,  2012, \mnras, p.~3144

\bibitem[\protect\citeauthoryear{{Hudson}, {Stevenson}, {Smith}, {Wegner},
  {Lucey} \& {Simard}}{{Hudson} et~al.}{2010}]{Hudson2010}
{Hudson} M.~J.,  {Stevenson} J.~B.,  {Smith} R.~J.,  {Wegner} G.~A.,  {Lucey}
  J.~R.,    {Simard} L.,  2010, \mnras, 409, 405

\bibitem[\protect\citeauthoryear{{Kauffmann}, {Heckman}, {White}, {Charlot},
  {Tremonti}, {Brinchmann}, {Bruzual}, {Peng}, {Seibert}, {Bernardi},
  {Blanton}, {Brinkmann} et~al.,}{{Kauffmann} et~al.}{2003}]{Kauffmann2003b}
{Kauffmann} G.,  {Heckman} T.~M.,  {White} S.~D.~M.,  {Charlot} S.,  {Tremonti}
  C.,  {Brinchmann} J.,  {Bruzual} G.,  {Peng} E.~W.,  {Seibert} M.,
  {Bernardi} M.,  {Blanton} M.,  {Brinkmann} J.,    et~al., 2003, \mnras, 341,
  33

\bibitem[\protect\citeauthoryear{{Kauffmann}, {White}, {Heckman}, {M{\'e}nard},
  {Brinchmann}, {Charlot}, {Tremonti} \& {Brinkmann}}{{Kauffmann}
  et~al.}{2004}]{Kauffmann2004}
{Kauffmann} G.,  {White} S.~D.~M.,  {Heckman} T.~M.,  {M{\'e}nard} B.,
  {Brinchmann} J.,  {Charlot} S.,  {Tremonti} C.,    {Brinkmann} J.,  2004,
  \mnras, 353, 713

\bibitem[\protect\citeauthoryear{{Kormendy}}{{Kormendy}}{1977}]{Kormendy77}
{Kormendy} J.,  1977, \apj, 218, 333

\bibitem[\protect\citeauthoryear{{Kormendy}}{{Kormendy}}{1993}]{Kormendy93}
{Kormendy} J.,  1993, in {H.~Dejonghe \& H.~J.~Habing} ed., Galactic Bulges
  Vol.~153 of IAU Symposium, {Kinematics of extragalactic bulges: evidence that
  some bulges are really disks}.
pp 209--+

\bibitem[\protect\citeauthoryear{{Kormendy} \& {Kennicutt} Jr.}{{Kormendy} \&
  {Kennicutt}}{2004}]{Kormendy04}
{Kormendy} J.,  {Kennicutt} Jr. R.~C.,  2004, \araa, 42, 603

\bibitem[\protect\citeauthoryear{{Kregel}, {van der Kruit} \& {de
  Grijs}}{{Kregel} et~al.}{2002}]{Kregel02}
{Kregel} M.,  {van der Kruit} P.~C.,  {de Grijs} R.,  2002, \mnras, 334, 646

\bibitem[\protect\citeauthoryear{{Kroupa}}{{Kroupa}}{2002}]{Kroupa2002}
{Kroupa} P.,  2002, Science, 295, 82

\bibitem[\protect\citeauthoryear{{Lackner} \& {Gunn}}{{Lackner} \&
  {Gunn}}{2012}]{Lackner2012}
{Lackner} C.~N.,  {Gunn} J.~E.,  2012, \mnras, p.~2423

\bibitem[\protect\citeauthoryear{{Larson}, {Tinsley} \& {Caldwell}}{{Larson}
  et~al.}{1980}]{Larson80}
{Larson} R.~B.,  {Tinsley} B.~M.,  {Caldwell} C.~N.,  1980, \apj, 237, 692

\bibitem[\protect\citeauthoryear{{Leauthaud}, {George}, {Behroozi}, {Bundy},
  {Tinker}, {Wechsler}, {Conroy}, {Finoguenov} \& {Tanaka}}{{Leauthaud}
  et~al.}{2012}]{Leauthaud2012}
{Leauthaud} A.,  {George} M.~R.,  {Behroozi} P.~S.,  {Bundy} K.,  {Tinker} J.,
  {Wechsler} R.~H.,  {Conroy} C.,  {Finoguenov} A.,    {Tanaka} M.,  2012,
  \apj, 746, 95

\bibitem[\protect\citeauthoryear{{Lewis}, {Balogh}, {De Propris}, {Couch},
  {Bower}, {Offer}, {Bland-Hawthorn}, {Baldry}, {Baugh}, {Bridges}, {Cannon}
  et~al.,}{{Lewis} et~al.}{2002}]{Lewis2002}
{Lewis} I.,  {Balogh} M.,  {De Propris} R.,  {Couch} W.,  {Bower} R.,  {Offer}
  A.,  {Bland-Hawthorn} J.,  {Baldry} I.~K.,  {Baugh} C.,  {Bridges} T.,
  {Cannon} R.,    et~al., 2002, \mnras, 334, 673

\bibitem[\protect\citeauthoryear{{Lintott}, {Schawinski}, {Bamford}, {Slosar},
  {Land}, {Thomas}, {Edmondson}, {Masters}, {Nichol}, {Raddick}, {Szalay},
  {Andreescu}, {Murray} \& {Vandenberg}}{{Lintott} et~al.}{2011}]{Lintott2011}
{Lintott} C.,  {Schawinski} K.,  {Bamford} S.,  {Slosar} A.,  {Land} K.,
  {Thomas} D.,  {Edmondson} E.,  {Masters} K.,  {Nichol} R.~C.,  {Raddick}
  M.~J.,  {Szalay} A.,  {Andreescu} D.,  {Murray} P.,    {Vandenberg} J.,
  2011, \mnras, 410, 166

\bibitem[\protect\citeauthoryear{{MacArthur}, {Ellis}, {Treu} \&
  {Moran}}{{MacArthur} et~al.}{2010}]{MacArthur2010}
{MacArthur} L.~A.,  {Ellis} R.~S.,  {Treu} T.,    {Moran} S.~M.,  2010, \apjl,
  709, L53

\bibitem[\protect\citeauthoryear{{MacArthur}, {Ellis}, {Treu}, {U}, {Bundy} \&
  {Moran}}{{MacArthur} et~al.}{2008}]{MacArthur08}
{MacArthur} L.~A.,  {Ellis} R.~S.,  {Treu} T.,  {U} V.,  {Bundy} K.,    {Moran}
  S.,  2008, \apj, 680, 70

\bibitem[\protect\citeauthoryear{{Macci{\`o}}, {Dutton} \& {van den
  Bosch}}{{Macci{\`o}} et~al.}{2008}]{Maccio2008}
{Macci{\`o}} A.~V.,  {Dutton} A.~A.,  {van den Bosch} F.~C.,  2008, \mnras,
  391, 1940

\bibitem[\protect\citeauthoryear{{Maller}, {Berlind}, {Blanton} \&
  {Hogg}}{{Maller} et~al.}{2009}]{Maller09}
{Maller} A.~H.,  {Berlind} A.~A.,  {Blanton} M.~R.,    {Hogg} D.~W.,  2009,
  \apj, 691, 394

\bibitem[\protect\citeauthoryear{{McIntosh}, {Rix} \& {Caldwell}}{{McIntosh}
  et~al.}{2004}]{McIntosh2004}
{McIntosh} D.~H.,  {Rix} H.-W.,  {Caldwell} N.,  2004, \apj, 610, 161

\bibitem[\protect\citeauthoryear{{Merritt}}{{Merritt}}{1984}]{Merritt1984}
{Merritt} D.,  1984, \apj, 276, 26

\bibitem[\protect\citeauthoryear{{Moore}, {Katz} \& {Lake}}{{Moore}
  et~al.}{1996}]{Moore96}
{Moore} B.,  {Katz} N.,  {Lake} G.,  1996, \apj, 457, 455

\bibitem[\protect\citeauthoryear{{Moore}, {Lake} \& {Katz}}{{Moore}
  et~al.}{1998}]{Moore98}
{Moore} B.,  {Lake} G.,  {Katz} N.,  1998, \apj, 495, 139

\bibitem[\protect\citeauthoryear{{Moore}, {Quinn}, {Governato}, {Stadel} \&
  {Lake}}{{Moore} et~al.}{1999}]{Moore99}
{Moore} B.,  {Quinn} T.,  {Governato} F.,  {Stadel} J.,    {Lake} G.,  1999,
  \mnras, 310, 1147

\bibitem[\protect\citeauthoryear{{Moorthy} \& {Holtzman}}{{Moorthy} \&
  {Holtzman}}{2006}]{Moorthy06}
{Moorthy} B.~K.,  {Holtzman} J.~A.,  2006, \mnras, 371, 583

\bibitem[\protect\citeauthoryear{{Moran}, {Loh}, {Ellis}, {Treu}, {Bundy} \&
  {MacArthur}}{{Moran} et~al.}{2007}]{Moran2007}
{Moran} S.~M.,  {Loh} B.~L.,  {Ellis} R.~S.,  {Treu} T.,  {Bundy} K.,
  {MacArthur} L.~A.,  2007, \apj, 665, 1067

\bibitem[\protect\citeauthoryear{{Mulchaey} \& {Zabludoff}}{{Mulchaey} \&
  {Zabludoff}}{1999}]{Mulchaey1999}
{Mulchaey} J.~S.,  {Zabludoff} A.~I.,  1999, \apj, 514, 133

\bibitem[\protect\citeauthoryear{{Muldrew}, {Croton}, {Skibba}, {Pearce},
  {Ann}, {Baldry}, {Brough}, {Choi}, {Conselice}, {Cowan}, {Gallazzi}, {Gray},
  {Gr{\"u}tzbauch}, {Li} et~al.,}{{Muldrew} et~al.}{2012}]{Muldrew2012}
{Muldrew} S.~I.,  {Croton} D.~J.,  {Skibba} R.~A.,  {Pearce} F.~R.,  {Ann}
  H.~B.,  {Baldry} I.~K.,  {Brough} S.,  {Choi} Y.-Y.,  {Conselice} C.~J.,
  {Cowan} N.~B.,  {Gallazzi} A.,  {Gray} M.~E.,  {Gr{\"u}tzbauch} R.,  {Li}
  I.-H.,    et~al., 2012, \mnras, 419, 2670

\bibitem[\protect\citeauthoryear{{Oemler} Jr.}{{Oemler}}{1974}]{Oemler1974}
{Oemler} Jr. A.,  1974, \apj, 194, 1

\bibitem[\protect\citeauthoryear{{Peletier} \& {Balcells}}{{Peletier} \&
  {Balcells}}{1996}]{Peletier96}
{Peletier} R.~F.,  {Balcells} M.,  1996, \aj, 111, 2238

\bibitem[\protect\citeauthoryear{{Peletier}, {Balcells}, {Davies},
  {Andredakis}, {Vazdekis}, {Burkert} \& {Prada}}{{Peletier}
  et~al.}{1999}]{Peletier99}
{Peletier} R.~F.,  {Balcells} M.,  {Davies} R.~L.,  {Andredakis} Y.,
  {Vazdekis} A.,  {Burkert} A.,    {Prada} F.,  1999, \mnras, 310, 703

\bibitem[\protect\citeauthoryear{{Peng}, {Lilly}, {Renzini} \&
  {Carollo}}{{Peng} et~al.}{2011}]{Peng2011}
{Peng} Y.,  {Lilly} S.~J.,  {Renzini} A.,    {Carollo} M.,  2011, preprint,
  (arXiv:1106.2546)

\bibitem[\protect\citeauthoryear{{Peng}, {Lilly}, {Kova{\v c}}, {Bolzonella},
  {Pozzetti}, {Renzini}, {Zamorani}, {Ilbert}, {Knobel}, {Iovino}, {Maier},
  {Cucciati}, {Tasca}, {Carollo}, {Silverman} et~al.,}{{Peng}
  et~al.}{2010}]{Peng2010}
{Peng} Y.-j.,  {Lilly} S.~J.,  {Kova{\v c}} K.,  {Bolzonella} M.,  {Pozzetti}
  L.,  {Renzini} m.,  {Zamorani} G.,  {Ilbert} O.,  {Knobel} C.,  {Iovino} A.,
  {Maier} C.,  {Cucciati} O.,  {Tasca} L.,  {Carollo} C.~M.,  {Silverman} J.,
   et~al., 2010, \apj, 721, 193

\bibitem[\protect\citeauthoryear{{Postman}, {Franx}, {Cross}, {Holden}, {Ford},
  {Illingworth}, {Goto}, {Demarco}, {Rosati}, {Blakeslee}, {Tran}
  et~al.,}{{Postman} et~al.}{2005}]{Postman05}
{Postman} M.,  {Franx} M.,  {Cross} N.~J.~G.,  {Holden} B.,  {Ford} H.~C.,
  {Illingworth} G.~D.,  {Goto} T.,  {Demarco} R.,  {Rosati} P.,  {Blakeslee}
  J.~P.,  {Tran} K.,    et~al., 2005, \apj, 623, 721

\bibitem[\protect\citeauthoryear{{Postman} \& {Geller}}{{Postman} \&
  {Geller}}{1984}]{Postman1984}
{Postman} M.,  {Geller} M.~J.,  1984, \apj, 281, 95

\bibitem[\protect\citeauthoryear{{Roediger}, {Courteau}, {MacArthur} \&
  {McDonald}}{{Roediger} et~al.}{2011}]{Roediger2011}
{Roediger} J.~C.,  {Courteau} S.,  {MacArthur} L.~A.,    {McDonald} M.,  2011,
  \mnras, 416, 1996

\bibitem[\protect\citeauthoryear{{Schawinski}, {Lintott}, {Thomas}, {Sarzi},
  {Andreescu}, {Bamford}, {Kaviraj}, {Khochfar}, {Land}, {Murray}, {Nichol},
  {Raddick}, {Slosar}, {Szalay}, {Vandenberg} \& {Yi}}{{Schawinski}
  et~al.}{2009}]{Schawinski2009}
{Schawinski} K.,  {Lintott} C.,  {Thomas} D.,  {Sarzi} M.,  {Andreescu} D.,
  {Bamford} S.~P.,  {Kaviraj} S.,  {Khochfar} S.,  {Land} K.,  {Murray} P.,
  {Nichol} R.~C.,  {Raddick} M.~J.,  {Slosar} A.,  {Szalay} A.,  {Vandenberg}
  J.,    {Yi} S.~K.,  2009, \mnras, 396, 818

\bibitem[\protect\citeauthoryear{{Schlegel}, {Finkbeiner} \&
  {Davis}}{{Schlegel} et~al.}{1998}]{SFD1998}
{Schlegel} D.~J.,  {Finkbeiner} D.~P.,  {Davis} M.,  1998, \apj, 500, 525

\bibitem[\protect\citeauthoryear{{Simard}, {Mendel}, {Patton}, {Ellison} \&
  {McConnachie}}{{Simard} et~al.}{2011}]{Simard2011}
{Simard} L.,  {Mendel} J.~T.,  {Patton} D.~R.,  {Ellison} S.~L.,
  {McConnachie} A.~W.,  2011, \apjs, 196, 11

\bibitem[\protect\citeauthoryear{{Skibba}}{{Skibba}}{2009}]{Skibba2009A}
{Skibba} R.~A.,  2009, \mnras, 392, 1467

\bibitem[\protect\citeauthoryear{{Skibba}, {Bamford}, {Nichol}, {Lintott},
  {Andreescu}, {Edmondson}, {Murray}, {Raddick}, {Schawinski}, {Slosar},
  {Szalay}, {Thomas} \& {Vandenberg}}{{Skibba} et~al.}{2009}]{Skibba2009}
{Skibba} R.~A.,  {Bamford} S.~P.,  {Nichol} R.~C.,  {Lintott} C.~J.,
  {Andreescu} D.,  {Edmondson} E.~M.,  {Murray} P.,  {Raddick} M.~J.,
  {Schawinski} K.,  {Slosar} A.,  {Szalay} A.~S.,  {Thomas} D.,    {Vandenberg}
  J.,  2009, \mnras, 399, 966

\bibitem[\protect\citeauthoryear{{Skibba} \& {Sheth}}{{Skibba} \&
  {Sheth}}{2009}]{SkibbaSheth2009}
{Skibba} R.~A.,  {Sheth} R.~K.,  2009, \mnras, 392, 1080

\bibitem[\protect\citeauthoryear{{Skibba}, {van den Bosch}, {Yang}, {More},
  {Mo} \& {Fontanot}}{{Skibba} et~al.}{2011}]{Skibba2010}
{Skibba} R.~A.,  {van den Bosch} F.~C.,  {Yang} X.,  {More} S.,  {Mo} H.,
  {Fontanot} F.,  2011, \mnras, 410, 417

\bibitem[\protect\citeauthoryear{{Smith}, {Treu}, {Ellis}, {Moran} \&
  {Dressler}}{{Smith} et~al.}{2005}]{Smith2005}
{Smith} G.~P.,  {Treu} T.,  {Ellis} R.~S.,  {Moran} S.~M.,    {Dressler} A.,
  2005, \apj, 620, 78

\bibitem[\protect\citeauthoryear{{Smith}, {Hudson}, {Lucey}, {Nelan} \&
  {Wegner}}{{Smith} et~al.}{2006}]{Smith2006}
{Smith} R.~J.,  {Hudson} M.~J.,  {Lucey} J.~R.,  {Nelan} J.~E.,    {Wegner}
  G.~A.,  2006, \mnras, 369, 1419

\bibitem[\protect\citeauthoryear{{Strateva}, {Ivezi{\'c}}, {Knapp},
  {Narayanan}, {Strauss}, {Gunn}, {Lupton}, {Schlegel}, {Bahcall}, {Brinkmann},
  {Brunner}, {Budav{\'a}ri} et~al.,}{{Strateva} et~al.}{2001}]{Strateva01}
{Strateva} I.,  {Ivezi{\'c}} {\v Z}.,  {Knapp} G.~R.,  {Narayanan} V.~K.,
  {Strauss} M.~A.,  {Gunn} J.~E.,  {Lupton} R.~H.,  {Schlegel} D.,  {Bahcall}
  N.~A.,  {Brinkmann} J.,  {Brunner} R.~J.,  {Budav{\'a}ri} T.,    et~al.,
  2001, \aj, 122, 1861

\bibitem[\protect\citeauthoryear{{Tanaka}, {Goto}, {Okamura}, {Shimasaku} \&
  {Brinkmann}}{{Tanaka} et~al.}{2004}]{Tanaka2004}
{Tanaka} M.,  {Goto} T.,  {Okamura} S.,  {Shimasaku} K.,    {Brinkmann} J.,
  2004, \aj, 128, 2677

\bibitem[\protect\citeauthoryear{{Tasca}, {Kneib}, {Iovino}, {Le F{\`e}vre},
  {Kova{\v c}}, {Bolzonella}, {Lilly}, {Abraham}, {Cassata}, {Cucciati},
  {Guzzo}, {Tresse}, {Zamorani}, {Capak}, {Garilli} et~al.,}{{Tasca}
  et~al.}{2009}]{Tasca2009}
{Tasca} L.~A.~M.,  {Kneib} J.-P.,  {Iovino} A.,  {Le F{\`e}vre} O.,  {Kova{\v
  c}} K.,  {Bolzonella} M.,  {Lilly} S.~J.,  {Abraham} R.~G.,  {Cassata} P.,
  {Cucciati} O.,  {Guzzo} L.,  {Tresse} L.,  {Zamorani} G.,  {Capak} P.,
  {Garilli} B.,    et~al., 2009, \aap, 503, 379

\bibitem[\protect\citeauthoryear{{Thomas}, {Maraston}, {Bender} \& {Mendes de
  Oliveira}}{{Thomas} et~al.}{2005}]{Thomas05}
{Thomas} D.,  {Maraston} C.,  {Bender} R.,    {Mendes de Oliveira} C.,  2005,
  \apj, 621, 673

\bibitem[\protect\citeauthoryear{{Thomas}, {Maraston}, {Schawinski}, {Sarzi},
  {Joo}, {Kaviraj} \& {Yi}}{{Thomas} et~al.}{2007}]{Thomas2007}
{Thomas} D.,  {Maraston} C.,  {Schawinski} K.,  {Sarzi} M.,  {Joo} S.-J.,
  {Kaviraj} S.,    {Yi} S.~K.,  2007, in {Vazdekis} A.,  {Peletier} R.~F.,
  eds, IAU Symposium Vol.~241 of IAU Symposium, {Environment and the epochs of
  galaxy formation in the SDSS era}.
pp 546--550

\bibitem[\protect\citeauthoryear{{Tonnesen} \& {Bryan}}{{Tonnesen} \&
  {Bryan}}{2009}]{Tonnesen2009}
{Tonnesen} S.,  {Bryan} G.~L.,  2009, \apj, 694, 789

\bibitem[\protect\citeauthoryear{{Trager}, {Faber} \& {Dressler}}{{Trager}
  et~al.}{2008}]{Trager2008}
{Trager} S.~C.,  {Faber} S.~M.,  {Dressler} A.,  2008, \mnras, 386, 715

\bibitem[\protect\citeauthoryear{{Trager}, {Faber}, {Worthey} \&
  {Gonz{\'a}lez}}{{Trager} et~al.}{2000}]{Trager2000}
{Trager} S.~C.,  {Faber} S.~M.,  {Worthey} G.,    {Gonz{\'a}lez} J.~J.,  2000,
  \aj, 120, 165

\bibitem[\protect\citeauthoryear{{Tremonti}, {Heckman}, {Kauffmann},
  {Brinchmann}, {Charlot}, {White}, {Seibert}, {Peng}, {Schlegel}, {Uomoto},
  {Fukugita} \& {Brinkmann}}{{Tremonti} et~al.}{2004}]{Tremonti2004}
{Tremonti} C.~A.,  {Heckman} T.~M.,  {Kauffmann} G.,  {Brinchmann} J.,
  {Charlot} S.,  {White} S.~D.~M.,  {Seibert} M.,  {Peng} E.~W.,  {Schlegel}
  D.~J.,  {Uomoto} A.,  {Fukugita} M.,    {Brinkmann} J.,  2004, \apj, 613, 898

\bibitem[\protect\citeauthoryear{{Treu}, {Ellis}, {Kneib}, {Dressler}, {Smail},
  {Czoske}, {Oemler} \& {Natarajan}}{{Treu} et~al.}{2003}]{Treu2003}
{Treu} T.,  {Ellis} R.~S.,  {Kneib} J.-P.,  {Dressler} A.,  {Smail} I.,
  {Czoske} O.,  {Oemler} A.,    {Natarajan} P.,  2003, \apj, 591, 53

\bibitem[\protect\citeauthoryear{{Tuffs}, {Popescu}, {V{\"o}lk}, {Kylafis} \&
  {Dopita}}{{Tuffs} et~al.}{2004}]{Tuffs2004}
{Tuffs} R.~J.,  {Popescu} C.~C.,  {V{\"o}lk} H.~J.,  {Kylafis} N.~D.,
  {Dopita} M.~A.,  2004, \aap, 419, 821

\bibitem[\protect\citeauthoryear{{van den Bosch}, {Aquino}, {Yang}, {Mo},
  {Pasquali}, {McIntosh}, {Weinmann} \& {Kang}}{{van den Bosch}
  et~al.}{2008}]{vandenBosch2008}
{van den Bosch} F.~C.,  {Aquino} D.,  {Yang} X.,  {Mo} H.~J.,  {Pasquali} A.,
  {McIntosh} D.~H.,  {Weinmann} S.~M.,    {Kang} X.,  2008, \mnras, 387, 79

\bibitem[\protect\citeauthoryear{{Weinmann}, {van den Bosch}, {Yang} \&
  {Mo}}{{Weinmann} et~al.}{2006}]{Weinmann2006}
{Weinmann} S.~M.,  {van den Bosch} F.~C.,  {Yang} X.,    {Mo} H.~J.,  2006,
  \mnras, 366, 2

\bibitem[\protect\citeauthoryear{{Weinzirl}, {Jogee}, {Khochfar}, {Burkert} \&
  {Kormendy}}{{Weinzirl} et~al.}{2009}]{Weinzirl09}
{Weinzirl} T.,  {Jogee} S.,  {Khochfar} S.,  {Burkert} A.,    {Kormendy} J.,
  2009, \apj, 696, 411

\bibitem[\protect\citeauthoryear{{Whitmore} \& {Gilmore}}{{Whitmore} \&
  {Gilmore}}{1991}]{Whitmore1991}
{Whitmore} B.~C.,  {Gilmore} D.~M.,  1991, \apj, 367, 64

\bibitem[\protect\citeauthoryear{{Worthey}}{{Worthey}}{1994}]{Worthey1994}
{Worthey} G.,  1994, \apjs, 95, 107

\bibitem[\protect\citeauthoryear{{Wyse}, {Gilmore} \& {Franx}}{{Wyse}
  et~al.}{1997}]{Wyse1997}
{Wyse} R.~F.~G.,  {Gilmore} G.,  {Franx} M.,  1997, \araa, 35, 637

\bibitem[\protect\citeauthoryear{{Yang}, {Mo}, {van den Bosch}, {Pasquali},
  {Li} \& {Barden}}{{Yang} et~al.}{2007}]{Yang2007}
{Yang} X.,  {Mo} H.~J.,  {van den Bosch} F.~C.,  {Pasquali} A.,  {Li} C.,
  {Barden} M.,  2007, \apj, 671, 153

\bibitem[\protect\citeauthoryear{{Zabludoff} \& {Mulchaey}}{{Zabludoff} \&
  {Mulchaey}}{1998}]{Zabludoff1998}
{Zabludoff} A.~I.,  {Mulchaey} J.~S.,  1998, \apj, 496, 39

\end{thebibliography}
\label{lastpage}

\appendix

%%classification appendix
\section{Classification of Galaxies}
\label{app:Class}

Below, we describe the galaxy classification scheme briefly outlined in \S\ref{sssec:classify}. We explain the classification of each type of galaxy, paying special attention to the probabilistic separation of ellipticals, classical bulge$+$disc galaxies, and pseudo-bulge$+$disc galaxies.

\figMags

\begin{enumerate}
\item {\it Bulge-less galaxies:} 
First, we identify galaxies which are 
best fit by a disc-only 
model. We assign them a probability of being fit by a single exponential
equal to one. These bulge-less galaxies are selected because they have
an de Vaucouleurs bulge-to-total luminosity ratio in the $r$ band
($(B/T)_r$) smaller 
than $10$ per cent,
or because there is no statistically significant improvement in
the fit by adding a bulge component. These galaxies account for $37$
per cent of the L12 sample ($26,882$ galaxies). These galaxies are not
strictly bulge-less, but  
include galaxies with bulges too small to detect using bulge$+$disc
decompositions. Fig.~\ref{fig:mags} shows that these galaxies are predominately
intrinsically (and apparently) faint, as expected for disc
galaxies with, at most, very small bulges. Because we use a
bright ($M_r < -19.77$) subset of galaxies from L12, bulge-less
galaxies are a smaller fraction ($28$ per cent) of our sample than of the whole L12 sample.

\item {\it Ellipticals and red disc galaxies: }
Next, we select a sample of elliptical galaxies. As above, galaxies
with $(B/T)_r>0.9$ are assumed to be disc-less elliptical
galaxies.  As discussed in L12 
\citep[see also][]{Allen06}, 2-dimensional bulge+disc models of
elliptical galaxies usually consist of a de Vaucouleurs profile and a
low surface brightness ``disc''. This model ``disc''
component has several possible origins: the outer halo of ellipticals,
a S\'ersic index larger than $4$, and/or inadequate sky subtraction
around bright galaxies in SDSS. This model ``disc'' component makes
elliptical galaxies indistinguishable from dustless, face-on, red
classical bulge$+$disc galaxies  based on photometry alone. {\rm Therefore, we must use other information to separate red ellipticals from face-on red classical bulge$+$disc galaxies. Below, we describe a method for separating face-on bulge$+$disc galaxies from ellipticals. We only apply this separation to galaxies with red colours for both the model de Vaucouleurs bulge and model disc: $u-r > 2.22$ \citep{Strateva01}. This colour cut ensures that the outer exponential halo around elliptical galaxies will not be star-forming. However, we cannot identify blue ellipticals in our sample using this colour cut. }
%In subsequent sections, we denote red classical bulge$+$disc galaxies
%as S0s. 

Despite their similarities, we can separate red bulge$+$disc galaxies and 
ellipticals  based on the
statistics of the modelled disc axis ratios, $q_d$. If the discs are randomly
oriented, then $q_d$ should be uniformly distributed for values larger
than the  disc scale height to
scale length ratio, $q_z$. The probability density function is given
by $f(q_d) = q_d/\sqrt{(q_d^2-q_z^2)(1-q_z^2)}$. We use $q_z=0.1$, which is
  smaller than the average value $q_z = 0.14 \pm 0.04$
  \citep{Kregel02}. However, any $q_z$ less than 0.2 has a negligible
  effect on the results. If 
the bulge$+$disc models are applied to ellipticals, then the
distribution of $q_d$ will be skewed to higher values, mimicking
face-on discs. Although it is impossible to distinguish a dustless,
face-on classical bulge$+$disc galaxy from an elliptical, we can assign each galaxy a
probability of being an elliptical or bulge$+$disc galaxy based on its model $q_d$.
\figSOEllip

Fig.~\ref{fig:S0E} shows the statistical separation of red bulge$+$disc and
ellipticals for galaxies divided into three stellar mass bins. The
upper panels show the 
distribution of $q_d$ for galaxies with red ($u-r>2.22$) model de Vaucouleurs 
bulge and exponential disc colours, and $0.1 < (B/T)_r < 0.9$. Each
distribution of axis 
ratios has two contributions, one  from disc galaxies, which follows the
distribution for randomly oriented discs, and 
one from ellipticals, which we model as a Gaussian distribution in
$q_d$ with an arbitrary centroid and width. We separately fit each
distribution of 
$q_d$ in Fig.~\ref{fig:S0E} with a linear combination of these two probability density
functions. Since small measured axis
ratios are often due to poor fits, we only use galaxies with
$q_d>0.25$ for the fitting. The fractional contribution from each function gives
the probability that a galaxy with a given $q_d$ is an elliptical
or a red classical bulge$+$red disc galaxy (lower panels Fig.~\ref{fig:S0E}). For low mass galaxies, the
distribution of axis ratios does not show any contribution from
ellipticals. At any stellar mass, a small value for $q_d$ implies a galaxy
has a real disc.  For $0.6 \lesssim  q_d
\lesssim 0.8$, the probability that a galaxy is an elliptical rises to
$\sim 50$ per cent for the highest mass galaxies. 

This method of
distinguishing between bulge$+$disc galaxies and ellipticals does not allow us to
definitively assign a galaxy to either category, but we can still
examine the properties of the whole sample, simply by weighting each
galaxy by its likelihood of being an red bulge$+$red disc galaxy or an elliptical. Counting all
galaxies with $B/T >0.9$ as ellipticals, we find that $9.4$ per cent
($6800$ galaxies) of the L12 sample are red bulge$+$red disc galaxies and $7.3$ per cent ($5300$ galaxies) are ellipticals.  

When selecting galaxies with red bulge and disc components for Fig.~\ref{fig:S0E}, we first
correct the bulge and disc colours for inclination in order to
eliminate edge-on, dusty galaxies. However, if the inclination
corrections are too large, there will be fewer galaxies with small
$q_d$, and the fraction of ellipticals will be
enhanced. Therefore, we iteratively adjust the inclination corrections
in $u$ and 
$r$ such that, after identifying elliptical galaxies, there is no
residual trend in corrected disc colour with disc inclination for red bulge$+$red disc 
galaxies. The final adjustments make the inclination corrections in
$u$ and $r$ smaller than those in L12, but the adjustments are minimal. They change 
the number of ellipticals in sample by less than $1$ per cent of the
total sample ($500$ galaxies). If we used no inclination correction,
the number of ellipticals in the sample would decrease by $1000$
galaxies.

\item {\it Classical and pseudo-bulges:}
After identifying ellipticals and red classical bulge$+$disc galaxies, we classify the remaining
galaxies as either classical bulge hosts or pseudo-bulge hosts. We
separate classical bulges and pseudo-bulges based on 
the age of the central stellar populations. \citet{Fisher06} show that
pseudo-bulges selected by morphology have higher central star
formation rates than classical bulges, in agreement with the notion
that pseudo-bulges form via secular processes in discs which drive gas
inwards and enhance the central star formation
\citep{Kormendy04}. Fig.~\ref{fig:classpseudo} shows the 
distribution of  
the $4000$\AA{} break strength for the classical and pseudo-bulge
galaxies.  These
values are taken from the MPA/JHU catalogue of measured spectroscopic
quantities from SDSS
\citep{Tremonti2004,SDSSDR82011}\footnote{http://www.mpa-garching.mpg.de/SDSS/DR7/raw\textunderscore
  data.html}. The value plotted is the narrow definition of the
$4000$\AA{} break, D$_n(4000)$ \citep{Balogh1999}. The SDSS
spectrograph uses $3$ arcsec fibres; 
%eighty per cent of  
%bulges have half-light radii smaller than $3$ arcsec, while only $16$
%per cent of discs have half-light radii smaller than $3$
%arcsec. 
two-thirds of the galaxies shown have a bulge-to-total flux ratio within 3
arcsec larger than $0.5$. Therefore, the fibre spectroscopic
quantities are typically dominated by  
the stellar light from the bulge. 

Fig.~\ref{fig:classpseudo} shows
there are  two populations of
bulges, those with recent star formation (within $\sim 1$Gyr) and
small $D_n(4000)$ and those without recent star formation and large
$D_n(4000)$. 
%D$_n(4000)$ correlates well with $B/T$ 
%ratio for the $n_b=4$ model; galaxies with $D_n(4000) < 1.6$ have a
%median $B/T$ of $zz$, while the remainder of the sample has a median
%of $yy$.
\figClassPseudo 
This suggests that $D_n(4000)$ can be used to separate
quiescent classical bulges from star-forming pseudo-bulges. We fit the distribution of
$D_n(4000)$ with two Gaussians and assign each galaxy a probability of
having a classical bulge or pseudo-bulge based on the ratio of the
Gaussians at a given $D_n(4000)$. Once again, an individual galaxy's
type is unknown, but the statistical properties of the whole
sample can be examined. Sixty per cent of the bulges in
Fig.~\ref{fig:classpseudo} are classical bulges. Classical bulge
galaxies make up $18$ per cent of the total L12 sample, while
pseudo-bulges represent $12$ percent.

The right panel of Fig.~\ref{fig:classpseudo} shows the distribution
of disc axis ratios for classical and pseudo-bulge host discs. This is one test of our separation of pseudo-bulges and classical bulges, as we
expect the distributions to match the flat distribution for randomly
oriented discs. Although the distribution of $q_d$ for classical bulge host
discs together with pseudo-bulge host discs is flat, the separate
distributions for classical and pseudo-bulge hosts are not. The pseudo-bulge
host discs have axis ratios which are too large. This is expected,
since pseudo-bulges are typically flattened, making them difficult to
detect in edge-on discs. Edge-on pseudo bulge hosts are more likely to be
considered bulge-less galaxies. 

The slight excess of low axis ratio discs
around classical bulges is due to the inclusion of inclined, but dust-poor, red bulge$+$red disc galaxies. When selecting red bulge$+$disc galaxies and ellipticals above, we included an
inclination correction, although we expect that some bulge$+$disc (and most 
ellipticals) will be essentially dust-free. Correcting these
galaxies for inclination shifts their colours  bluewards,
removing them from the red sample.  This is especially true
for highly-inclined (low $q_d$) galaxies, where the colour corrections
are largest.  Therefore, the
excess of small $q_d$ classical bulges 
galaxies consists mainly of red classical bulge$+$red disc galaxies. Since all these galaxies host classical bulges, it is not important to accurately separate them based on colour.

The physical origins of classical and pseudo-bulges are very
different. Classical bulges are thought to have formed by the same
mechanisms as ellipticals, while pseudo-bulges arise from secular
processes in discs (disc instabilities, bars, etc)
\citep[e.g.][]{Kormendy04}. Furthermore, the 
absolute magnitude distribution of pseudo-bulge galaxies in Fig.
\ref{fig:mags} is more similar to that of bulge-less and unclassifiable
galaxies than it is to classical bulge hosts. In this work, we will
consider galaxies with pseudo-bulges as a subset of bulge-less disc
galaxies. Because pseudo-bulges galaxies are relatively faint,
  they are a   small fraction ($8$ per cent) of the bright ($M_r \lesssim
  -19.77$)   subsample from L12  used below. Therefore, the   exclusion of pseudo-bulge hosts from bulge$+$disc
  galaxies does   significantly affect our results.

\item {\it Unclassifiable galaxies:}
The unclassifiable galaxies in our sample are modelled by a single
S\'ersic profile. These are galaxies which are not well fit by any of
the other models, i.e. they have model bulges which are
larger than their discs, or they have fluxes in $g$, $r$, or $i$ consistent
with zero for the bulge or disc (or both). They are given a
probability of being fit by a S\'ersic profile equal to unity. Of the
$72658$ galaxies in L12, $12459$ ($17$ per cent) are deemed 
unclassifiable. The average unclassifiable galaxy is 
$0.4$ magnitudes fainter than the average galaxy in the sample (see
Fig.~\ref{fig:mags}). Seventy-five percent of unclassifiable galaxies
have a S\'ersic  index less than $2.3$ and the same fraction 
lie in the blue cloud ($u-r<2.22$). These galaxies are probably
disc-like irregulars, which are unlikely to have a
well-defined bulge and disc. The remaining $25$ per cent of
unclassifiable galaxies are mostly merger remnants, starbursts, and other
complicated morphologies. None the less, because the majority of
unclassifiable galaxies exhibit 
disc-like properties, we group them with other
bulge-less galaxies described above. Since the sample used here is a bright subsample of the L12 sample, the unclassifiable
galaxies make up a smaller fraction (less than $10$ per cent) of the
bright subsample than of the whole L12 sample.

\end{enumerate}

\end{document}